\documentclass[11pt]{article}
\pdfoutput=1
\usepackage{physics}
\usepackage{graphicx}
\usepackage{amsthm, amssymb}
\usepackage[UKenglish]{babel}
\usepackage{listings}
\usepackage{amsfonts}
\usepackage{comment}
\usepackage{float}
\usepackage{enumitem}
\usepackage{appendix}
\usepackage{url}
\usepackage{relsize}
\usepackage{tabularx}
\usepackage{geometry}
\usepackage{subcaption}
\usepackage{algorithm}
\usepackage{tikz}
\usepackage{authblk}
\usepackage{algpseudocode}
\usepackage{bm}
\usepackage{multirow}
\usepackage{makecell}
\usepackage{cleveref}
\graphicspath{ {imageslatex/} }
\usetikzlibrary{calc,tikzmark}
\usetikzlibrary{arrows.meta}

\newcount\Comments  
\usepackage[authoryear]{natbib}
\graphicspath{ {Figures/}}
\begin{document}

\title{Multi-resolution Dynamic Mode Decomposition for \\ Damage Detection in Wind Turbine Gearboxes}

\author[1]{Paolo Climaco}
\author[1,2]{Jochen Garcke}
\author[2]{Rodrigo Iza-Teran}
\affil[1]{Institut f\"ur Numerische Simulation, Universit\"at Bonn, Bonn, Germany }
\affil[2]{Fraunhofer SCAI, Sankt Augustin, Germany}
\maketitle

  \abstract{
    We introduce an approach for damage detection in gearboxes based on the analysis of sensor
  data with the multi-resolution dynamic mode decomposition (mrDMD). The
  application focus is the condition monitoring of wind turbine gearboxes under
  varying load conditions, in particular irregular and stochastic wind
  fluctuations. We analyze data stemming from a simulated vibration response of
  a simple nonlinear gearbox model in a healthy and damaged scenario and under
  different wind conditions. With mrDMD applied on time-delay snapshots of the
  sensor data, we can extract components in these vibration signals that
  highlight features related to damage and enable its identification. A
  comparison with Fourier analysis, Time Synchronous Averaging and Empirical Mode Decomposition shows the
  advantages of the proposed mrDMD-based data analysis approach for damage
  detection.
    }

\section{Introduction}
\label{the problem}
The condition monitoring of wind turbine gearboxes presents a challenging
scenario that is often not amenable to classical data analysis techniques due to
the existing operating conditions. Note that wind turbines are one of the main
sources of renewable energy~\citep{euro} nowadays and gearbox-related defects
are among those problems with the highest amount of downtime hours per
failure~\citep{Dao2019}. Therefore, damage detection in gearboxes is a
significant aspect for keeping wind energy production high. To cope with the
difficulties in the sensor signals, arising due to the prevailing wind
fluctuations, we propose an approach for damage detection in gearboxes based on
the analysis of sensor data with the multi-resolution dynamic mode decomposition
(mrDMD). 

Sensor data analysis is widely used to determine the health condition of mechanical devices. 
Analyzing the data generated by sensors installed on
these devices, it is possible to continually monitor their integrity and detect
anomalous behaviours, damages, and faults when they arise. Optimizing the
monitoring process is of great interest because detecting damages in their early
stage empowers companies to prevent more severe problems that could lead to a 
significant loss in terms of energy production and money. 

Damage detection in gearbox vibration signals is a relatively easy task in
steady operating conditions, as for example in a laboratory, see~\cite{esta} for an overview of established methods. Unfortunately, when we deal
with wind turbines, steady operating conditions represent an unrealistic
scenario that does not take into account weather condition, temperature
variation, and, most importantly, wind turbulence, all of which results in varying load
conditions on the blades of the wind turbine. These factors strongly affect gearbox
vibration signals, introducing a stochastic component that makes them  non-deterministic and far from smooth. 
In this scenario damage detection is a much more difficult task since the data analysis strategies 
have not only to deal with signal anomalies induced by damage, but also by wind turbulence.

In this work, we consider the gearbox model and the scenario developed
in~\cite{Kah} and further refined in~\cite{anto}. The numerically produced
signals include a stochastic component simulating the varying load condition
caused by wind turbulence. Several kinds of damage can appear in gearboxes, here
we focus on the crack of a gear's tooth. The crack of a gear's tooth is a
bending fatigue failure. Bending fatigue failures usually occur in three
progressive stages: crack initiation, crack propagation and tooth fracture or
breakage \citep{Robert2002}. According to \cite{Me2014} ``Tooth fracture is the most serious fault in
gearbox and it may cause complete failure of the gear". Thus, the crack of a
gear's tooth is a damage in its early stage because the continuous interaction
of the cracked tooth with the other gear leads to a progressive propagation of
the crack until the breakage of the tooth occurs, leading to more severe damage
that may even cause the total failure of the gear. 
 
Historically speaking, fault detection in gearboxes was first performed using
methods that focus on analyzing the spectral properties of sensor
signals~\citep{Kumar2021}, such as the Fast Fourier transform~\citep{Rao2011}
and Cepstrum analysis~\citep{Bada2004}.  However, in the scenario we consider,
these methods are strongly challenged by weather conditions, such as wind
turbulence and temperature variation, which induce a stochastic component to the
signals studied, changing their spectra non-deterministically over time. This
stochastic change of the signals' spectral properties strongly affects the
effectiveness and reliability of spectral methods~\citep{Kumar2021}. 
\textit{Empirical mode decomposition} (EMD)~\citep{nor} was proposed as an
alternative to overcome the issues arising when studying spectral
characteristics of signals. In the case of a damaged gearbox, EMD separates the
time-domain signals in several modes that may contain damage features and may be
used to detect faults. Unfortunately, EMD is affected by the so called
``\textit{mode mixing problem}". The problem consists of the fact that the modes
produced by EMD either contain information of the signal relative to widely
disparate time-scales or similar time-scale information resides in different
modes~\citep{hua, gui}. The main implications are that the physical meaning of
each mode is unclear and that a proper interpretation of the information they
contain is not a trivial task. This does not mean that such modes cannot be
studied or selected according to some specific strategy in order to extract
meaningful insights about the data, it rather means that there is no principled
way to do that.

In this work we propose multi-resolution dynamic mode decomposition
(mrDMD)~\citep{mrDMD}, a variation of dynamic mode decomposition
(DMD)~\citep{schmidt}, as an alternative to EMD that incorporates all its
advantages and overcomes the mode mixing problem as well. mrDMD is a data-driven
algorithm that has been used to perform different tasks, such as:
spatio-temporal filtering of video or multi-scale separation of complex weather
systems~\citep{mrDMD}. We propose it as a novel approach to detect the crack of
a tooth in a wind turbine gearbox by analyzing gearbox vibration signals
generated under varying load conditions. mrDMD is able to decompose the signal
in spatio-temporal modes that capture its geometrical characteristics in the
time-domain at different time-scales. The difference to EMD is that it
associates to each of its modes a complex number that could either represent a
frequency or an energy content, according to the point of view one wants to
consider, and thanks to this additional information high/low-frequency or energy
content separation of the modes is possible. On the one hand, low-frequency
modes represent those structures that vary slowly over time, which therefore are not
affected by the non-deterministic variation caused by wind turbulence that mainly
influences the signal at smaller time-scales. On the other hand, high-frequency
modes represent those structures that evolve at higher frequencies and are
strongly affected by wind turbulence and sudden changes in the gear stiffness
caused by the cracked tooth we want to identify. We will see that computing the
instantaneous amplitude of high-frequency modes, via the Hilbert Transform (HT),
will enable us to easily identify anomalies arising in vibration signals from
the crack of a gear's tooth.
 
We aim to show the potential of mrDMD as a fault detection method,
in a context in which the analyzed vibration signals are affected by stochastic
perturbations. We study the proposed approach on simulated signals representing the vibration response
of a simple gearbox model with two spur gears, as well on experimental data under controlled situations from a benchmark study.

The next sections are organized as follows: Section~\ref{related_work}
presents an overview of related work on the topic of early damage detection in
wind turbine gearboxes under varying load condition. Section~\ref{multiresolution_dmd} introduces our mrDMD based strategy for tooth damage
identification. In Section~\ref{gearboxmodelll} the gearbox model is
introduced and numerical simulations of acceleration signals representing the
vibration response of the modelled gearbox are considered. Finally, Section
$\ref{faultdetection}$ demonstrates the performance of our method, and shows the
issues the fast Fourier transform as a spectral
method, an approach based on Time Synchronous Averaging, and an
EMD-based approach experience by the presence of wind fluctuations. 

\section{Related work}
~\label{related_work}
Generally speaking, condition monitoring methods consist of damage detection
methods, with the difference that one knows a priori about the frequency bands
related to damage for the different monitored components. Reviews of methods
used for damage detection in wind turbines can be found in~\cite{esta,
Kumar2021, Salameh2018, Sharma2021}. The most classical techniques used for
condition monitoring, and more in general for damage detection, mainly focus on
analyzing the spectral characteristics of vibration signals, such as the
spectral Kurtosis~\citep{Antoni2006}, Fourier analysis and modulation
sidebands~\citep{Inal2010} and Cepstrum analysis~\citep{Bada2004}. These methods
have proven their capability to provide good results in identifying damages
under simplified conditions, such as steady loading of the gearboxes. However,
when considering time-varying load conditions, the spectral characteristics of
gearbox vibration signals change non-deterministically with time. This
stochastic behaviour can not be handled using Fourier analysis as the Fourier
transform expands a signal as a linear combination of wave functions with
constant frequency over time. To overcome this issue, many time-frequency
methods have been employed in damage detection, such as: the Wigner–Ville
distribution~\citep{wig}, the Vold–Kalman filter~\citep{Feng2019}, wavelet
analysis~\citep{wav} and cyclostationary analysis~\citep{Dia,Xin2020}.  
Probably, the most popular and successful of these methods is wavelet
analysis~\citep{ Hu2018, Peng}. Nonetheless, in the context we are considering,
such a method has the disadvantage of relying on a fixed set of basis functions.
This fact has a detrimental effect on its effectiveness and reliability in
processing signals affected by perturbations of stochastic nature and that have
spectral properties that change randomly with time. An additional approach to
perform fault detection is given by entropy analysis~\citep{Sharma2016}.
However,  many entropy features are available for different types of faults, and
it is not clear what are the right features for a given fault. Moreover, under
varying load condition, the stochastic behaviour of the signal affects the
performance of entropies which may show insignificant results, hence making them
non-reliable~\citep{Sharma2021, Sharma2016}. Another major contribution to the
gear diagnostics filed is Time Synchronous Averaging (TSA) \citep{Randall2021}. TSA is used to
extract signal components associated with certain frequencies, as the meshing
frequency, that may highlight damage features. Unfortunately, also this
technique rely on being able to analyse signals recorded under steady speed and
load conditions and it is strongly affected by the fluctuating operating
conditions caused by wind turbulence.
 
Other strategies for damage detection, which try to overcome the
problems experienced with TSA, entropy analysis and time-frequency methods, are based on the
empirical mode decomposition (EMD)~\citep{nor}. The main advantage of
using EMD with respect to entropy analysis and time-frequency methods is that it decomposes the
signal into several components, or modes, which may contain meaningful information
related to damage. After decomposing the vibration signal, the instantaneous
frequency and amplitude of each component can be estimated, most commonly by
applying the Hilbert Transform (HT), in order to extract features of the signal
that allow a better determination of whether there is damage or not.
The decomposition's procedure is based on local geometric characteristics of the
signal in the time-domain, and it acts at different time-scales. It is
completely data driven and there is no need of an a priori choice of a set of
functions or a mother wavelet. Moreover, EMD-based methods for machinery damage
identification do not rely on any kind of a priori chosen window function or any
assumption about the regularity of the signal for any time-span~\citep{man,
Dfen, jun}. As a consequence of that, these methods lead, in some cases, to a
better estimation of even small variations in the signals. In general, EMD-based
methods have the promise of high-quality results at a low computational cost.
In~\cite{anto} an EMD-based strategy is developed to  identify the crack of a
gear's tooth in a scenario that, similar to this work, takes into account the
effects of the wind turbulence.  

Unfortunately, EMD-based strategies have some limitations too. The major problem
is the mode mixing problem studied in~\cite{hua} and further
analyzed in the wind turbine damage detection context in~\cite{anto}. The mode
mixing problem consists in the fact that each of the modes obtained after the
signal's decomposition, either contains information of the signal relative to widely
disparate time-scales, or similar time-scale information  resides in different
modes~\citep{hua, gui}. In recent years work has been carried out to provide a
primary theoretical framework for the development of EMD. Despite the steps
forward that give us a  theoretical understanding of the algorithm when it is employed
in a simplified scenario, e.g., simple signals with only a pure oscillation
component~\citep{Ge2018}, there is still no full mathematical justification for the general
application of the EMD procedure, as it was already pointed out in~\cite{anto}. This further affects the interpretability
and reliability of the method.
 
In this work we propose mrDMD as an alternative to EMD. An mrDMD-based procedure
for damage detection in wind turbines was already developed in~\cite{Dang2018}.
However, in~\cite{Dang2018}, the authors focused on the rolling bearing fault rather than the crack
of a tooth, and they did not consider turbulent wind conditions as we do here.
Moreover, there is a fundamental difference between the method we develop and
the method they proposed: they analyze the spectral characteristics of the
signals' structures associated with the slow modes computed via mrDMD,
while here, we focus on studying the information contained in high-frequency
modes in their time-domain, avoiding all the issues related to
examining the spectra of signals generated under turbulent wind conditions.
This difference between the two approaches can be related to the different
nature of the damages analyzed. On the one hand, we have the crack of a tooth
that primarily affects gear mesh frequency and high order
harmonics of the signals~\citep{anto, Alf2012}. On the
other hand, the rolling bearing fault affects different frequency bands,
including the low-frequencies~\citep{Hu2019}.

\section{Multi-resolution dynamic mode decomposition for damage detection}
\label{multiresolution_dmd}
The DMD algorithm was developed in the fluid dynamic community for the purpose
of feature extraction~\citep{schmidt}. Since its creation numerous variants of
the original algorithm have been developed~\citep{ergoticDMD,HODMD,cDMD,
debiasingDMD, leastDMD}. Moreover, DMD has been applied in several contexts, such as
finance~\citep{finaDMD}, epidemiology~\citep{infectDMD}, highway traffic
forecasting~\citep{trafficDMD} and biomedical informatics~\citep{HealthDMD}.
It is important to mention that DMD  has a solid theoretical interpretation through Koopman
operator theory~\citep{onconv}, which makes the method understandable and
interpretable. We now introduce the basics of DMD.
 
\subsection{Dynamic mode decomposition}
Let $\mathbf{y}(t_i)\in \mathbb{R}^n$ be a column vector containing data from sensors,
evaluated at $n$ different points, at time $t_i \in \mathbb{R}^+$. Assuming we have vectors
representing $T+1$  consecutive equispaced time steps, we arrange them in two matrices:
 \begin{equation}
\label{snapshots_wind}
 \mathbf{Y}=[\mathbf{y}(t_0),\mathbf{y}(t_1),\dots,\mathbf{y}(t_{T-1})] \quad \text{and} \quad \mathbf{\bar{Y}}=[\mathbf{y}(t_1),\mathbf{y}(t_2),\dots,\mathbf{y}(t_T)]  .
\end{equation}
These matrices are  the so called snapshots matrices and each one of their
columns represents sensor values at a certain time step. The columns are
arranged chronologically and those of $\mathbf{\bar{Y}}$ are shifted one time step
forward in the future with respect to those of $\mathbf{Y}$. Notice that, in this work, we always assume that the snapshots
we analyze are taken at equispaced intervals in time, i.e., $ \Delta t = t_{i+1}-t_i$ for each $i=0,1,\dots,T-1$ with $\Delta t \in \mathbb{R}^+$.

Dynamic mode decomposition relies on the assumption that there exists a matrix
$\mathbf{A}\in \mathbb{R}^{n,n}$ such that
\begin{equation}
\mathbf{y}(t_{i+1})=\mathbf{A}\mathbf{y}(t_{i}) \quad \text{for} \;i=0,1,\dots,T-1,
\end{equation}
which written in a matrix form yields 
\begin{equation}
\mathbf{\bar{Y}}=\mathbf{AY}.
\end{equation}
The algorithm's goal is to find the matrix $A$ that best  maps $\mathbf{y}(t_{i})$ into
$\mathbf{y}(t_{i+1})$, for $i=0,1,\dots,T-1$, by solving the following minimization problem in
the Frobenius norm
\begin{equation}
\label{minporb_wind}
\min_{\mathbf{A}}\|\mathbf{\bar{Y}}-\mathbf{AY}\|_F.
\end{equation}
Thus, the matrix $A$, solution of (\ref{minporb_wind}), is the matrix that
best advances the sensor values in time, therefore best maps $\mathbf{Y}$ into $\mathbf{\bar{Y}}$. The
solution of the minimization problem (\ref{minporb_wind}) is given by the
matrix
\begin{equation}
\label{invetransowind}
\mathbf{A}=\mathbf{\bar{Y}Y^{\dagger}},
\end{equation}
where $\mathbf{Y^{\dagger}}$ is the Moore-Penrose inverse of the matrix $\mathbf{Y}$. However, for
computational reasons connected to the data's dimension, computing the matrix
$\mathbf{A}$ using Formula (\ref{invetransowind}) becomes computationally difficult and
unstable~\citep{dmd}. Instead, a low-rank approximation $\mathbf{\tilde{A}}$ of $\mathbf{A}$ is
computed. The eigenvalues and eigenvectors of $\mathbf{\tilde{A}}$ are used to obtain an
approximation of the eigenvalues and eigenvectors of $\mathbf{A}$, called DMD eigenvalues and
DMD modes, respectively. 
\begin{algorithm}[t]
  \caption{Dynamic Mode Decomposition (DMD)}\label{alg:DMD}
\begin{algorithmic}[1]

    \State Arrange the data into matrices $\mathbf{\bar{Y}}$ and $\mathbf{Y}$ as in
    (\ref{snapshots_wind}). \State Compute the reduced singular value
    decomposition (SVD) of $\mathbf{Y}$, i.e., $\mathbf{Y}\approx \mathbf{U}_r \mathbf{\Sigma}_r \mathbf{V^*}_r$. Here
    $\mathbf{U}_r$, $\mathbf{\Sigma}_r$ and $ \mathbf{V^*}_r$ are the rank-$r$ truncation of the
    matrices $\mathbf{U},\; \mathbf{\Sigma}$ and $ \mathbf{V^*}$ computed via SVD and such that  $\mathbf{Y}= \mathbf{U
    \Sigma V^*}$. Further, $\mathbf{V^*}$ is the conjugate transpose of $\mathbf{V}$ and $r$ is
    a parameter to be chosen. \State Construct the matrix $\mathbf{\tilde{A}}:=
    \mathbf{U^*}_r\mathbf{\bar{Y}V}_r \mathbf{\Sigma}^{-1}_r$. \State Compute eigenvalues and eigenvectors
    of $\mathbf{\tilde{A}}$, solving $\mathbf{\tilde{A}W}= \mathbf{\Lambda W}$. With $\mathbf{\Lambda}$ a diagonal
    matrix of DMD eigenvalues and $\mathbf{W}$ a matrix where the columns are eigenvectors
    of $\mathbf{\tilde{A}}$. \State The DMD modes corresponding to DMD eigenvalues
    $\mathbf{\Lambda}$ are given by the columns of the matrix $\mathbf{\Phi}= \mathbf{\bar{Y}V}_r \mathbf{\Sigma}_r^{-1}\mathbf{W}$.
  
    \State The reconstruction of the dynamics is given  by the linear evolution
    \begin{equation}
    \label{prediction_wind}
    \mathbf{\hat{y}}(t_{i})=\sum_{k=1}^M b_k\pmb{\phi}_k   \text{exp}(\omega_k t_{i}),
    \quad i=0,\dots,T,
    \end{equation}
    where $M$ is the number of computed DMD modes, which depends on  the size of
$\mathbf{\tilde{A}}$, $\phi_k$ is the $k$-th DMD mode, $b_k$ is the $k$-th entry
of the vector $\mathbf{b}= \mathbf{\Phi}^{\dagger}\mathbf{y}(t_{0})$ and
$\mathbf{\hat{y}}(t_{i})$ is the reconstruction of the snapshot $\mathbf{y}(t_{i})$.
Moreover, $\omega_k$ is defined as the frequency value associated with the
$k$-th DMD mode and obtained via the DMD eigenvalue $\lambda_k$:
\begin{equation}
  \omega_k= \log(\lambda_k)/\Delta t.
  \label{def:omega_frequency}
\end{equation} 
Here, $\Delta t$ represents
the time difference between two consecutive measurements. Since in this work we
only consider equispaced measurements we have that $\Delta t = t_{i+1}-t_{i}$ for
$i=0,1,\dots,T-1$.

\end{algorithmic}
\end{algorithm}

Algorithm~\ref{alg:DMD} shows the main numerical steps of the DMD. The first
important observation is that the temporal evolution of the analyzed signal can
be reconstructed as a weighted sum of the DMD modes, as it can be seen
in (\ref{prediction_wind}). Thus, DMD modes are signal's components that encode
meaningful spatio-temporal information. Notice that, since the DMD eigenvalue
$\lambda_k$ is a complex number, in (\ref{prediction_wind}) the weight
$\text{exp}(\omega_k t_i)$, associated to the $k$-th DMD mode, evolves
periodically with the index `` $i$ ". Consequently, each DMD mode contribution to
the dynamic's reconstruction changes periodically over time with oscillation
frequency and growth/decay rate determined by its associated DMD eigenvalue.
Furthermore, we can associate DMD modes with a concept of speed by looking at
the frequencies associated with them. Given two DMD modes
$\pmb{\phi}_i,\pmb{\phi}_j, \; i \neq j$, we say that $\pmb{\phi}_i$ is faster
than $\pmb{\phi}_j$ (or equivalently $\pmb{\phi}_j$ is slower than
$\pmb{\phi}_i$) if $| \text{exp}(\omega_i)|>| \text{exp}(\omega_j)| $.
On the one hand, slow modes represent those signal's components whose contribution to
the signal's reconstruction vary slowly over time~\citep{mrDMD}. On the other hand, fast modes
represent those structures whose contribution to the signal's reconstruction
vary with higher frequency.

Notice that, DMD operates at a single time-resolution, i.e., DMD modes and
eigenvalues  fully characterize the signal's evolution over all the given
sampling window, as shown in (\ref{prediction_wind}). 
However, we aim to study signals that are locally affected by stochastic
perturbations, and that can show a very different behavior at different
time-scales. Thus, we aim to use an improved DMD that enable us to sift out
information at different time-scales.
 
\subsection{Multi-resolution dynamic mode decomposition}
Multi-resolution dynamic mode decomposition (mrDMD)~\citep{mrDMD} is a variation
of DMD, and it produces modes that extract spatial features at different
time-scales. mrDMD basically consists of iteratively applying the DMD algorithm
at different time-ranges. To be more precise, it starts by analyzing the largest
sampling window, which consists of all the available snapshots. After that
it computes DMD modes, identifies the slow ones and extracts them. Subsequently, it
reduces the duration of the observation window by half, determines again slower
modes in each half and extracts those modes from the residual signal. mrDMD
iterates this process until the desired resolution or level of decomposition is
reached. The final residual signal will be composed of fast modes representing
spatial structures that characterize the signal locally, with a time-resolution
determined by the number of iterations performed in the mrDMD algorithm. We now
shortly recapture the formulation of mrDMD from~\cite{mrDMD}, where  
further details can be found. 

In its first iteration, mrDMD reconstructs the time-domain signal  as follows
\begin{equation}
\label{multideco_wind}
\mathbf{\hat{y}}(t_{i})=\sum_{k=1}^M b_k\pmb{\phi}_k   \text{exp}(\omega_k t_i)=\underbrace{\sum_{k=1}^{m_1} b_k\pmb{\phi}_k^{(1)} \text{exp}(\omega_k t_i)}_{\text{slow modes}} +\underbrace{\sum_{k=m_1+1}^M b_k\pmb{\phi}_k^{(1)}   \text{exp}(\omega_k t_i)}_{\text{fast modes}},
\end{equation}
where $\mathbf{\phi}_k^{(1)}$ represent the modes computed using the entire set of
$T+1$ snapshots, $M$ is the total number of modes and $m_1$ the number of slow
modes at the first iteration level. The first sum in the right-hand side of (\ref{multideco_wind})
represents the slow-modes dynamics, whereas the second sum is everything else. 
At the second iteration level, DMD is now performed after a split of the second sum
\begin{equation}
  \mathbf{Y}_{(T+1)/2}= \mathbf{Y}_{(T+1)/2}^{(1)}+\mathbf{Y}_{(T+1)/2}^{(2)},
\end{equation}
where the columns of the matrices $\mathbf{Y}_{(T+1)/2}^{(1)},\mathbf{ Y}_{(T+1)/2}^{(2)}$
represent the fast modes reconstruction of the first $(T+1)/2$ and last $(T+1)/2$
snapshots, respectively. The iteration process continues by recursively removing
slow frequency components computed separately on each half of the snapshots. One
builds the new matrices $\mathbf{Y}_{(T+1)/2},\;\mathbf{Y}_{(T+1)/4},\;\mathbf{Y}_{(T+1)/8},\dots$ until a
suitable multi-resolution decomposition has been achieved.
 
The representation (\ref{multideco_wind}) can be made more precise. Specifically, one
must account for the number of levels $(L)$ of the decomposition, the number of
time bins $(J)$ for each level, and the number of modes retained at each level
$(m_L)$. Thus, the solution is parametrized by the following three indices
\begin{subequations}
  \label{parameters_mrdmd}
\begin{align}
& l=1,2,\dots,L: \; \text{number of decomposition levels,}\\
& j=1,2,\dots,J:\; \text{number of time bins per level}(J=2^{(l-1)}),\\
& k=1,2,\dots,m_L : \; \text{number of modes extracted at level L.}
\end{align}
\end{subequations}
To formally determine the reconstructed snapshots $\hat{y}(t_{i})$, the following
indicator function is defined:
\begin{equation}
\label{indicator}
    f^{l,j}(t):= \left\{\begin{array}{lr}
        1, & \text{for }t \in[t^{l}_j,t^{l}_{j+1}]\\
        0, & \text{elsewhere } \\
       
        \end{array}\right\}, \; \text{with} \; j=1,2,\dots,J \; \text{and} \; J=2^{(l-1)},
\end{equation}
where $t^{l}_j, t^{l}_{j+1} \in \mathbb{R}^+$ determine 
the interval of the $j$-th time bin at the $l$-th decomposition level.
The above indicator function is only nonzero in the interval, or time bin,
associated with the value of $j$. The three indices and the indicator function in (\ref{indicator}) are used to
formally write equation (\ref{multideco_wind}) as
\begin{equation}
\label{formal_version}
\mathbf{\hat{y}}(t_{i})= \sum_{l=1}^L \sum_{j=1}^J\sum_{k=1}^{m_L} f^{(l,j)}(t_i)b_{k}^{(l,j)}\pmb{\phi}_{k}^{(l,j)}\exp(\omega_{k}^{(l,j)} t_i ).
\end{equation}
\begin{figure}
    \begin{tikzpicture}[scale=0.8]
    
      \draw[black,  thick] (0,-0.2) rectangle (8,0.5) node[pos=0.5]{$\pmb{\phi}_k^{1,1}$};
      \node[below] at (4,-0.1){Time ($t$)};
      \draw[black,  thick] (0,0.5) rectangle (4,1.5)node[pos=0.5]{$\pmb{\phi}_k^{2,1}$};
      \node[ rotate=90] at (-0.3, 2.5) {Frequency ($\omega$)};
      \draw[black,  thick] (4,0.5) rectangle (8,1.5)node[pos=0.5]{$\pmb{\phi}_k^{2,2}$};
      \draw[black,  thick] (0,1.5) rectangle (2, 3)node[pos=0.5]{$\pmb{\phi}_k^{3,1}$};
      \draw[black,  thick] (2,1.5) rectangle (4,3)node[pos=0.5]{$\pmb{\phi}_k^{3,2}$};
      \draw[black,  thick] (4,1.5) rectangle (6, 3)node[pos=0.5]{$\pmb{\phi}_k^{3,3}$};
      \draw[black,  thick] (6,1.5) rectangle (8,3)node[pos=0.5]{$\pmb{\phi}_k^{3,4}$};
      \draw[black,  thick] (0,3) rectangle (1, 5.5)node[pos=0.5]{$\pmb{\phi}_k^{4,1}$};
      \draw[black,  thick] (1,3) rectangle (2,5.5)node[pos=0.5]{$\pmb{\phi}_k^{4,2}$};
      \draw[black,  thick] (2,3) rectangle (3, 5.5)node[pos=0.5]{$\pmb{\phi}_k^{4,3}$};
      \draw[black,  thick] (3,3) rectangle (4,5.5)node[pos=0.5]{$\pmb{\phi}_k^{4,4}$};
      \draw[black,  thick] (4,3) rectangle (5, 5.5)node[pos=0.5]{$\pmb{\phi}_k^{4,5}$};
      \draw[black,  thick] (5,3) rectangle (6,5.5)node[pos=0.5]{$\pmb{\phi}_k^{4,6}$};
      \draw[black,  thick] (6,3) rectangle (7, 5.5)node[pos=0.5]{$\pmb{\phi}_k^{4,7}$};
      \draw[black,  thick] (7,3) rectangle (8,5.5)node[pos=0.5]{$\pmb{\phi}_k^{4,8}$};
      \node[text width=15cm,  anchor=west, right] at (12,3)
      {\LARGE$\pmb{\phi}_{\tikzmark{a}k}^{(\tikzmark{b}l,\tikzmark{c}j)}$};
   
\end{tikzpicture}

\begin{tikzpicture}[overlay,remember picture, scale=0.8]
  \draw[-{Stealth[length=3mm, width=2mm]}] 
  ( $ (pic cs:a) +(14pt,-5.5ex) $ ) -- 
  ( $ (pic cs:a) +(5pt,-0.5ex) $ );
  \node[anchor=north]
  at ( $ (pic cs:a) +(9pt,-6ex) $ )
  {mode number of level $l$};
  \draw[-{Stealth[length=3mm, width=2mm]}] 
  ( $ (pic cs:b) +(-15pt,8.5ex) $ ) -- 
  ( $ (pic cs:b) +(3pt,4ex) $ );
  \node[anchor=north]
  at ( $ (pic cs:b) +(-50pt,+12ex) $ )
  { decomposition level};
  \draw[-{Stealth[length=3mm, width=2mm]}] 
  ( $ (pic cs:c) +(+21pt,8.5ex) $ ) -- 
  ( $ (pic cs:c) +(7pt,4ex) $ );
  \node[anchor=north]
  at ( $ (pic cs:c) +(+40pt,+12ex) $ )
  {time bin };

\end{tikzpicture}
\caption{Illustration of the mrDMD hierarchy. Represented are the modes
$\pmb{\phi}_{k}^{(l,j)}$ and their position in the decomposition structure. The
triplet of integer values $l$, $j$, and $k$ uniquely expresses the level,
bin, and mode of the decomposition}
\label{mrdmd}
\end{figure}
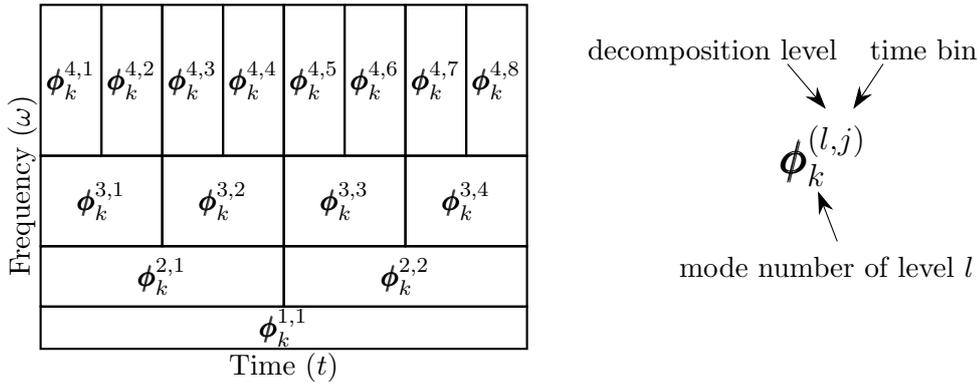
This concise definition of the mrDMD solution includes the information on the
level, time bin location, and the number of modes extracted. Fig.~\ref{mrdmd}
demonstrates mrDMD in terms of (\ref{formal_version}). In
particular, each mode is represented in its respective time bin and level. 

\paragraph{Choice of parameters}

The parameters in (\ref{parameters_mrdmd}) play a relevant role in determining
the effectiveness of the mrDMD algorithm in a specific application. Notice that,
in the setting we consider, if the number of decomposition levels $L$ is
determined, the number of time bins considered at each level is known. Thus,
only the number of decomposition levels $L$ and the slow modes to consider at
each level must be chosen.

To our knowledge, no principled method for choosing the parameters has been
formulated so far. We follow~\cite{mrDMD} and define general guidelines. First,
notice that the choice of the parameter $L$ is related to the number of sampling
points or snapshots available. For instance, if we know a priori that we want to
compute $k\in \mathbb{N}$ decomposition levels, and that in each time
bin at the $k$-th decomposition level a certain amount $m \in \mathbb{N}$ of
sampling points is required, then $m2^{k-1}$ sampling points are needed at the
first level in order to be able to reach the $k$-th level of decomposition with
the desired amount of snapshots in each time bin. Thus, the number of available
sampling points determines the number of decomposition levels, $L$, that can be
attained with the mrDMD. Another important factor to consider in the choice of
the parameter $L$ is that the higher the decomposition level is, the higher is
the frequency associated with the signal components represented by the modes
computed at that level~\citep{mrDMD}. Thus, if in a given time window we are
only interested in the slow modes, a high number of decomposition levels and a
high snapshots' sampling rate in that time window are unnecessary since there is
no intention of extracting high-frequency temporal content. On the contrary, if
we aim to analyze high-frequency components of the signal, we compute a higher
number of decomposition levels so that the modes computed at the last level
represent signal's components related to higher frequencies.  

Regarding the strategy to select the slow modes at each time bin we can consider
a simple threshold metric. Concretely, we define the threshold to choose
DMD eigenvalues (and associated DMD modes) whose temporal behavior allows for a
single wavelength or less to fit into the sampling window.  As pointed out
in~\cite{mrDMD}, given a specific application there could be more optimal
threshold values that could be derived from the knowledge of the system
analyzed. For a more detailed analysis on how the choice of these parameters
affects the analysis performed by the mrDMD see~\cite{mrDMD}.

\subsection{mrDMD-based approach for damage detection}

We now describe a mrDMD-based strategy to extract features that highlight the
presence of a cracked tooth in a gearbox by analyzing signals representing its
vibration response.

\subsubsection*{First step}

Given snapshots $y(t_{i}) \in \mathbb{R}^1$ for $i=0,\dots,T$, representing the
temporal evolution of a sensor signal, the first step consists of constructing
time-delay snapshots as follows
\begin{equation}\label{eq:time-delay-snapshots}
\mathbf{\tilde{y}}(t_{i})=[y(t_{i}),\dots,y(t_{i+d})],
\end{equation}
and arranging them into matrices
\begin{equation}
  \label{eq:time-delay-matrices}
  \mathbf{\bar{Y}}=[\mathbf{\tilde{y}}(t_{1})\dots,\mathbf{\tilde{y}}(t_{m})] \quad \text{and} \quad Y=[\mathbf{\tilde{y}}(t_{0}),\dots,\mathbf{\tilde{y}}(t_{m-1})].
\end{equation}
Here $d$ is the length of the delay, and it has to be chosen a priori, and $m$
has to be chosen consequently according to the data availability. 

Time-delay embedding is often applied when DMD-based methods are used to process
univariate signals~\citep{chao,  Clainche2018, uniDMD}. In this work, the
reasons why the time-delay embedding is used are connected to the fact that the
spatial resolution of the signals we study is much lower than the temporal
resolution, i.e., we have several one-dimensional snapshots $y(t_{i}) \in
\mathbb{R}^1$. It was first observed in~\cite{uniDMD} that DMD is incapable of
accurately representing even very simple one-dimensional signals. For instance,
if we collect one-dimensional measurements representing the temporal evolution
of a single sine wave, DMD reconstructs the dynamics using only a single real
eigenvalue, which does not capture periodic oscillations~\citep[Chapter
7]{DMDbook}. Fortunately, this issue can be addressed by considering the time
delay embedding. See~\cite{DMDbook} for a detailed and technical discussion on
the motivations and implications of using DMD with the time-delay embedding.

\subsubsection*{Second step}
The second step consists of applying the mrDMD algorithm to the time-delay
snapshots to obtain the slow-modes reconstruction of the signal in the
time-delay coordinates domain. The reconstruction of the time-delay snapshot
$\mathbf{\tilde{y}}(t_{i})$ at the $i$-th time step is given by (\ref{formal_version}) and is
represented by $\mathbf{\hat{y}}(t_{i})$. Since the information related to damage is assumed to be in 
high-frequency structures, as we will see in the next section, the general guideline is to choose
the parameter $L$ large enough to enable the modes computed at the last decomposition level
to represent the frequencies of interest. 

\subsubsection*{Third step}
In the third and final step, we want to obtain information about the signal's
geometrical structures related to perturbations caused by the damage we are
considering, which is assumed to affect high frequencies of the signal. To do
that, we first pick a time-range by selecting a time-delay snapshot, i.e., the
$i$-th snapshot contains the evolution of the signal from time step $i$ to time
step $i+d$. After that, we subtract from the original time-delay snapshot,
$\mathbf{\tilde{y}}(t_{i})$, the slow modes reconstruction, $\mathbf{\hat{y}}(t_{i})$, obtained using the
mrDMD algorithm. After the subtraction, we are left with a residual, i.e.,
\begin{equation}
\label{residual}
\mathbf{r}_i=\mathbf{\tilde{y}}(t_{i})-\mathbf{\hat{y}}(t_{i}).
\end{equation}
The residual $\mathbf{r}_i$ gives us the information contained in the fastest modes
computed at the deepest decomposition level for each time bin. In the residual,
anomalies related to damage are
emphasized, making it possible to visually determine whether there is a
cracked tooth or not in the time-range we are considering.

Algorithm~\ref{alg:PNP} summarizes the main steps of the numerical procedure we
just presented.

\begin{algorithm}
  \caption{mrDMD-based approach for damage detection}\label{alg:PNP}
  \hspace*{\algorithmicindent} \textbf{Input} snapshots $y(t_{i}) \in \mathbb{R}^1$ for $i=0,\dots,T$, delay's length $d$, number of decomposition levels $L$. \\
  \hspace*{\algorithmicindent} \textbf{Output} Residuals $\mathbf{r}_i$.
\begin{algorithmic}[1]
\State Construct time-delay snapshots (\ref{eq:time-delay-snapshots})
and arrange them into matrices (\ref{eq:time-delay-matrices})
 \State Apply the mrDMD algorithm to the time-delay
 snapshots to obtain the slow-modes reconstruction of the signal in the
 time-delay coordinates domain. The reconstruction of the time-delay snapshot
 $\mathbf{\tilde{y}}(t_{i})$ at the $i$-th time step is represented by $\mathbf{\hat{y}}(t_{i})$.
 \State Compute the residuals
 $
 \mathbf{r}_i=\mathbf{\tilde{y}}(t_{i})-\mathbf{\hat{y}}(t_{i})
 $.
\end{algorithmic}
\end{algorithm}
\subsection{Computational cost}
Algorithm $\ref{alg:PNP}$ clearly shows that the computational cost of the
proposed procedure is determined by the computational cost of the mrDMD
algorithm performed in the second step, as the other steps consist of storing
data and subtracting vectors.  From work carried out in~\cite{mrDMD}, we know
that computational efforts required by the mrDMD are dominated by the SVD of the
matrix $\mathbf{Y}$ in the second step of Algorithm~\ref{alg:DMD} and by the number of
decomposition levels $L$. In particular, if at a given decomposition level and
time bin we have $T$  time-delay snapshots with delay's length $d$, the
computational cost of the SVD is $Q=O(T^2d + d^3)$ floating point operations.
Since in the mrDMD procedure the SVD is performed at each time bin of each
decomposition level, we have that the overall cost of the mrDMD is $O(2^LQ)$.
See~\cite{mrDMD} for more details.

\section{Gearbox model}
\label{gearboxmodelll}
Data representing the response of gearboxes in presence of damages, such as
cracks, can be either experimental using vibration measurements or numerical
using simulation models. A lot of work has been conducted to analyze
experimentally measured vibration signals in order to identify damages in
vibrating structures~\citep{anto, ahm}. The main advantage of experimental data
is that they reflect the behaviour of a real system. However, such data  are
expensive to obtain in terms of time and money, especially when repeated
measurements have to be performed for different damage scenarios~\citep{ahm}. 
Dynamic modelling and simulation of gearbox vibration signals
can overcome these issues and can be a good alternative for studying the
dynamic behaviour of a gear system more simply and economically, in particular for our goal of developing a data analysis approach for early damage detection. 
Dynamic
modelling and simulation also have the advantage of increasing the understanding
of the system's behaviour before the initiation of a measurement campaign.
\subsection{Theoretical setting and mathematical model}
\label{mathematical_model}

\begin{figure*}
  \begin{center}
    \includegraphics[width=6cm, height=5cm]{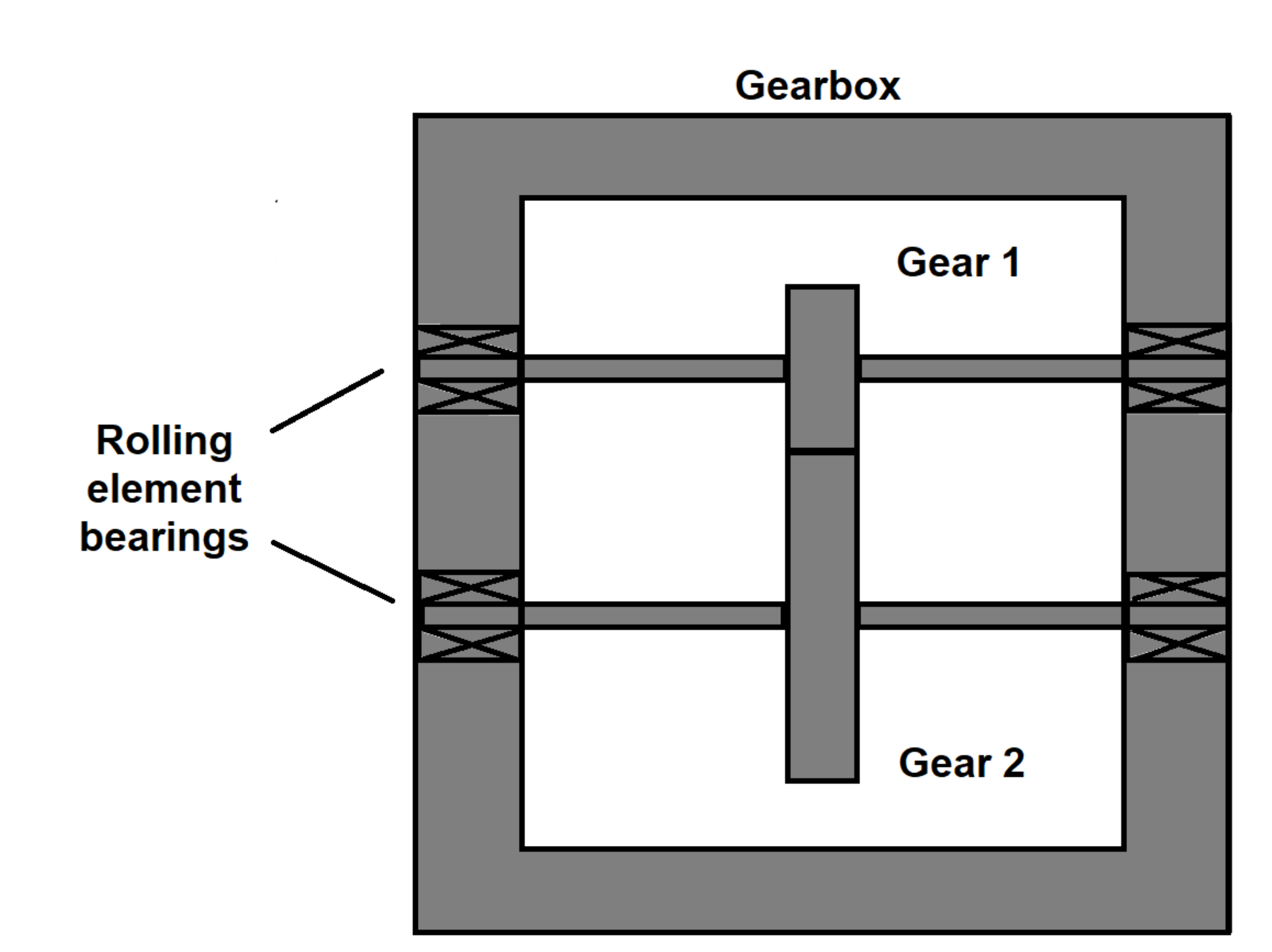}
    \caption{A generic gear rotor bearing system}
  \label{schematic_simulation_gear}
\end{center}
\end{figure*}
We consider a generic geared-rotor bearing system, shown in Fig.~\ref{schematic_simulation_gear}, consisting of a rigid gearbox
containing a spur gear pair mounted on flexible shafts, supported by rolling
element bearings. Such a model has been previously used in other works~\citep{anto, Kah} and its reliability has been
shown in~\cite{Kah}. We now introduce it briefly.
 
The vibration response of the gearbox model we consider can be modelled 
by the following dimensionless equation of motion~\citep{anto}:
\begin{equation}
  \label{dimensionless}
  \ddot{x}(t)+2z\dot{x}(t)+K(t)B(x(t))=F_m+F_{te}(t)+F_{var}(t).
\end{equation}
The coordinate $x(t)$ represents the difference between the dynamic
transmission error and the static transmission error. The quantity $z$
is a  dimensionless linear viscous dumping parameter. $F_m$ represents the mean
force excitations, while $F_{te}$ and $F_{var}$ pertain to internal excitations
related to the static transmission error and external excitations related to
wind turbulence, respectively. The quantity $F_{var}$ is derived from the
fluctuating component of external input torque that simulates the wind
turbulence's effect. The \textit{backlash} 
is simulated by the function 
\begin{equation}
  \label{backlash_function}
  B(x(t)) := \left\{\begin{array}{lr}
        x(t)-1, & \text{for }  x(t)\geq 1\\
        0, & \text{for } -1< x(t)< 1\\
        x(t)+1, & \text{for } x(t)\leq -1 
        \end{array}\right\},
\end{equation}
while the time varying \textit{mesh stiffness} is incorporated via the periodic
function $K(t)$. For a more detailed explanation of the specific structure of
the functions mentioned and, more in general,  for a
deeper understanding of the gearbox model, see Appendix A and~\cite{anto,Kah}.
Note that, this is a very simple wind turbine gearbox model, e.g., it ignores the bearing vibrations
and wind turbine gearboxes usually consist of three gear stages with one or two being planetary stages. 
According to~\cite{anto}, it is sufficient for studies such as ours, since 
``more gear stages in a vibration signal would mean more frequency components at different frequency bands. 
Damage at a specific gear stage would therefore be shown in the vibration signal associated with the meshing 
frequency and its harmonics of the gear stage examined."
What is important in~\cite{anto} and in our study, is to investigate the influence of the varying load conditions
on the vibration signals, which can be achieved.

\subsection{Including damage and wind turbulence.}
\paragraph{Damage simulation}
The damage we consider in this work is the crack of a gear's tooth that, like
other types of tooth damage, causes a reduction in the gear's stiffness. This
phenomenon can be represented by a periodic magnitude change in the mesh
stiffness function $K(t)$~\citep{Sta}. Thus, an amplitude change is periodically
applied to the mesh stiffness function to simulate the cracked tooth. In the
simulations used in this work, the crack of a gear's tooth was modelled by
decreasing the dimensionless mesh stiffness function  by 13\%  of its nominal
stiffness  for 5 degrees of the shaft rotation, periodically, for every rotation
of the damaged gear. Similar modelling was used in previous studies, see
~\cite{anto, ahm} for a more detailed explanation. Recapturing results reported
in~\cite{anto}, we notice that in the model we consider, one should expect
damage features to be evident at high-frequency components of the simulation
signals because it is in the harmonics of the meshing frequency of the damaged
gear pair that damage features occur. This is a phenomenon that was also
observed in the experimental data used in~\cite{anto}.
\paragraph{Wind turbulence's simulation} 
\label{wind turbolence simulation}
The representation of the wind as a smooth flow is not realistic because it does
not take into account all its irregular and stochastic fluctuations. Due to wind
turbulence, wind turbines experience transient and time-varying load conditions.
In order to build a realistic model, we need to be able to consider and simulate
these phenomena. To do that, it is possible to use a series of wind turbine
aerodynamics codes, developed by the National Renewable Energy Laboratory (NREL,
US)~\citep{FAST}. Specifically, the FAST design code has been used to simulate the
turbulent wind conditions, associated with wind speed of 5 m/s and 13 m/s. See
Appendix A for a deeper understanding of how  the different wind conditions have
been simulated.
\subsection{Numerical data}
~\label{numerical_data}
The vibration response's simulation of the gearbox model we considered in the
previous section is given by numerical approximation of the function
$\ddot{x}(t)$ obtained from the numerical solution of the dimensionless equation
of motion (\ref{dimensionless})~\citep{anto,Kah,ahm}. The numerical solution of
equation (\ref{dimensionless}) has been computed with the MATLAB $ode45$ differential
equation solver, with a fixed time step of 0.015. We originally intended
to reproduce and use the simulated scenario as given in~\cite{anto}.
Unfortunately, the data are not publicly available and could not be
provided. Moreover, using their stated model's parameters we could not
obtain similar signals to those illustrated in that work. Thus, we
choose the model parameters as to get data of at least qualitatively similar behavior. 
The details on how the numerical model has been built are
described in Appendix A, where also all the parameters used for the simulations
are reported in Table~\ref{simulation_parameters}. The MATLAB code
to perform the simulations will be made available after acceptance of the article and 
is in the supplementary material of the submission. 

\begin{figure*}
    \begin{subfigure}{0.5\textwidth}
      \includegraphics[width=\textwidth, height=3cm]{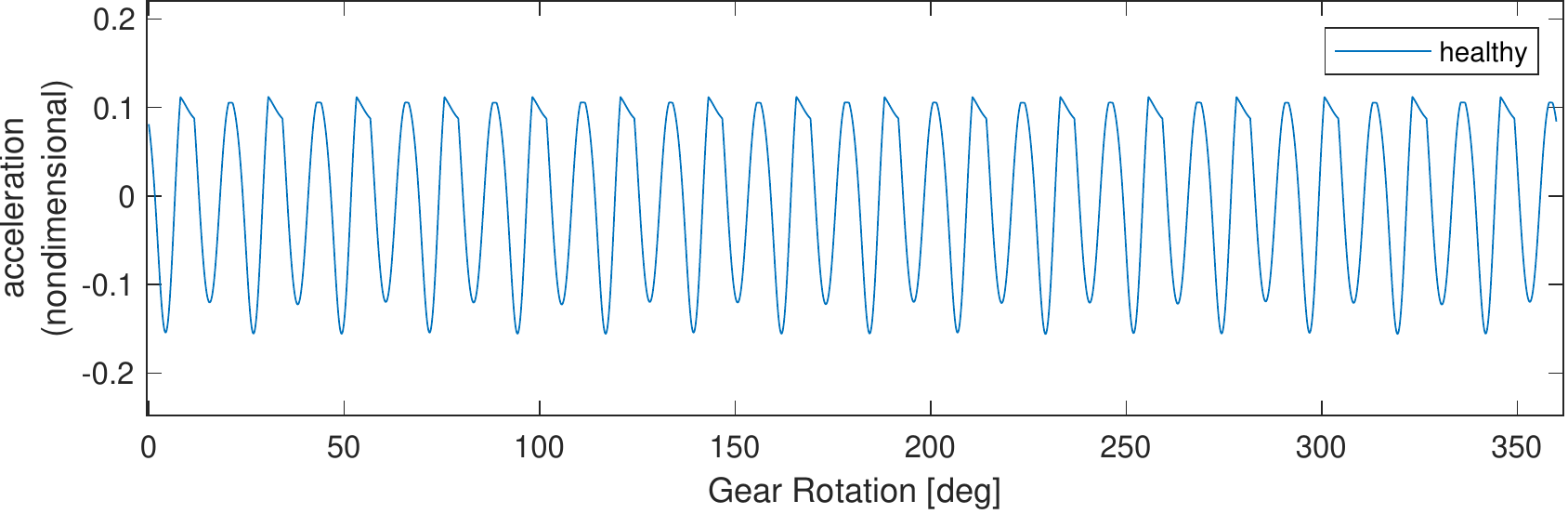}
      \caption{undamaged case}
      \label{signal_nowind_a}
    \end{subfigure}
    \begin{subfigure}{0.5\textwidth}
      \includegraphics[width=\textwidth, height=3cm]{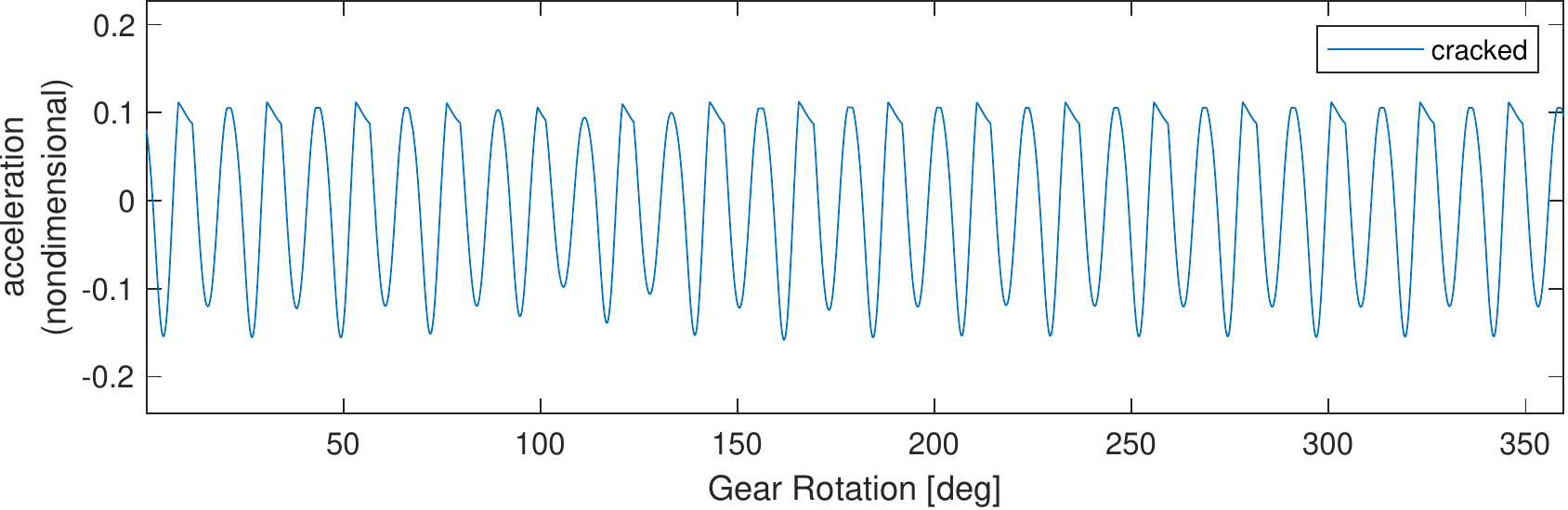}
      \caption{damaged case}
      \label{signal_nowind_b}
    \end{subfigure}
  \caption {Simulation of an acceleration signal  under a steady load condition} 
  \label{signal_nowind}
\end{figure*}

\begin{figure*}
  \begin{subfigure}{0.5\textwidth}
    \includegraphics[width=\textwidth, height=3cm]{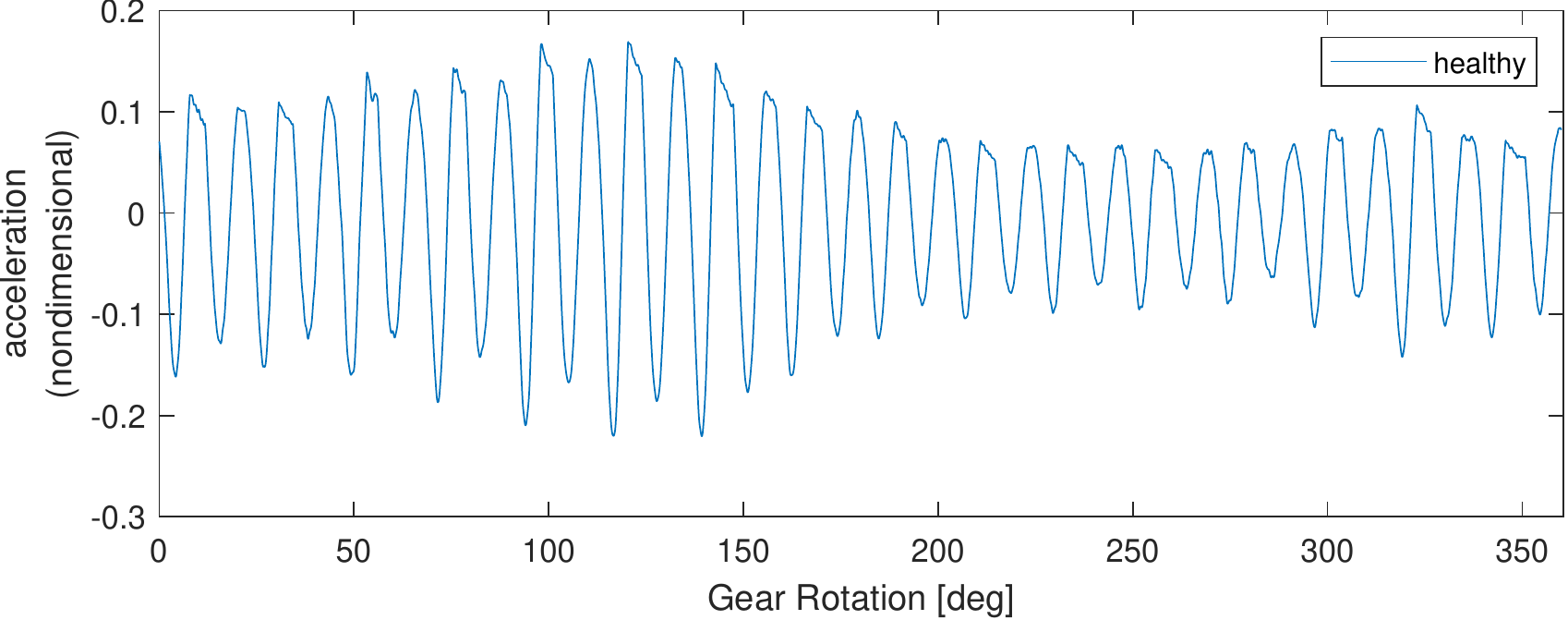}
    \caption{undamaged case}
    \label{signal_wind5_a}
  \end{subfigure}
  \begin{subfigure}{0.5\textwidth}
    \includegraphics[width=\textwidth, height=3cm]{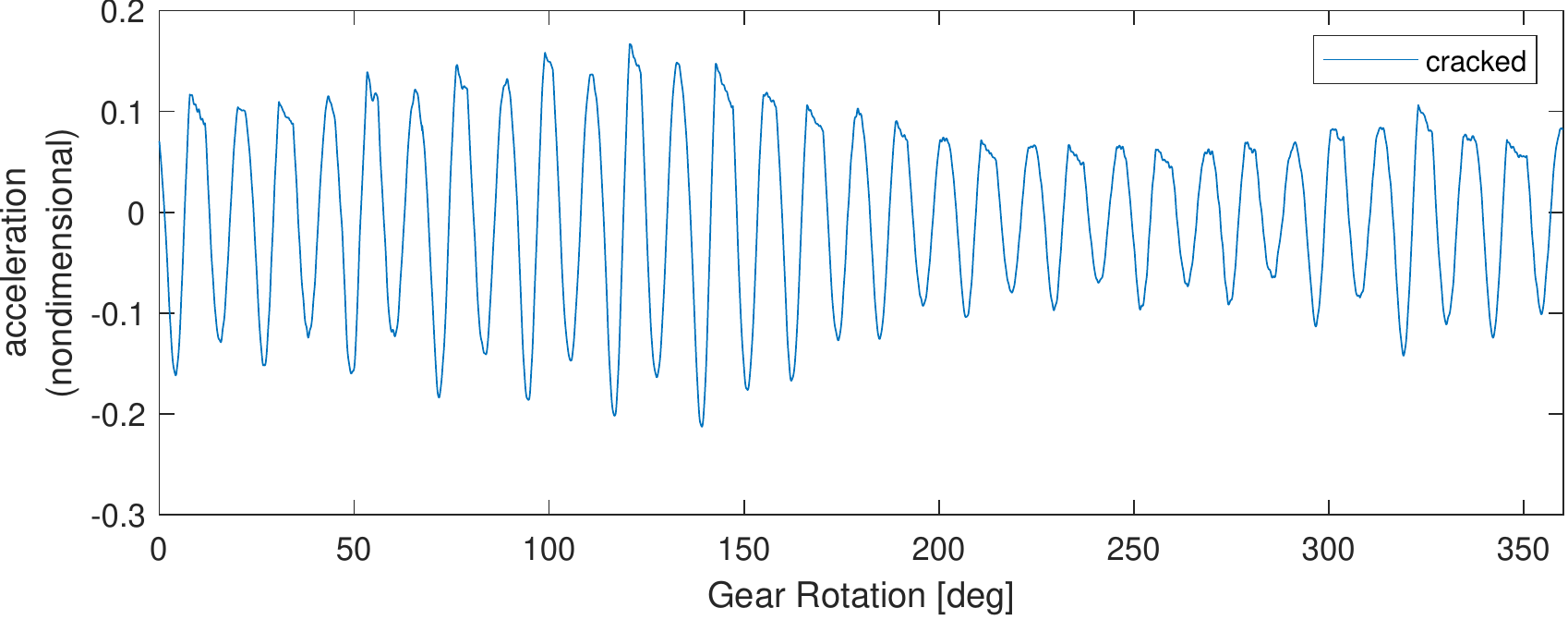}
    \caption{damaged case}
    \label{signal_wind5_b}
  \end{subfigure}
\caption {Simulation of an acceleration signal under varying load condition with wind speed 5 m/s} 
 \label{signal_wind5}
\end{figure*}

\begin{figure*}
    \begin{subfigure}{0.5\textwidth}
      \includegraphics[width=\textwidth, height=3cm]{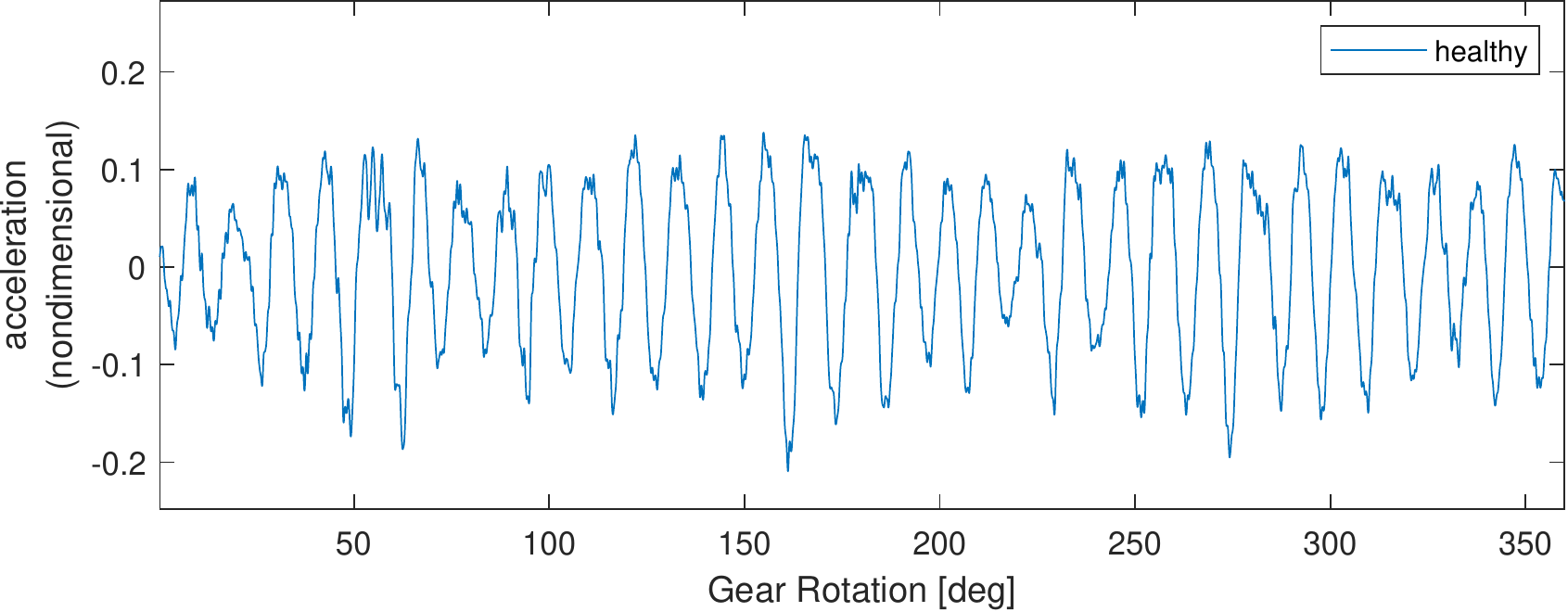}
      \caption{undamaged case}
      \label{signal_wind_a}
    \end{subfigure}
    \begin{subfigure}{0.5\textwidth}
      \includegraphics[width=\textwidth, height=3cm]{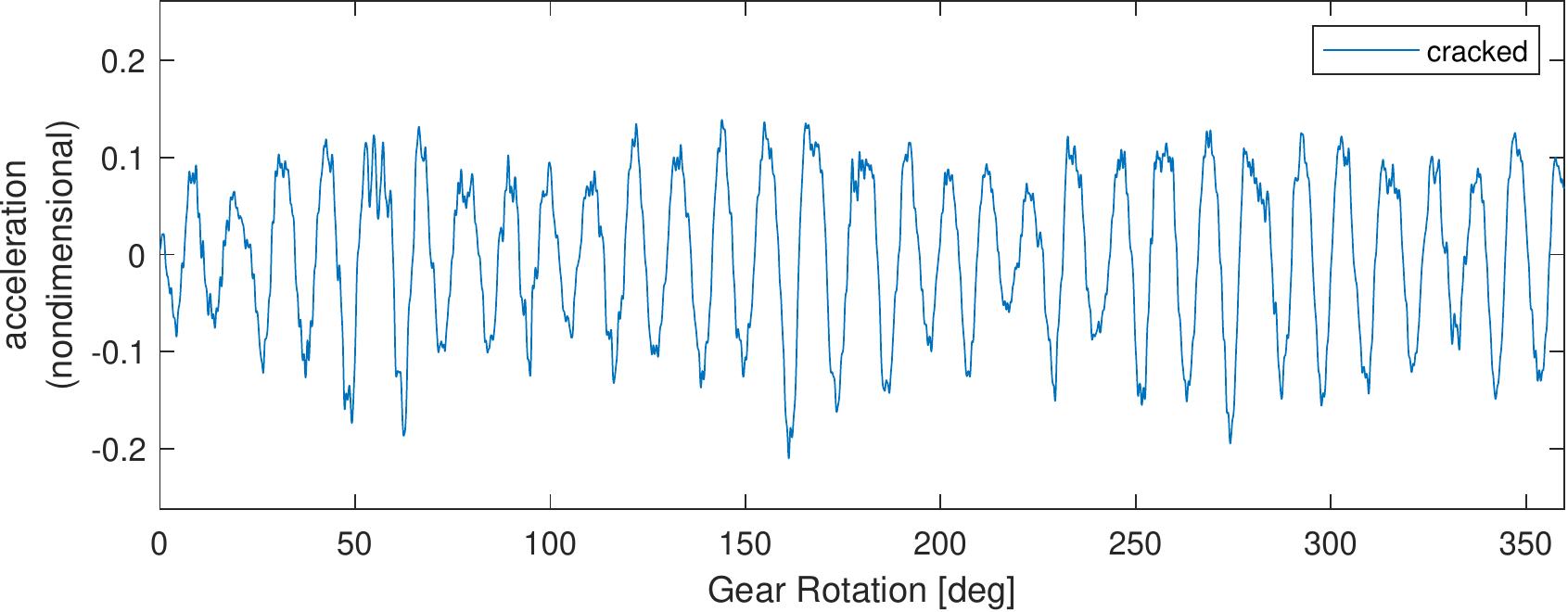}
      \caption{damaged case}
      \label{signal_wind_b}
    \end{subfigure}
  \caption {Simulation of an acceleration signal under varying load condition with wind speed 13 m/s} 
   \label{signal_wind}
\end{figure*}

Simulations can represent different scenarios. On the one hand, we can produce
numerical signals where the fluctuations of the input torque,
due to wind turbulence, are not taken into account, i.e.,
$F_{var}(t)=0$ (Fig. $\ref{signal_nowind}$). These simulations
represent an improbable scenario in which the load on the wind turbine blades is
constant. Such signals can be helpful to test different numerical strategies in
a simplified context, and we will refer to them as signals in \textit{steady
load conditions}.
On the other hand, we can produce simulations representing the more realistic
scenario where the wind is not considered as a uniform constant flow and its
stochastic behaviour is included in the model. These simulations represent a
situation in which the load on the wind turbine blades fluctuates and has a
stochastic behaviour that determines random variations in the input torque. We
will say that simulations generated in this scenario have \textit{varying load
conditions}. Fig.~\ref{signal_wind5} and Fig.~\ref{signal_wind} show the
acceleration signals of simulations with varying load conditions produced
considering two different wind speeds.
 
Independently of the scenario  we consider, with steady or varying load
conditions, we can produce signals representing a situation  in which a
perfectly healthy gearbox is modelled (Fig.~\ref{signal_nowind_a},
~\ref{signal_wind5_a},~\ref{signal_wind_a}) or we can simulate a situation in
which one of the gears is damaged
(Fig.~\ref{signal_nowind_b},~\ref{signal_wind5_b},~\ref{signal_wind_b}). The
simulated  damage consists of a cracked tooth in one of the two gears.
Recapturing some results from~\cite{anto}, we notice the signals simulated
taking into account the wind turbulence are less smooth and show a more chaotic
behaviour than the signals where the load condition is steady. This is motivated
by the fact that when wind turbulence is incorporated into the model it
introduces a stochastic component that perturbs the acceleration signals. The
higher the simulated wind condition, the noisier the signal. Note that,
independently of the wind condition considered, the damaged and healthy signals
do not show any relevant difference and look almost identical. This is because
only high-frequency components are affected by damage and just for a short time.
This fact underlines that it would be very difficult to address the damage
identification task without relying on a method that can extract only the
relevant information from the signals.

\section{Results}
\label{faultdetection}

\subsection{Results with simulation data}
\begin{figure*}
  \begin{center}
 \begin{subfigure}{0.7\textwidth}
   \includegraphics[width=\textwidth, height=3.5cm]{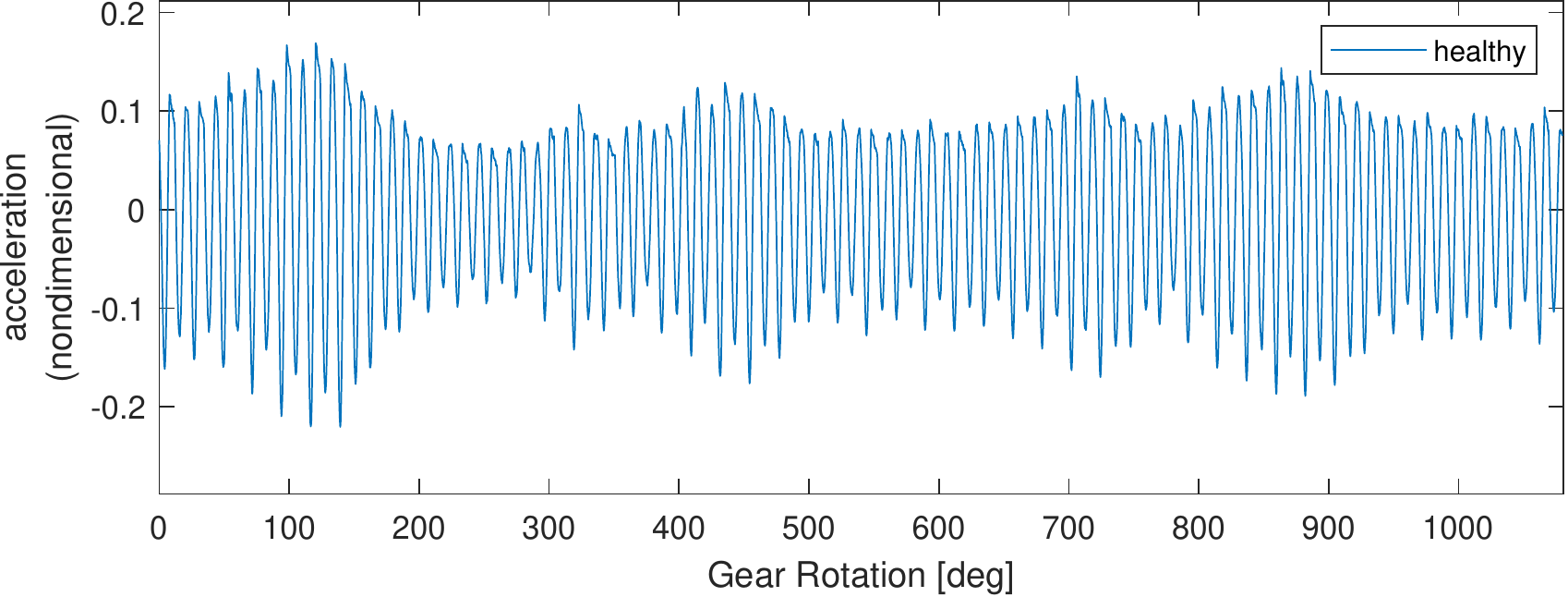}
   \caption{undamaged case}
   \label{wind_5health}
 \end{subfigure}
 \begin{subfigure}{0.7\textwidth}
   \includegraphics[width=\textwidth, height=3.5cm]{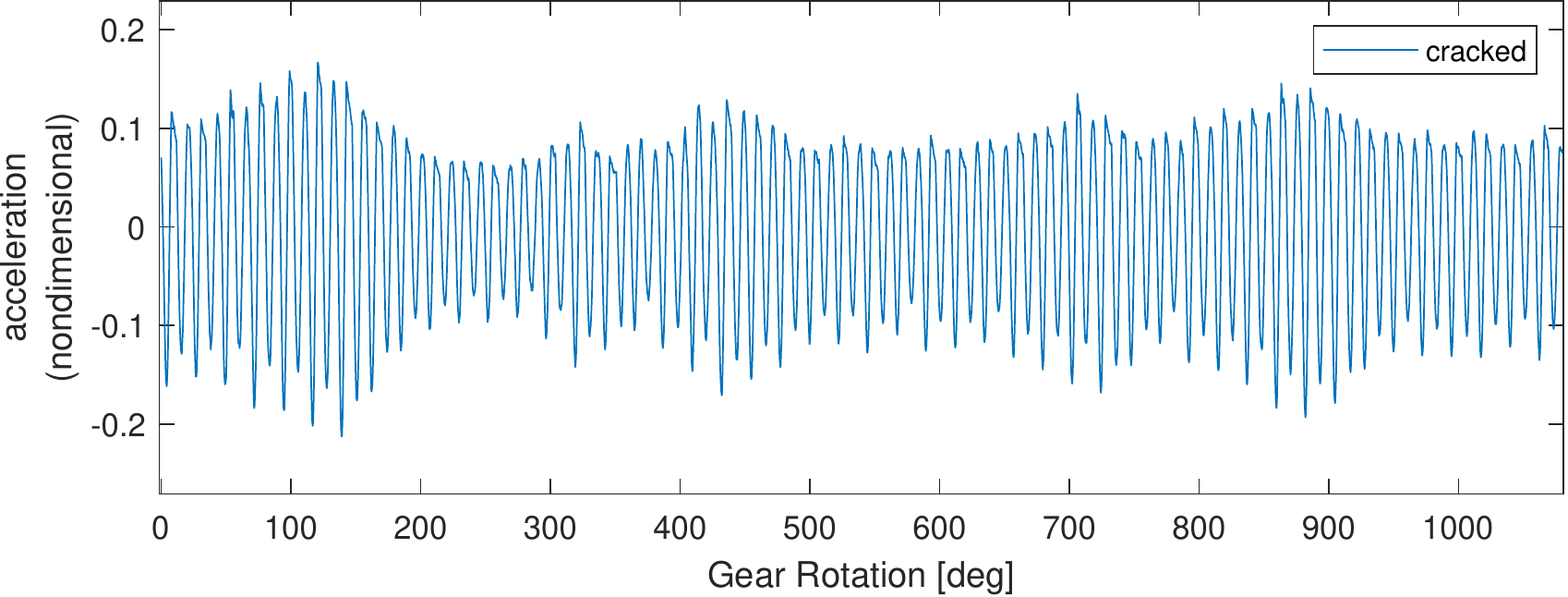}
   \caption{damaged case}
   \label{wind_5crack}
 \end{subfigure}
 \end{center}
\caption {Simulation of an acceleration signal for three gear rotations under varying load condition with wind speed 5 m/s} 
\label{wind_5_1080}
\end{figure*}

\begin{figure*}
     \begin{center}
    \begin{subfigure}{0.7\textwidth}
      \includegraphics[width=\textwidth, height=3.5cm]{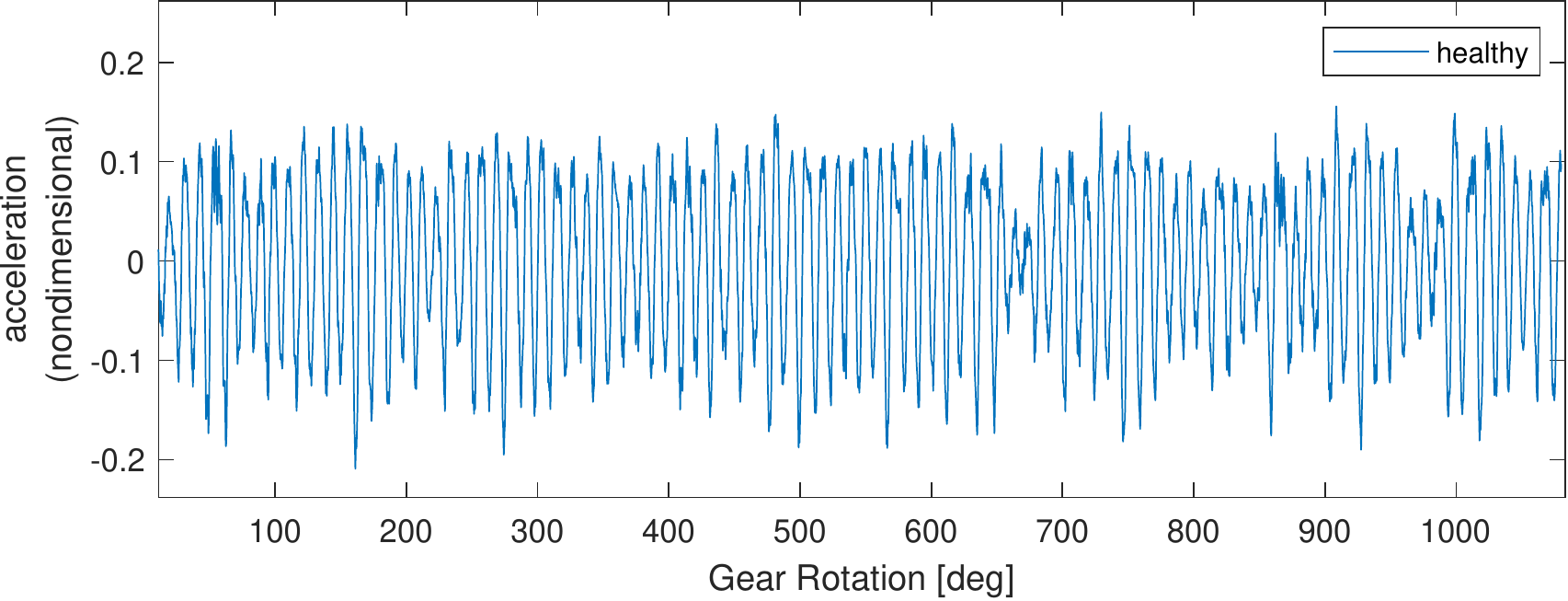}
      \caption{undamaged case}
      \label{wind_1080_a}
    \end{subfigure}
    \begin{subfigure}{0.7\textwidth}
      \includegraphics[width=\textwidth, height=3.5cm]{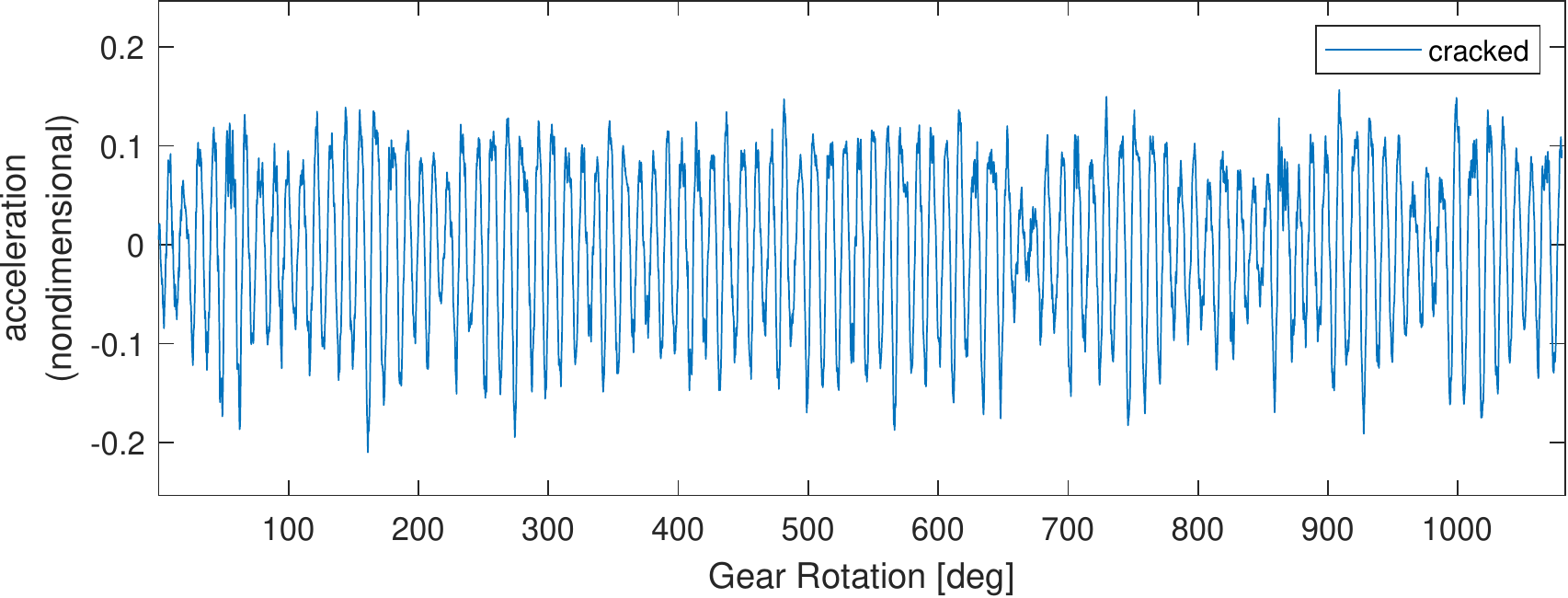}
      \caption{damaged case}
      \label{wind_1080_b}
    \end{subfigure}
    \end{center}
    \caption {Simulation of an acceleration signal for three gear rotations under varying load condition with wind speed 13 m/s} 
  \label{wind_1080}
 \end{figure*}

In this section, we study the numerical strategy for early damage detection
based on multi-resolution dynamic mode decomposition proposed in
Section~\ref{multiresolution_dmd}. We compare our method with classical
approaches used in this context, Fast Fourier transform as a frequency-domain
approach, Time Synchronous Averaging (TSA) and EMD as a time-domain approaches.
\paragraph{Dataset}
\label{Dataset}
We analyze the vibration response data of the gearbox from three gear
revolutions. We use the same two different health conditions of healthy and with
cracked tooth and the same two wind speeds of 5 m/s (Fig.~\ref{wind_5_1080})
and 13 m/s (Fig.~\ref{wind_1080}) from before. The simulated signals are
composed of 40201 one-dimensional snapshots. In the simulations that include the
damage, the presence of the cracked  tooth affects the signals  once per
rotation at $67^{\circ}$, $427^{\circ}$ and $787^{\circ}$. Note that we will
make the simulated data and the code for generating the data available after
acceptance of the article, it is part of the supplementary material of the
submission.
 
\subsubsection{Benchmark methods}

\begin{figure}
  \begin{subfigure}{0.5\textwidth}
  \includegraphics[width=\linewidth, height=7cm]{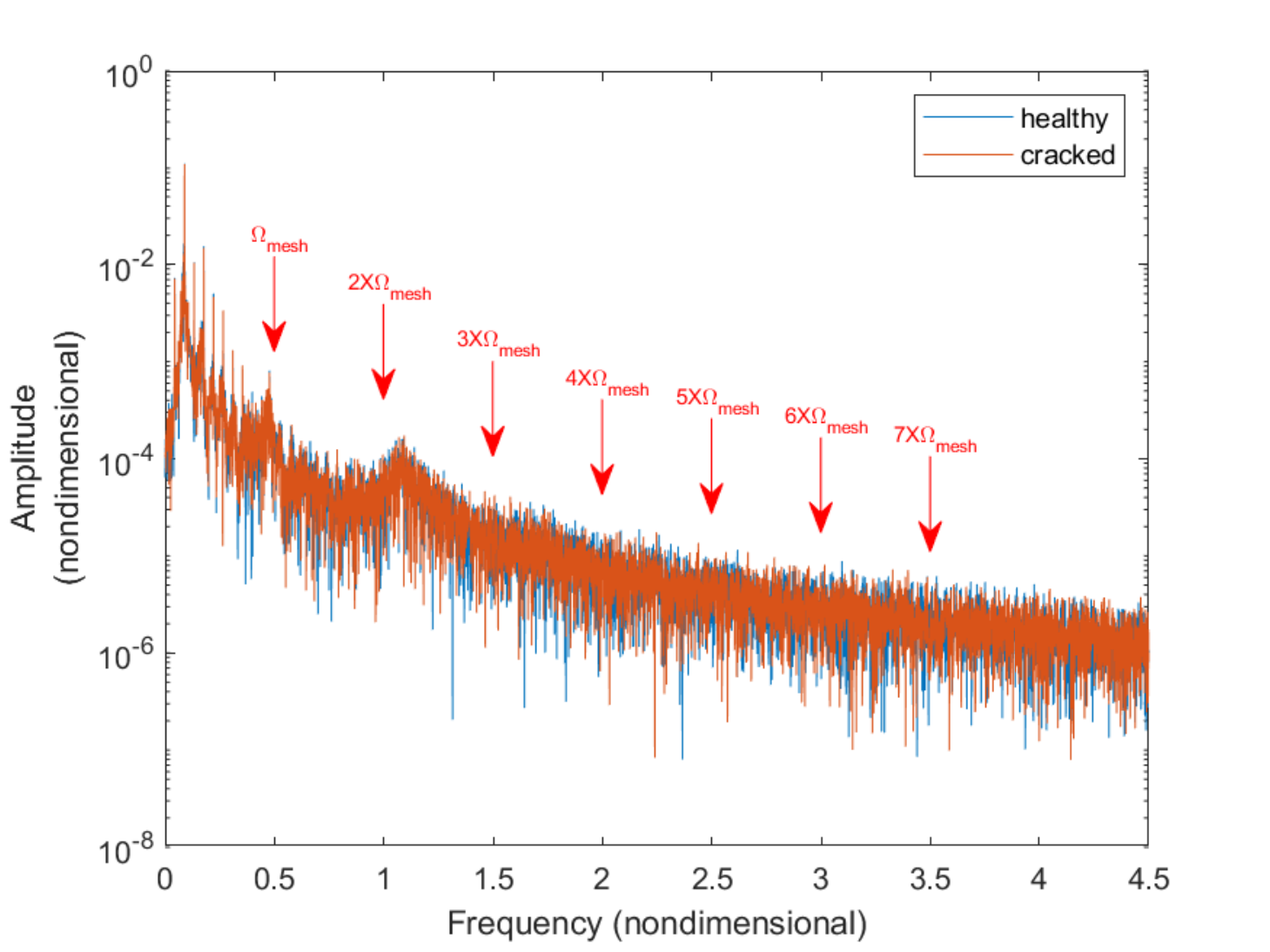} 
  \caption{wind speed 5 m/s}
  \label{spectrum_wind5}
  \end{subfigure}
  \begin{subfigure}{0.5\textwidth}
  \includegraphics[width=\linewidth, height=7cm]{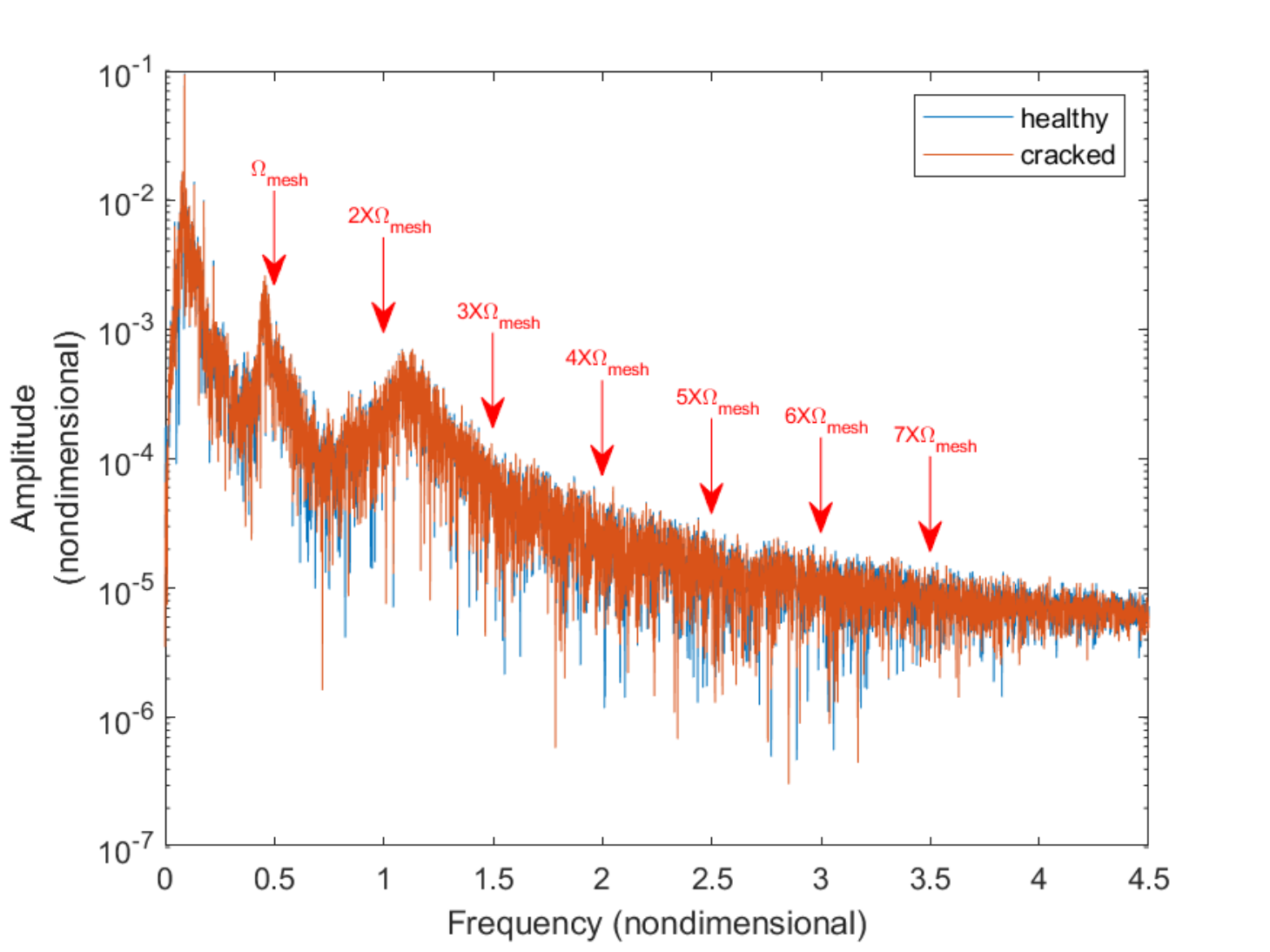}
  \caption{wind speed 13 m/s}
  \label{spectrum_wind}
  \end{subfigure}  
  \caption{Fourier spectra of the dimensionless acceleration signals of the
  healthy and cracked tooth cases for varying load condition. The arrow-shaped cursors point
  to the mesh frequency, $\Omega_{mesh}$, and its harmonics.}
  \label{spectrum_winds}
  \end{figure}

\paragraph{Fast Fourier transform (FFT)}
We now apply the FFT~\citep{Rao2011} to the vibration signals shown in
Fig.~\ref{wind_5_1080} and Fig.~\ref{wind_1080}. Fig.~\ref{spectrum_wind5} and
Fig.~\ref{spectrum_wind} show the Fourier spectra of the acceleration signals we
have considered, representing the damaged and undamaged scenarios in varying
load conditions associated with wind speed 5 m/s and 13 m/s, respectively.
Independently of the wind condition considered, the differences between the
healthy and cracked cases are very hard to identify just looking at the signals'
spectra. To be more specific, in Fig.~\ref{spectrum_wind5} and
Fig.~\ref{spectrum_wind} we highlight the meshing frequency and its harmonics,
which are damage indicators because they relate to the frequency at which the damaged
tooth has been excited~\citep{LiXu2019}. Unfortunately, also in the highlighted
frequencies, there is no significant or obvious magnitude change that allows us
to identify the presence of damage. 

This shows us an arduous difficulty that frequency methods have to overcome in
this scenario: when the effects of the wind turbulence are included, it is very
challenging to distinguish between the spectral characteristics of the signals
associated with damage and those associated with wind turbulence. In addition to
that, we have that due to wind turbulence, the spectra of the signals change
randomly with time, making the damage identification task even more difficult.
The shown results for the comparison between the healthy and damaged
signals spectra are rather basic. But note, we do not aim to show that conventional
gear monitoring techniques, for example Cepstrum or Kurtosis analysis, would be
ineffective. These results mainly show that the operating conditions we are
considering may have detrimental effects on the effectiveness of gear
diagnostics techniques based on the signal's spectral analysis.
\paragraph{Time Synchronous Averaging (TSA) and Hilbert transform (HT)}
Time Synchronous Averaging (TSA) is a well-established procedure in rotating machines
diagnostics~\citep{Randall2021}. One of the primary purposes of TSA analysis is to extract periodic
waveforms from signals. For instance, it may be used to extract a periodic
signal, such as the tooth meshing vibration of a single gear, from the whole
machine's vibration in order to detect damages by analyzing only the extracted
signal's component. Following along~\cite{Fadden1986}, we use TSA to demodulate the signals component
associated with the gear mesh. After that, we use the Hilbert transform to
demodulate the extracted TSA signal in order to detect local variations caused
by damage and not visible to the eye. See~\cite{Fadden1986}, for a more
detailed description of the TSA-based strategy we are going to review \medskip\\
  
  \subparagraph{Time Synchronous Averaging (TSA)}
   Given a
  signal $y(t) \in \mathbb{R}$, its time synchronous average $y_a(t)$ is defined
  as 
  \begin{equation}
    \label{def: TSA}
    y_a(t) = \frac{1}{N}\sum_{n=0}^{N-1} y(t+nT_r),
  \end{equation}
  where $N \in \mathbb{N}$ and $T_r$ is the so-called periodic time. Formula (\ref{def: TSA}) can be written as the convolution of $y(t)$ with a sequence of $N$ delta functions displaced by integer multiples of the periodic
  time $T_r$, i.e., 
   \begin{equation}
     y_a(t)= c(t) * y(t),
   \end{equation}
   with
   \begin{equation}
    c(t)= \frac{1}{N}\sum_{n=0}^{N-1} \delta(t+nT_r).
   \end{equation}
  
   It can be show that the desired signal component can be demodulated through
   TSA if the distance between the multiples of the periodic time is equal to
   its period. The main reason behind this fact is that (\ref{def: TSA}) is
   equivalent in the frequency domain to the multiplication of the Fourier
   transform of $y(t)$ by a comb filter, removing all the components
   which fall outside the fundamental and harmonic frequencies of the desired signal.
   For more details see~\cite{Fadden1987}.\smallskip\\

   \subparagraph{Hilbert transform (HT)}
The HT of a signal $y(t) $ is defined as
\begin{equation}
  \hat{y}(t):=H[y(t)]=\frac{1}{\pi}\int_{-\infty}^{\infty}\frac{y(s)}{t-s}ds = y(t) \ast \frac{1}{\pi t}, 
\end{equation}
where $\ast $ denotes the convolution operator. Given the HT of $y(t)$ we can
define the analytic signal
\begin{equation}
  \label{analytic signal}
  z(t):=y(t)+i\hat{y}(t)=a(t)e^{i\theta(t)},
\end{equation}
where  $i:=\sqrt{-1}$ is the imaginary unit, $a(t):=\sqrt{y^2(t)+\hat{y}^2(t)}$
is the instantaneous amplitude and $\theta(t)=\arctan(\frac{\hat{y}(t)}{y(t)})$
is the instantaneous phase. The instantaneous frequency is defined as $\omega:=
\frac{\theta}{dt}$. The HT does not change the domain of the
variable. Indeed, the HT of a time-dependent signal is also a
function of time. Given a signal $y(t)$, the combination of
(\ref{signal EMD}) and (\ref{analytic signal}) yields
\begin{equation}
  y(t)= Re\Bigg\{\sum_{k=1}^{N}a_k(t)e^{i \theta_k(t)}\Bigg\} + r_N(t),
\end{equation}
where $a_k(t)$ and $\theta_k(t)$ are the instantaneous amplitude and phase of the 
$k$-th IMF, respectively.
We compute the Hilbert Transform using the ``scipy.signal.function" in python.
  
   \paragraph{Application of a TSA-based Strategy}

  \begin{figure}[h!]
    \begin{subfigure}{0.33\textwidth}
      \includegraphics[width=\textwidth,height=3cm]{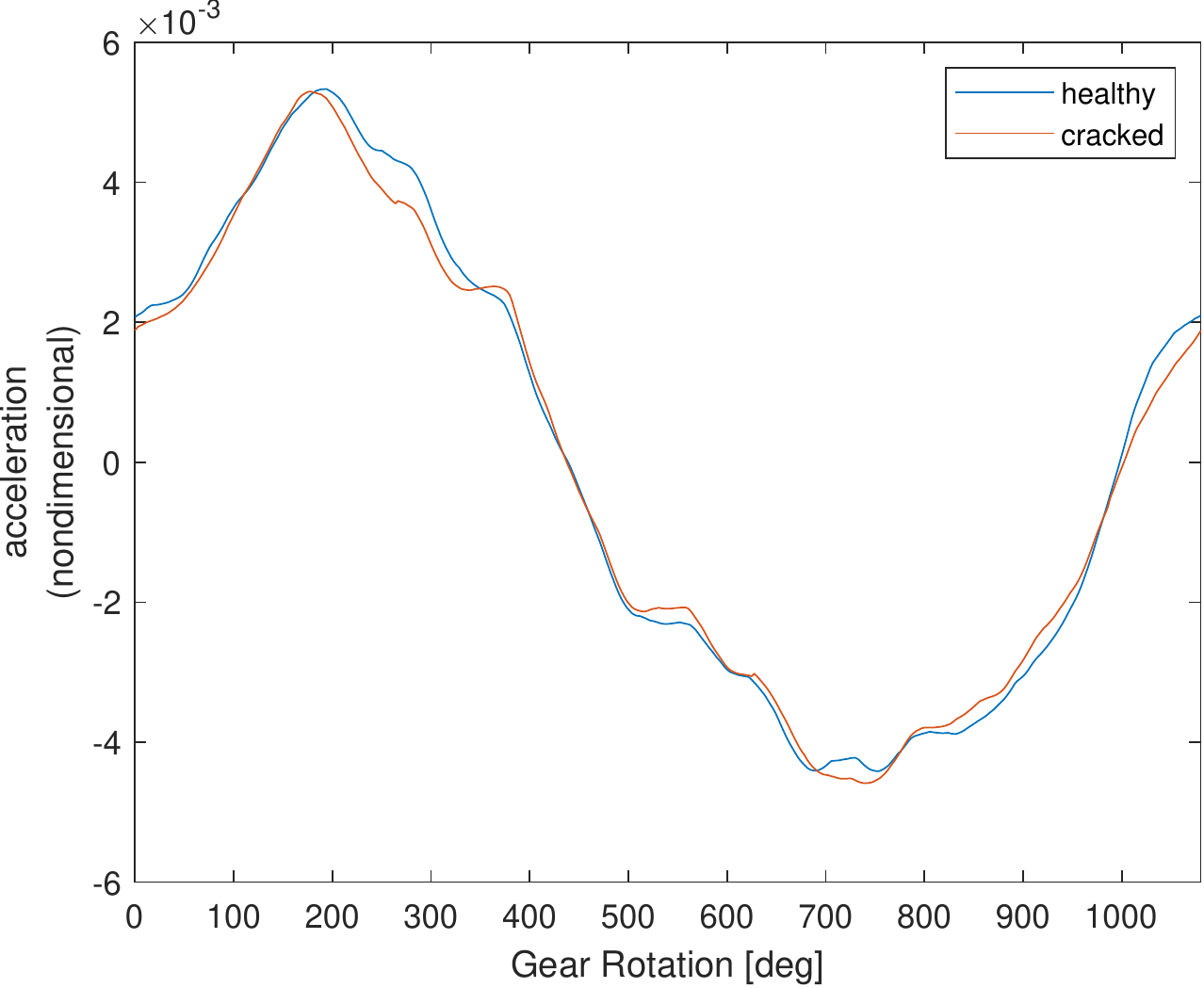}
      \caption{TSA signal}
      \label{TSA signal_5_a}
    \end{subfigure}
    \begin{subfigure}{0.33\textwidth}
    \includegraphics[width=\textwidth,height=3cm]{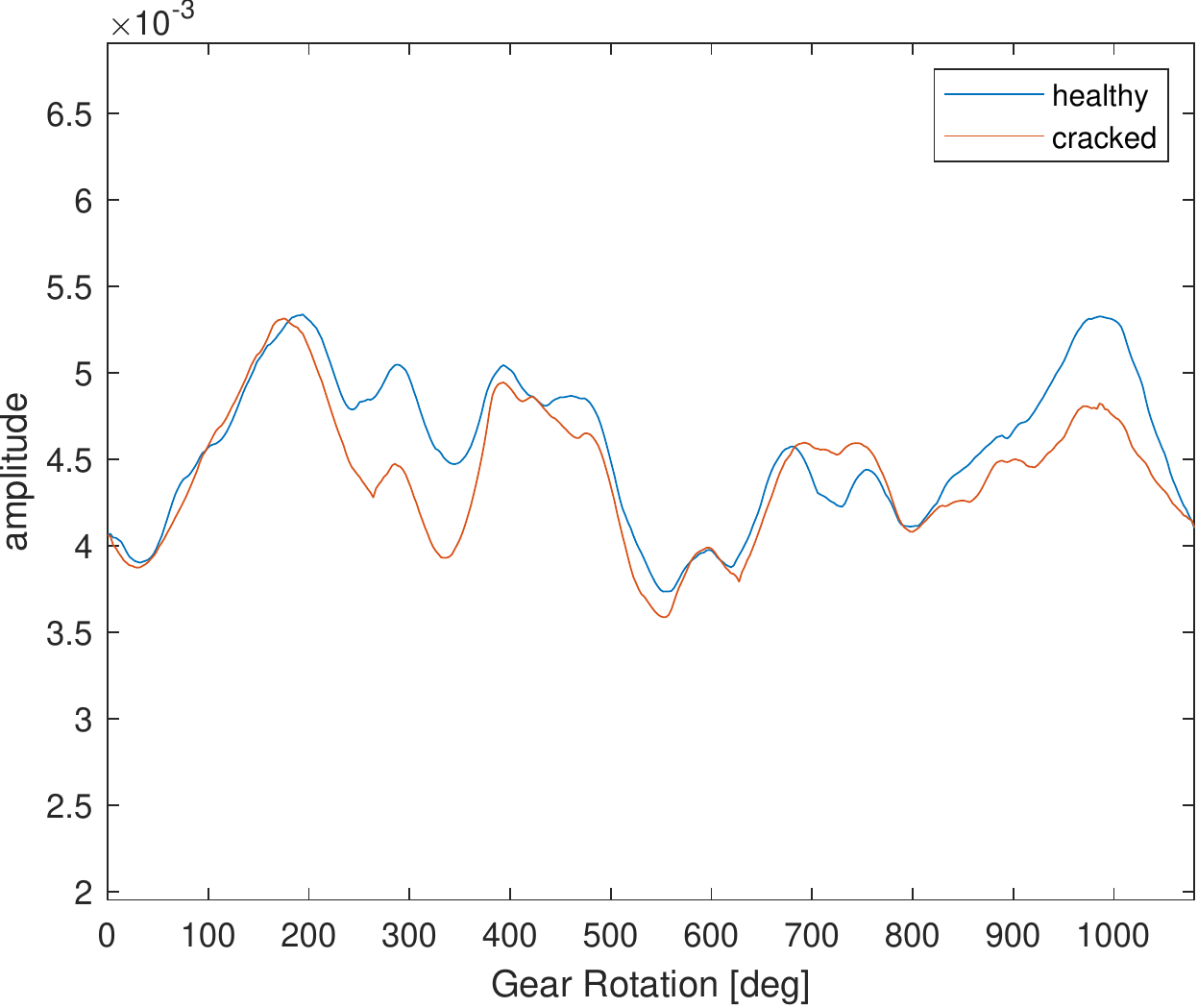}
    \caption{Instantaneous amplitude}
    \label{amplitude_tsa_5_b}
  \end{subfigure}
  \begin{subfigure}{0.33\textwidth}
  \includegraphics[width=\textwidth,height=3cm]{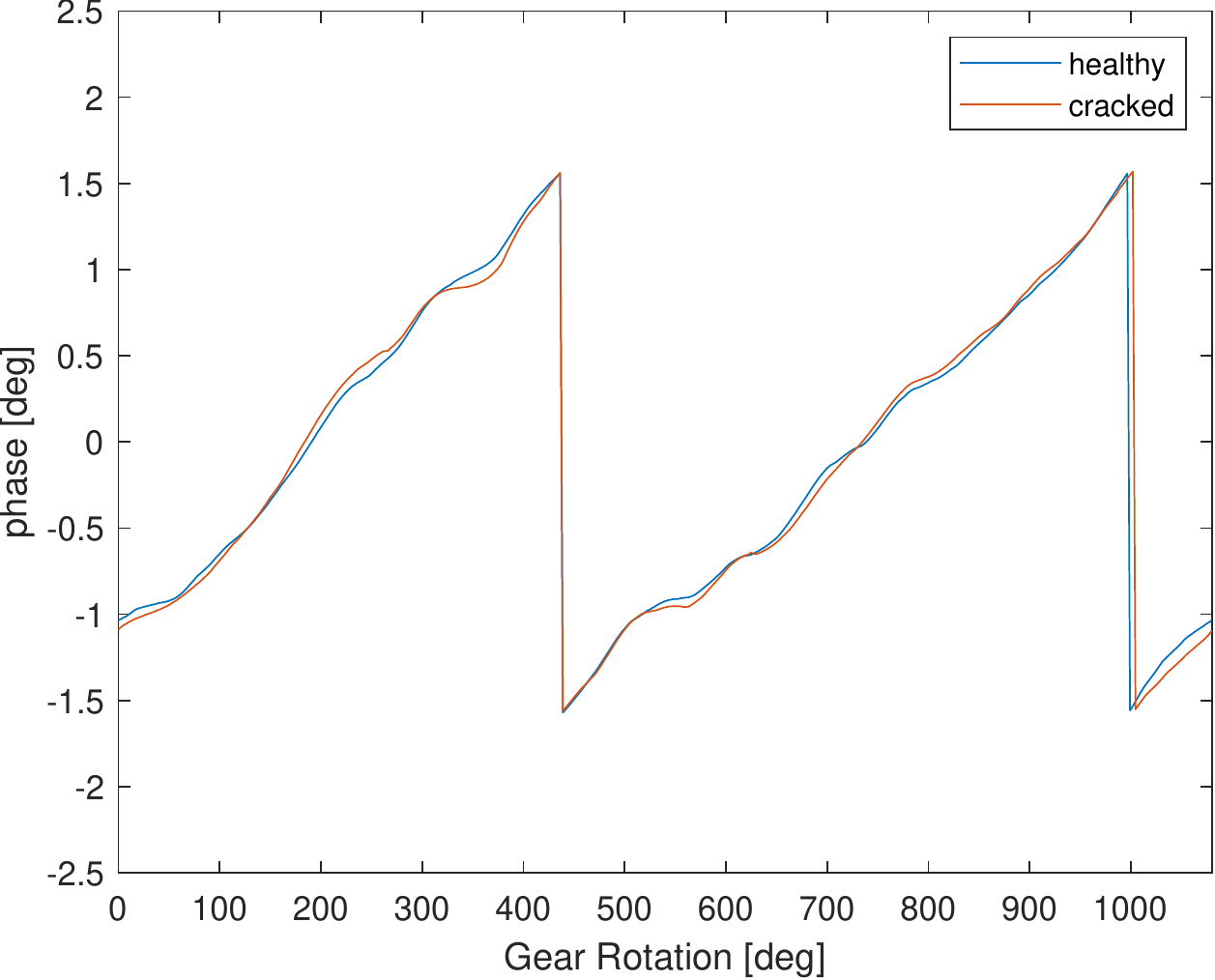}
  \caption{Instantaneous phase}
  \label{phase_tsa_5_c}
\end{subfigure}
 \caption{Results of TSA-based analysis, wind speed 5 m/s}
 \label{TSA_HT_5}
\end{figure}

\begin{figure}[h!]
  \begin{subfigure}{0.33\textwidth}
    \includegraphics[width=\textwidth,height=3cm]{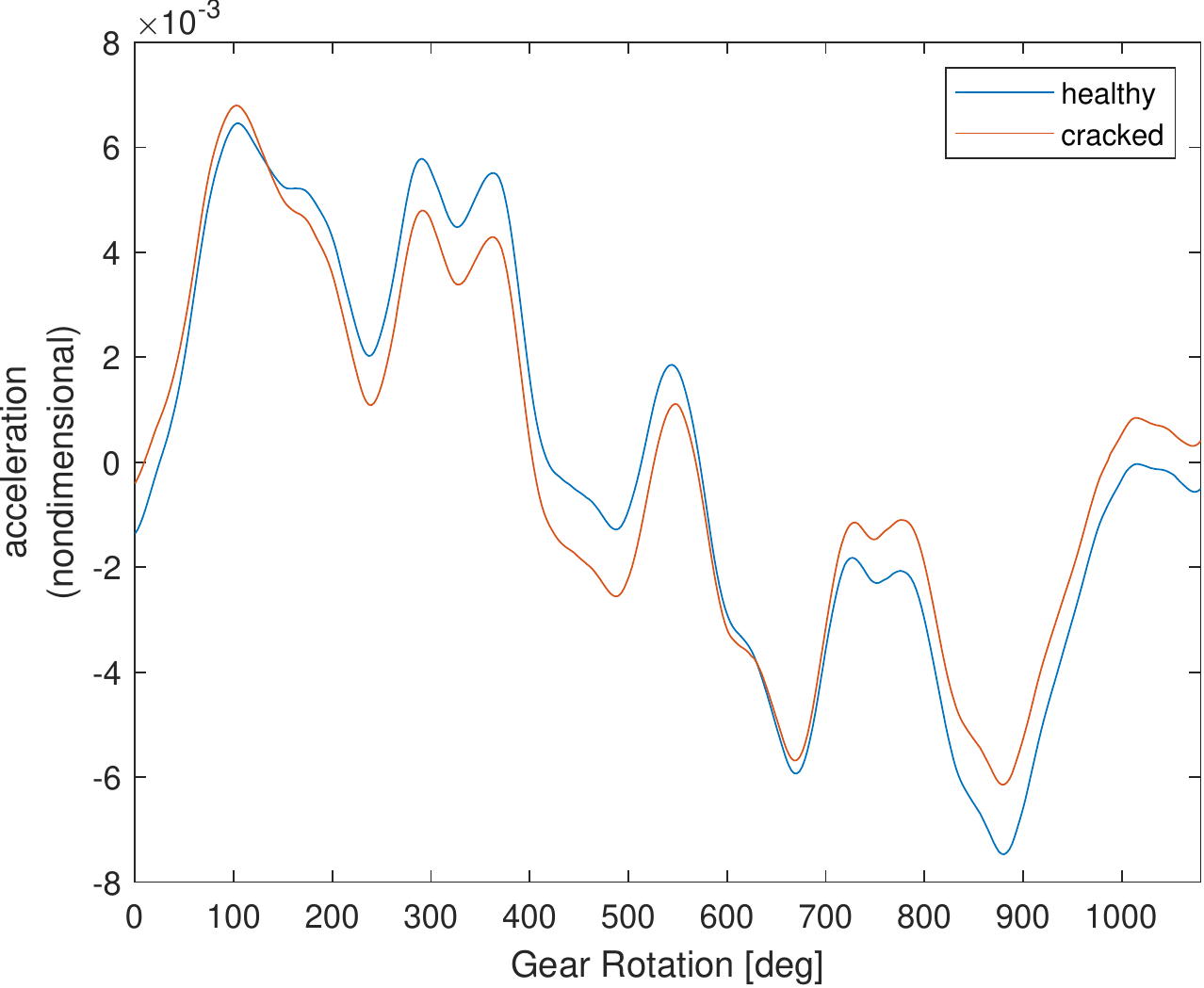}
    \caption{TSA signal}
    \label{TSA signal_13_a}
  \end{subfigure}
  \begin{subfigure}{0.33\textwidth}
  \includegraphics[width=\textwidth,height=3cm]{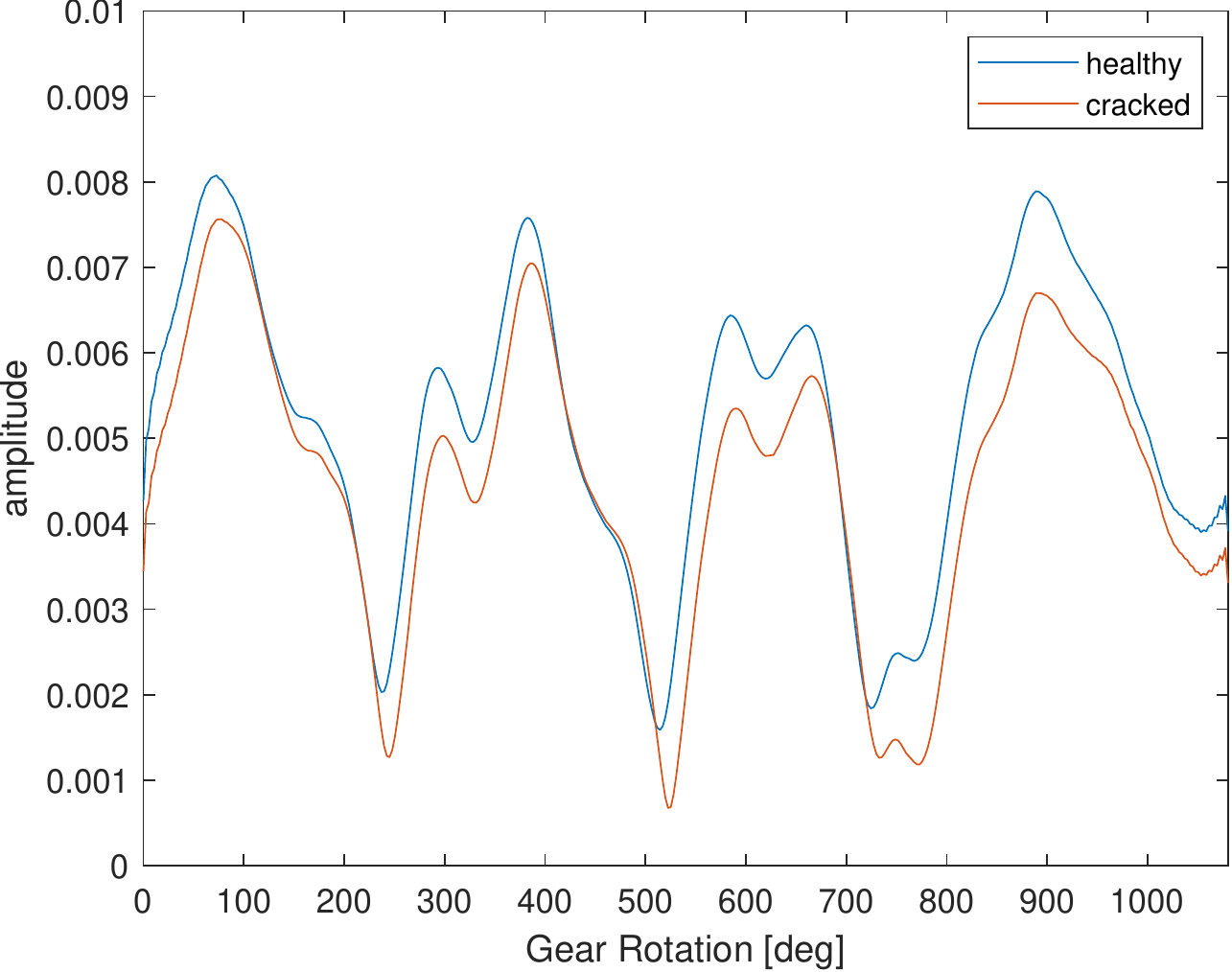}
  \caption{Instantaneous amplitude}
  \label{amplitude_tsa_13_b}
\end{subfigure}
\begin{subfigure}{0.33\textwidth}
\includegraphics[width=\textwidth,height=3cm]{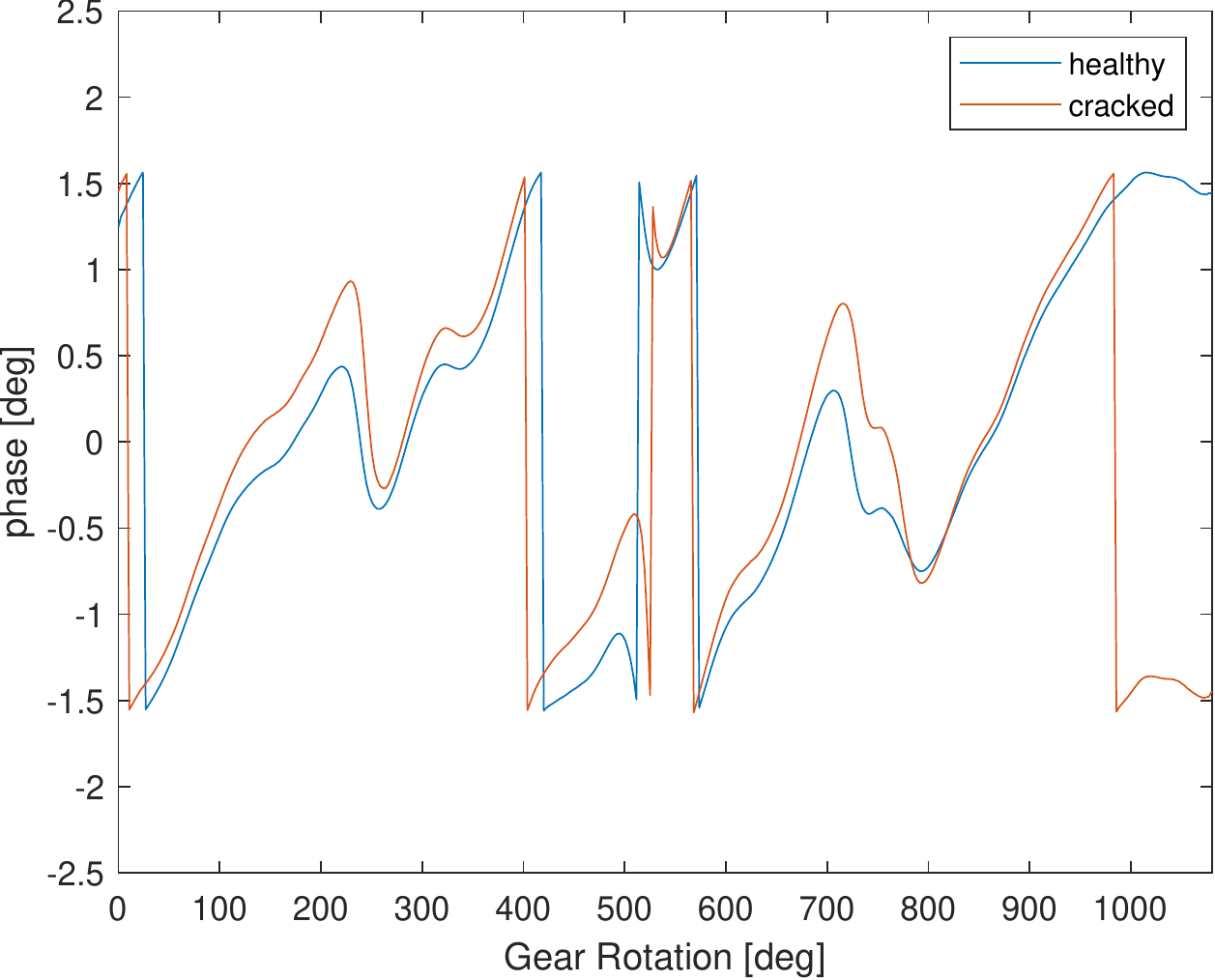}
\caption{Instantaneous phase}
\label{phase_tsa_13_c}
\end{subfigure}
\caption{Results of TSA-based analysis, wind speed 13 m/s}
\label{TSA_HT_13}
\end{figure}

Fig.~\ref{TSA signal_5_a} and Fig.~\ref{TSA signal_13_a} show the signals
average obtained applying the TSA to acceleration signals in
Fig.~\ref{wind_5_1080} and Fig.~\ref{wind_1080}, respectively. Since, our aim
was to use TSA to extract the vibration signal associated with the meshing
frequency, we initially set $T_r= \frac{1}{\Omega_{mesh}}=2$, so that the
distance between the multiples of the periodic time $T_r$ is equal to the
meshing period.  After that, we computed the instantaneous amplitude
(Fig.~\ref{amplitude_tsa_5_b} and~\ref{amplitude_tsa_13_b})  and phase
(Fig.~\ref{phase_tsa_5_c} and~\ref{phase_tsa_13_c}) of the TSA signals in order
to detect small magnitude variations and highlight the presence of damage. As
can be seen in Fig. \crefrange{amplitude_tsa_5_b}{phase_tsa_5_c} and
\crefrange{amplitude_tsa_13_b}{phase_tsa_13_c}, independently of the wind
condition, both the instantaneous phase and amplitude associated with the
healthy and cracked case show a very similar behavior. Thus, comparing the
instantaneous phase and amplitude computed in the different health scenarios,
we cannot emphasise any feature that indicates the presence of damage. The only
exception is in Fig.~\ref{phase_tsa_13_c}, which shows that in the instantaneous
phase associated with the cracked tooth, there is a sudden phase lag between
$900^\circ$ and $1000^\circ$ of the gear rotation. Such a phase lag marks a
substantial difference between the phase associated with healthy and cracked
health conditions, and may be considered a damage indicator. However,
Fig.~\ref{phase_tsa_13_c} shows us that phase lags also arise in the
instantaneous phase associated with the healthy signal and therefore may be
misleading to consider only them as damage indicators. Furthermore, shifting the
analysis focus only on the instantaneous amplitude and phase associated with the
cracked scenario, there is no particular anomaly arising at the angles at which
the cracked tooth interacts with the other gear ($67^{\circ}$, $427^{\circ}$ and
$787^{\circ}$) that provides us with evidence of damage. 

Notice  that, we performed the above TSA-based analysis also using different
values for the periodic time $T_r$. Specifically, we set $T_r = \frac{1}{m
\Omega}$ with $m \in \{1,2,\dots,10\}$, so that  the distance between the
multiples of the periodic time $T_r$ is equal to the periods of the meshing
frequency and its harmonics. Unfortunately, independently of the parameter choice, the
results were qualitatively similar to those attained by setting $T_r =
\frac{1}{\Omega_{mesh}}=2$. For the TSA approach we used the MATLAB
function ``tsa", which sets automatically the parameter $N$ in (\ref{def: TSA})
as the maximal amount of meshing periods in the time-span considered to analyze
the signal. 

In this work, we investigated TSA in combination with the HT, but there are other
TSA-based techniques~\citep{Randall2021}, relying for example on Kurtosis
analysis instead of the HT, which might give different results. 
Here we mainly want to show that TSA as other classical approaches may not
always be effective in the operating conditions we are considering.

\paragraph{Empirical mode decomposition (EMD) and Hilbert transform (HT)}
\label{empiricalmodedecomposition based }
The EMD-based strategy we present here, consists of applying EMD together with HT to detect
damages, and was introduced in the wind turbine damage
detection context to overcome issues experienced by studying the signals' 
spectral characteristics. See~\cite{anto} for a more exhaustive analysis of the 
EMD-based fault detection method we are going to review.\\
\subparagraph{Empirical mode decomposition (EMD)}
The essence of EMD is to decompose the signal into oscillatory functions, also
called intrinsic mode functions (IMFs). Each IMF  represents characteristics of
the signal associated with a  certain frequency band and are such that
\begin{equation}
  \label{signal EMD}
y(t)= \sum_{k=1}^N C_k(t)+r_N(t),
\end{equation}
where $y(t) \in \mathbb{R}$ is a given sensor signal, $C_k(t)$ is the $k$-th IMF
and $r_N(t)$ is the rest of the approximation performed by the IMFs. To identify
the desired information in the IMFs there is not a principled approach, but
it is important to know that the first IMFs represent structures in the signal
associated with the highest frequencies, and the ones that come after depict
lower frequency components~\citep{nor}. Generally, the first IMFs contain damage
indicators, because they represent the highest frequency structures in the
signal, which are those affected by the damage we are considering. 
For further details of the EMD algorithm see~\cite{nor}. We use the PyEMD python package\footnote{available at \url{https://github.com/laszukdawid/PyEMD}} for the following analysis. 
 
\paragraph{Application of an EMD-based strategy}
\begin{figure*}[p] 
    \begin{subfigure}{\textwidth}
      \includegraphics[width=\textwidth]{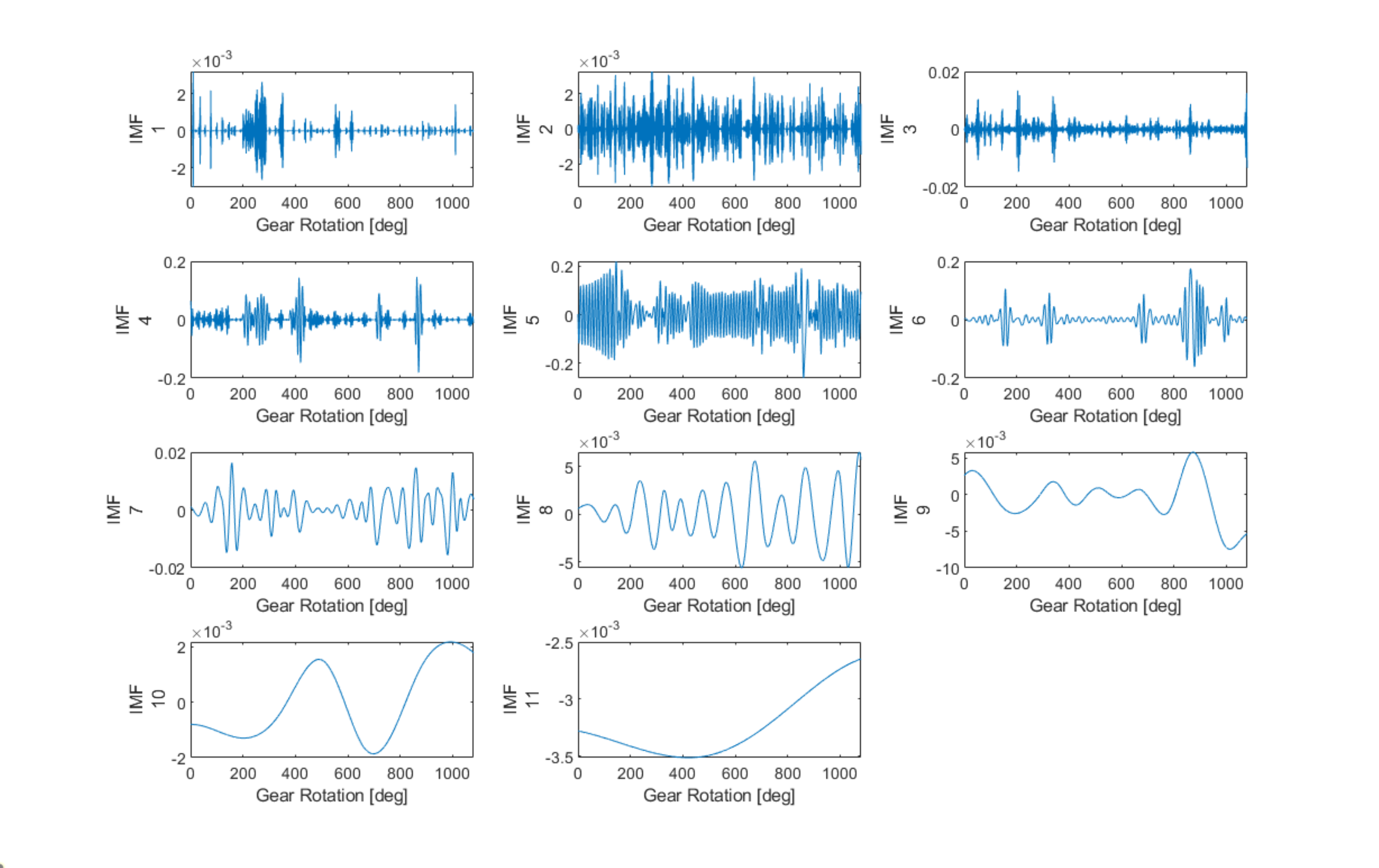}
      \caption{undamaged case}
       \label{IMF5}
    \end{subfigure}
    \begin{subfigure}{\textwidth}
      \includegraphics[width=\textwidth]{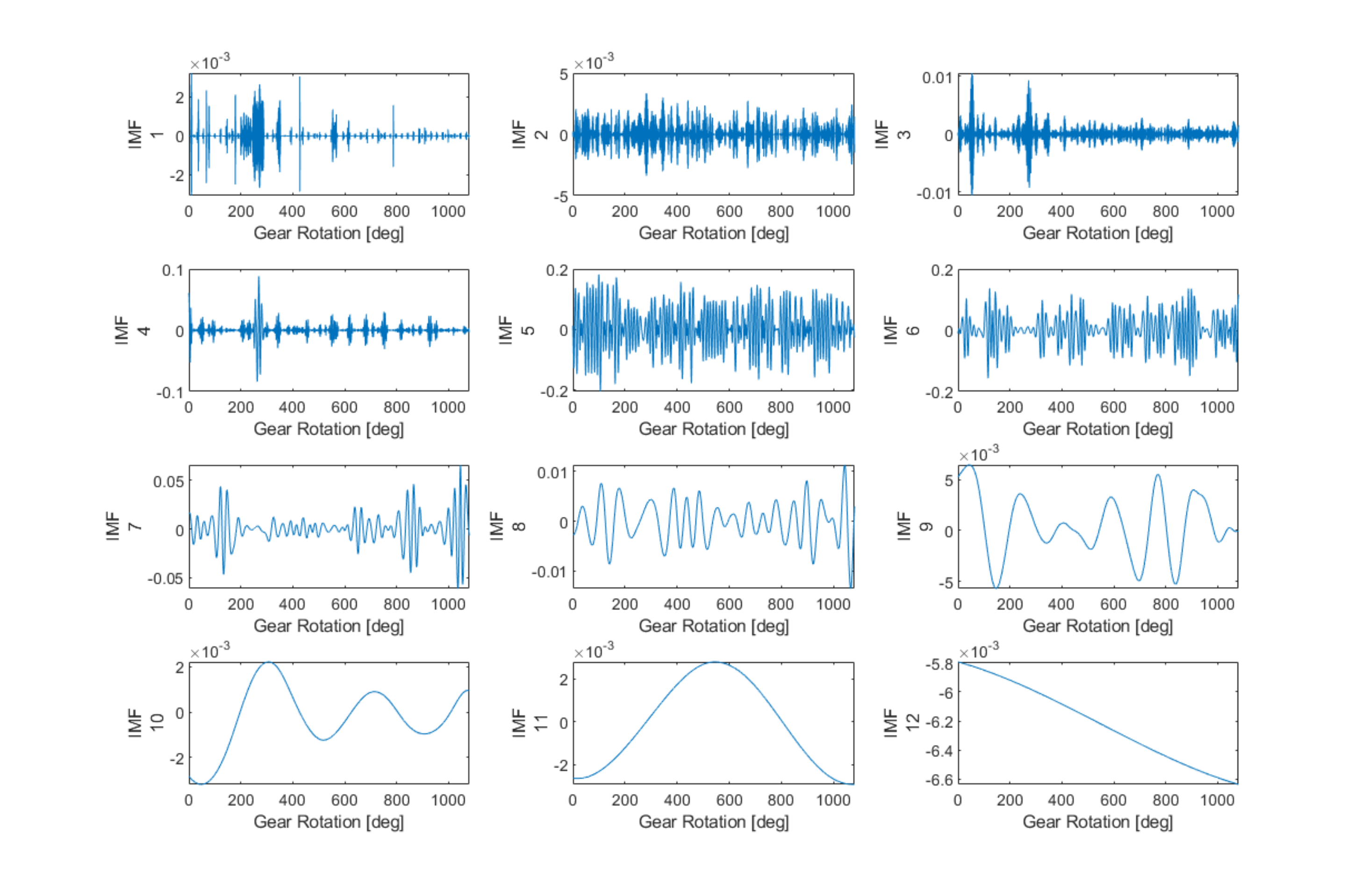}
      \caption{damaged case}
     \label{IMF5cr}
    \end{subfigure}  
  \caption {Intrinsic mode functions of simulations under  varying load condition with wind speed 5 m/s} 
  \label{IMFs5}
  \end{figure*}
Fig.~\ref{IMFs5}  represents the IMFs obtained applying EMD on the simulated
vibration signals in Fig.~\ref{wind_5_1080}. The signals in
Fig.~\ref{wind_5_1080} represent the vibration response of the modelled gearbox
under varying load condition associated with a wind speed of 5 m/s.
 
Looking at the IMFs in Fig.~\ref{IMFs5} one can observe that in the damaged case EMD has produced more IMFs
than in the healthy case. This happens because EMD perceived the effects of
damage as more signal components, and therefore it added IMFs to represent them.
Another explanation is that the additional IMF is related to the mode mixing
problem: some IMFs include damage features that are probably already contained
in another IMF. 
 
\begin{figure*}
  \begin{center}
   \includegraphics[width=0.7\linewidth, height=4cm]{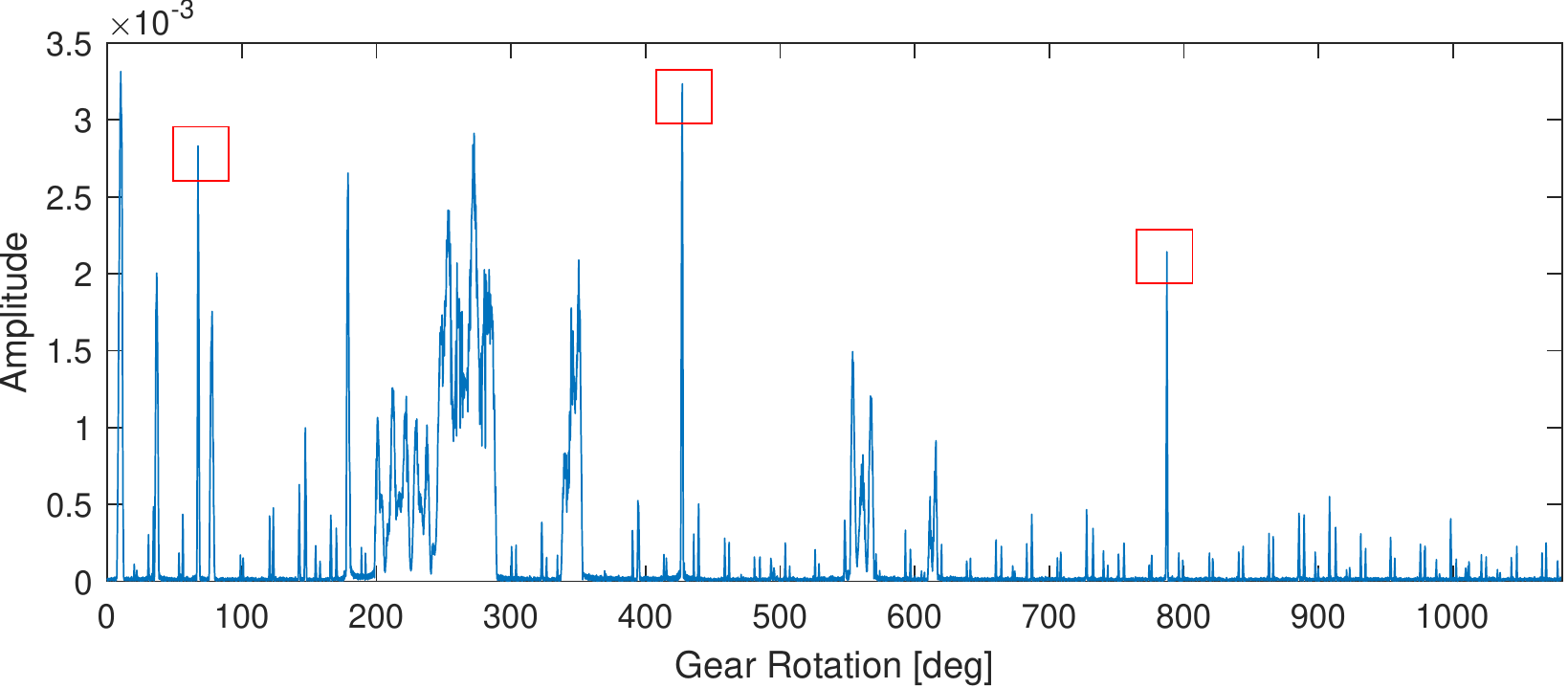}  
  \end{center}
\caption {Instantaneous amplitude of IMF 1 computed in the damaged case with wind condition 5 m/s. The pulses related
to the presence of damage are marked with red squares} 
\label{amplitude5}
\end{figure*}

Analyzing the IMFs in more detail, it can be seen that the IMF 1 in Fig.~\ref{IMF5cr} contains damage features. The damage features appear as increases
in magnitude or pulses in the IMF, arising at the angles in which the cracked
tooth affects the signal. Looking at the instantaneous amplitude of the first
IMF (Fig.~\ref{amplitude5}), it can be clearly seen that at the degrees where
the cracked tooth interacts with the other gear, there are sudden increases of
magnitude. Such increases of magnitude allow us to identify the presence of
damage and, therefore, to perform the damage identification task. Unfortunately,
from Fig.~\ref{amplitude5}, it is also clear that together with those related
to damage, there are other pulses in the IMF 1, probably related to the varying
wind condition, which make the identification of damage features a very
challenging task. In particular, without previous knowledge about where the
cracked tooth affected the signal, we would not have been able to identify it. A
possible explanation for such an event is that the mode mixing problem occurred.
Thus,  information related to the frequency bands affected by damage and by wind turbulence is represented in the
same IMF and can not be distinguished trivially. As a consequence of the mode mixing
problem, we could also have that different IMFs contain the same frequency band
information. For instance, using acceleration signals representing a scenario
related to those we consider in this work, we found that
sudden increases of magnitude caused by a cracked tooth could be identified in
different IMFs. Thus, information related to the frequency band affected by the
damage was contained in different IMFs. Such a problem was also experienced
in~\cite{anto}, where they use EMD to study signals similar to those analyzed
here. 
Analyzing the simulated data generated with the higher wind speed
of 13 m/s we could not successfully perform the damage identification task.
Specifically, applying EMD to signals in Fig.~\ref{wind_1080_b}, no IMFs could
visually highlight the different nature of the effects of the wind and of the
damage on the signal. 

After this brief analysis, it is clear that one of the disadvantages of EMD is
that information about the damage's effects in the signal is not localized: EMD
does not tell us which frequency-band each IMF exactly represents. Thus, we
do not know in which IMF damage features can be detected. The second main weakness
of EMD  is the mode mixing problem, which affects the interpretability of the
method and its capability to perform the fault detection task effectively. As we
will see, the numerical procedure we propose overcomes these problems thanks to
the characteristics of the mrDMD algorithm.

\subsubsection{mrDMD-based approach for damage detection }
For our numerical experiments with mrDMD, we use the simulated gearbox vibration
signals shown in Fig.~\ref{wind_5_1080} and Fig.~\ref{wind_1080} that
represent the vibration response of the modelled gearbox in the damaged and
undamaged scenarios under the two different wind conditions we are considering.
The same data was used to test the EMD-based method previously.
The length of the delay $d$ has been fixed to $d=32000$. Thus, looking at the residual $\mathbf{r}_0$ we
are analysing a portion of the signal representing its temporal evolution for
$860^{\circ}$ of the shaft rotation. The number of the  decomposition levels $L$
in the mrDMD algorithm has been set to $L=11$. We use the PyDMD python package\footnote{available at \url{https://mathlab.github.io/PyDMD/}}, where we use the previously mentioned
threshold strategy for the choice of parameters. 

\begin{figure*}
      \begin{subfigure}{0.5\textwidth}
        \includegraphics[width=\textwidth,height=3cm]{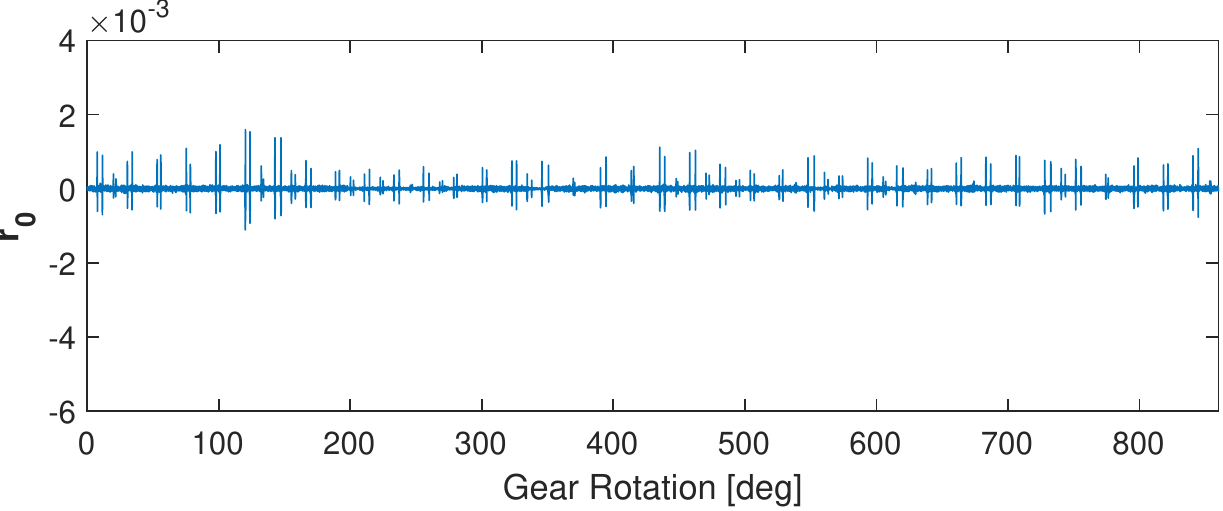}
        \caption{undamaged case}
        \label{Residual5_h}
      \end{subfigure}
      \begin{subfigure}{0.5\textwidth}
        \includegraphics[width=\textwidth,height=3cm]{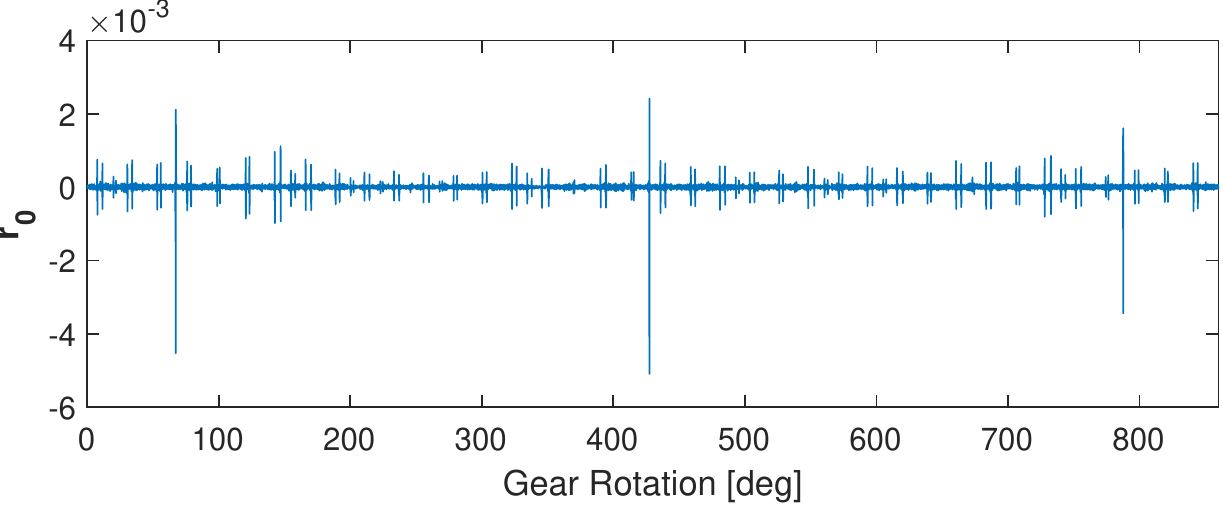}
        \caption{damaged case}
        \label{Residual5_cr}
      \end{subfigure}
      \caption {Residual of a simulation under varying load conditions with wind speed 5 m/s, computed calculating $L=11$ decomposition levels}  
      \label{residuals1}
    \end{figure*}
\begin{figure*}
    \begin{subfigure}{0.5\textwidth}
      \includegraphics[width=\textwidth,height=3cm]{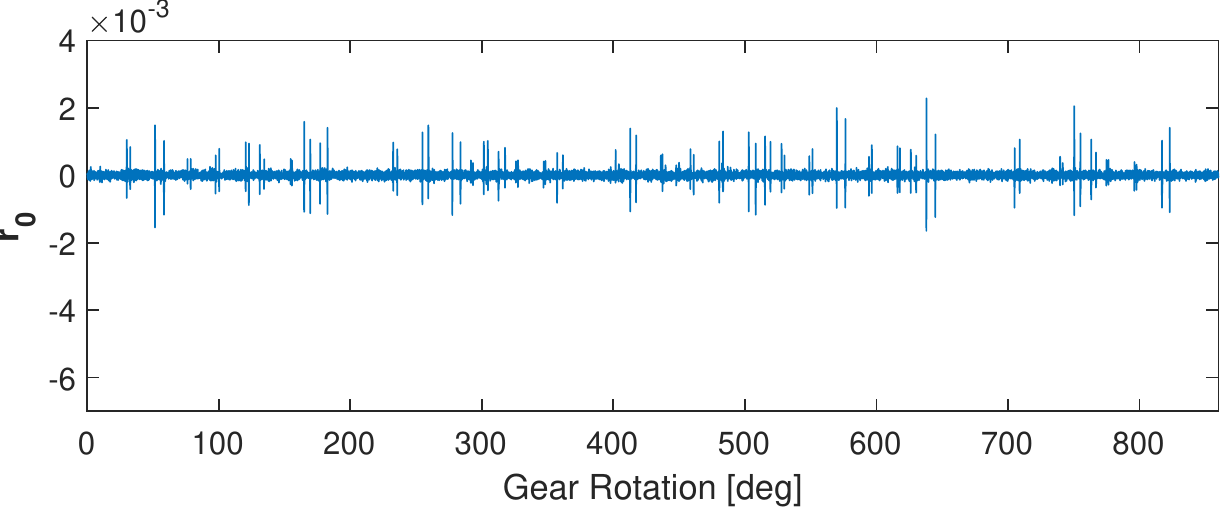}
      \caption{undamaged case}
      \label{Residual13_h}
    \end{subfigure}
    \begin{subfigure}{0.5\textwidth}
      \includegraphics[width=\textwidth,height=3cm]{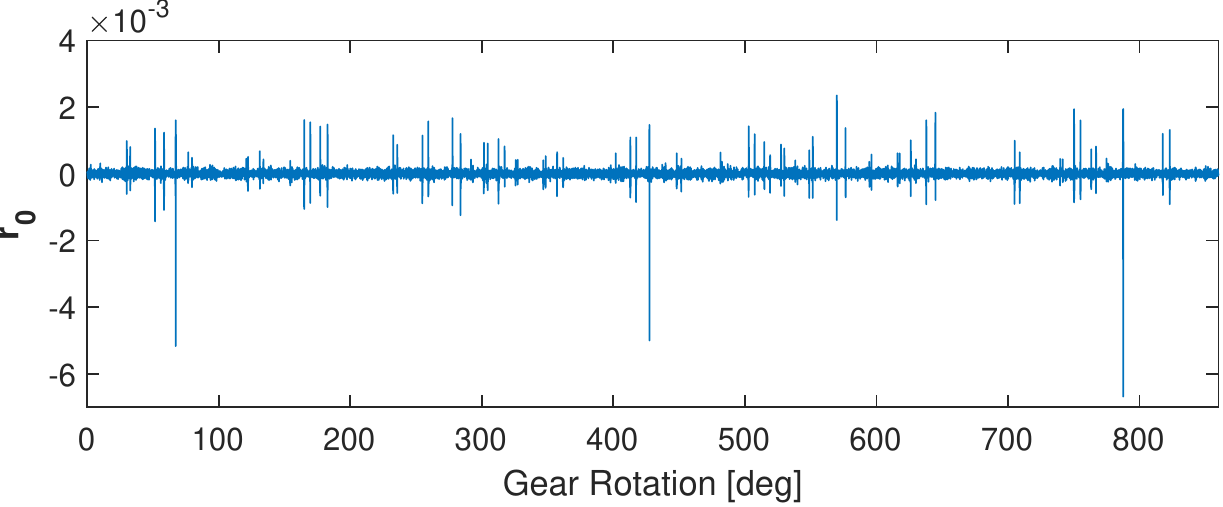}
      \caption{damaged case}
      \label{Residual13_cr}
    \end{subfigure}
  \caption {Residual of a simulation under varying load conditions with wind speed 13 m/s, computed calculating $L=11$ decomposition levels }  
  \label{residuals2}
\end{figure*}
\begin{figure}
    \begin{subfigure}{0.5\textwidth}
    \includegraphics[width=\textwidth,height=3cm]{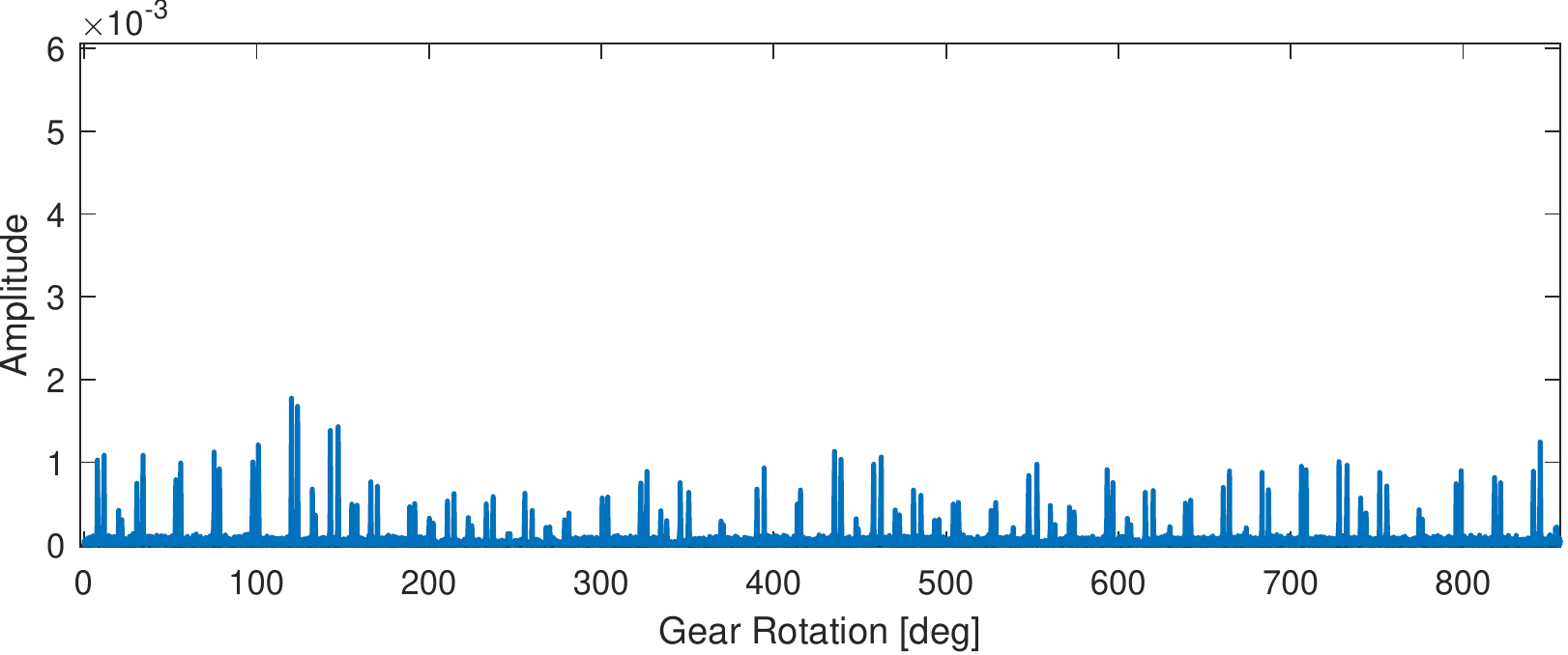}
    \caption{undamaged case}
    \label{amplitude_residuals5_a}
  \end{subfigure}
  \begin{subfigure}{0.5\textwidth}
  \includegraphics[width=\textwidth,height=3cm]{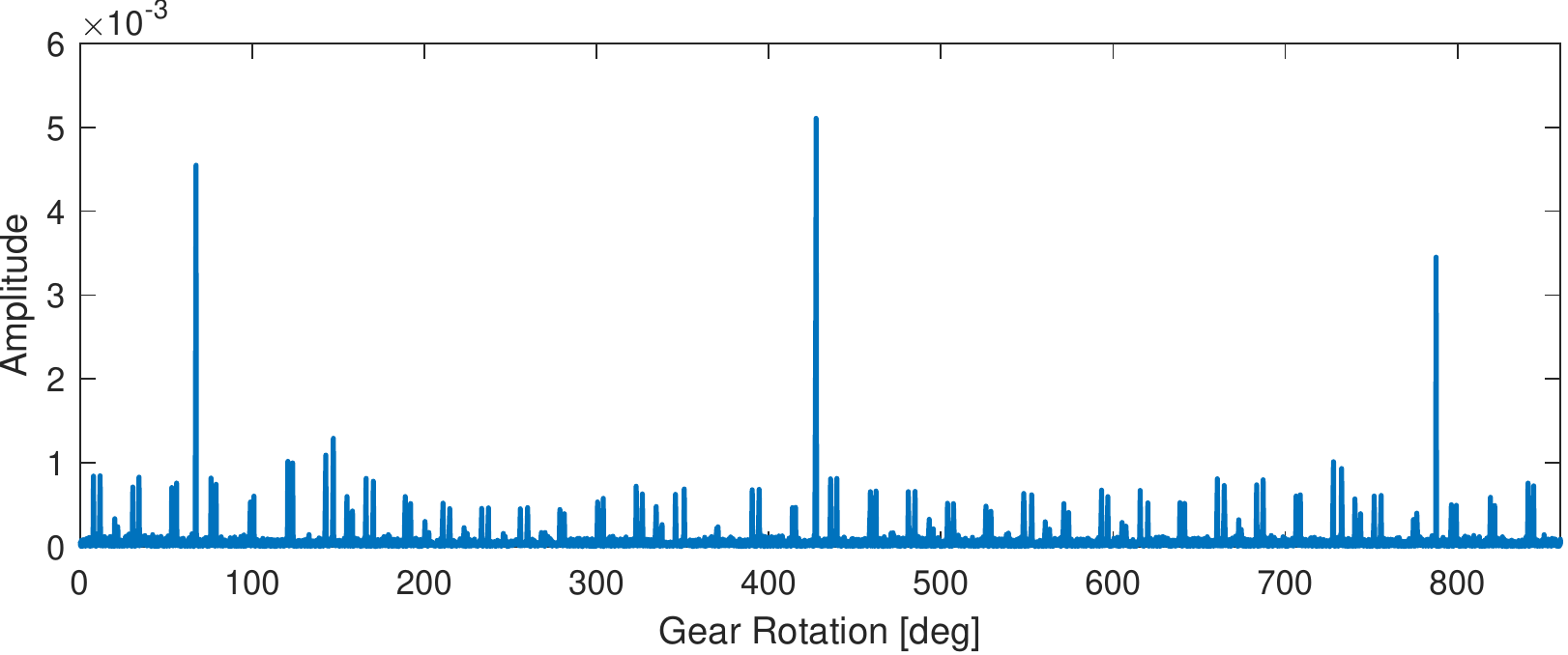}
  \caption{damaged case}
  \label{amplitude_residuals5_b}
\end{subfigure}
  \caption{Instantaneous amplitude of residual $\mathbf{r}_0$ ($L=11$), wind speed 5 m/s}
  \label{amplitude_residuals5}
\end{figure}
\begin{figure}
  \begin{subfigure}{0.5\textwidth}
  \includegraphics[width=\textwidth,height=3cm]{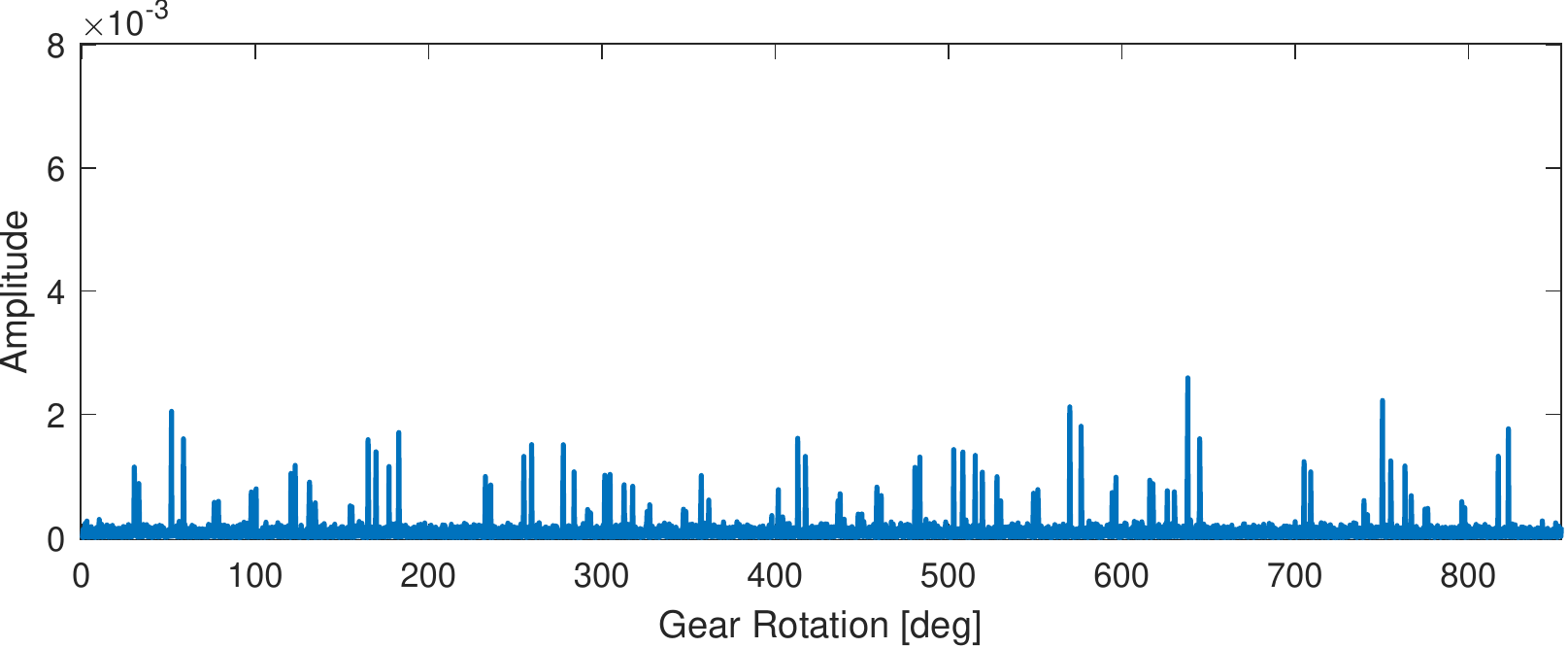}
  \caption{undamaged case}
  \label{amplitude_residuals13_a}
\end{subfigure}
\begin{subfigure}{0.5\textwidth}
\includegraphics[width=\textwidth,height=3cm]{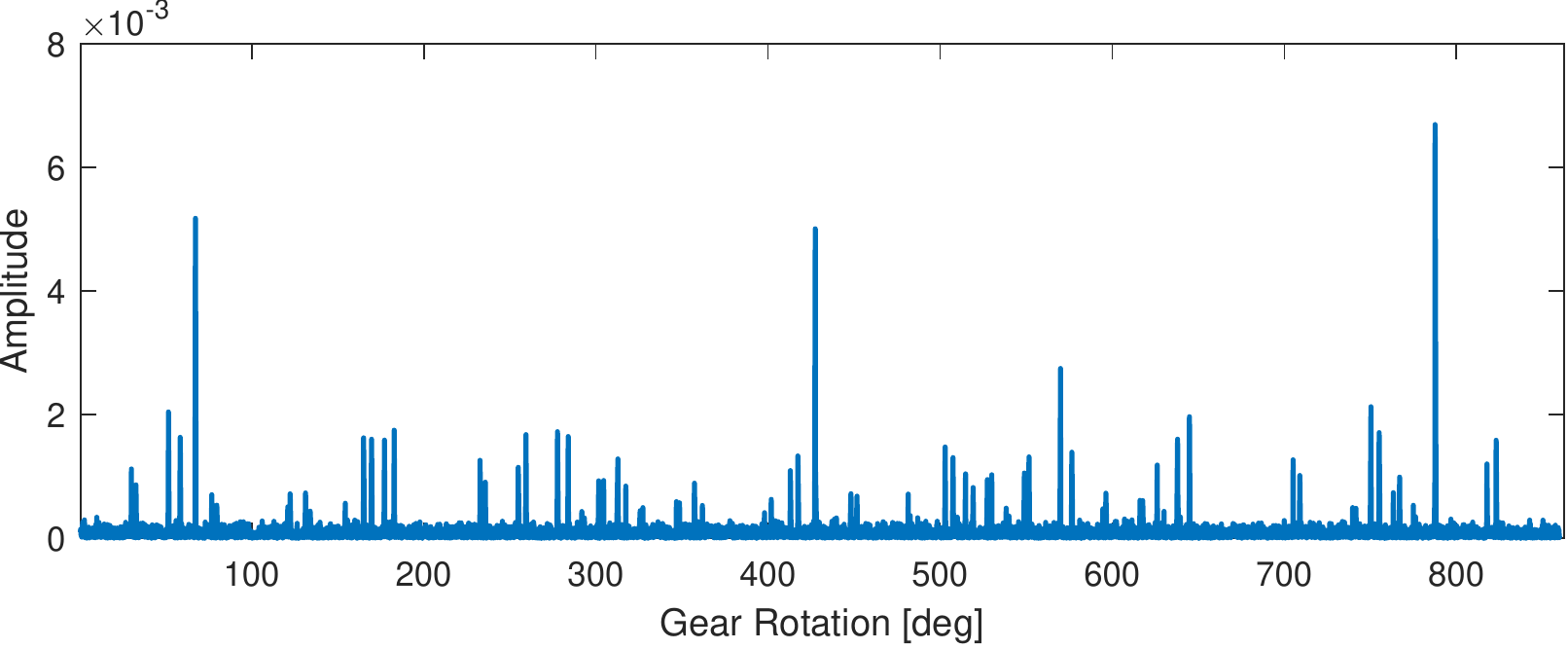}
\caption{damaged case}
\label{amplitude_residuals13_b}
\end{subfigure}
\caption{Instantaneous amplitude of residual $\mathbf{r}_0$ ($L=11$), wind speed 13 m/s}
\label{amplitude_residuals13}
\end{figure}

\paragraph{MrDMD-residual analysis  on simulation data}
Fig.~\ref{residuals1} and Fig.~\ref{residuals2} represent the results of the
mrDMD-based approach that we proposed in Section~\ref{multiresolution_dmd}.
The first observation we make, comparing the residuals obtained from simulations
produced with different wind conditions in the undamaged case (Fig.
~\ref{Residual5_h},~\ref{Residual13_h}), is that when the wind speed is higher
the high-frequency signals' structures represented by the residuals show a more
irregular behaviour. This is because high-frequency structures are affected by
wind turbulence, which is more relevant and volatile for higher wind speed
conditions. The second observation concerns simulations representing the damaged
and undamaged cases in the different wind conditions. Considering only the
residual $\mathbf{r}_0$, the effects of the damaged tooth are clearly visible as sudden
increases in magnitude arising at the angles in which the cracked tooth affects
the signal (Fig.~\ref{Residual5_cr},~\ref{Residual13_cr}). The effects of the
cracked tooth on the residuals can be further analyzed by considering their
instantaneous amplitudes (Fig.~\ref{amplitude_residuals5} and Fig.
~\ref{amplitude_residuals13}) computed through the Hilbert transform.
Independently of the wind condition considered and the residuals' mean breadth,
looking at the instantaneous amplitudes, the different characteristics of the
residuals associated with the damaged and undamaged cases are even more evident.
The instantaneous amplitudes related to both the healthy and damaged scenarios
present a spiky structure. However, in the undamaged cases (Fig.
~\ref{amplitude_residuals5_a},~\ref{amplitude_residuals13_a}), the spikes are of
similar magnitude over the time-span analyzed, while in damaged scenarios (Fig.
~\ref{amplitude_residuals5_b},~\ref{amplitude_residuals13_b}), the magnitudes of the
spikes are much higher at the angles in which the cracked tooth affects the
signal, making the damage identification even easier than it was considering
only the residuals.
 
It can be clearly seen that the numerical procedure we propose can be
effectively employed to highlight the effects of the cracked tooth on the
signal, distinguishing them from the effects of the varying load condition, and
enabling us to identify the damage independently of the wind condition.
Moreover, a consistent advantage of the proposed strategy with respect to EMD is
that the information related to damage is localized. With
the proposed strategy, all the information required is contained in the
residual. Note that, due to the mode mixing problem, such a residual
strategy is not feasible with EMD. To effectively employ EMD to detect damages, a user search
among the produced IMFs is needed to identify those structures containing information
related to damage.

We note that we also applied our procedure on signal simulated  with parameters
other than those represented in Table~\ref{table: simulation parameters}. Specifically, we considered two
additional internal excitation parameter settings and two crack lengths in the
numerical model. In addition to that, we performed each experiment in each of
the two wind conditions we considered so far. Overall, we performed
experiments in eighteen operating conditions, which gave results comparable to
those here presented. The specific parameter settings are given in Appendix A.


\begin{figure}
  \begin{subfigure}{0.5\textwidth}
  \includegraphics[width=\textwidth,height=3cm]{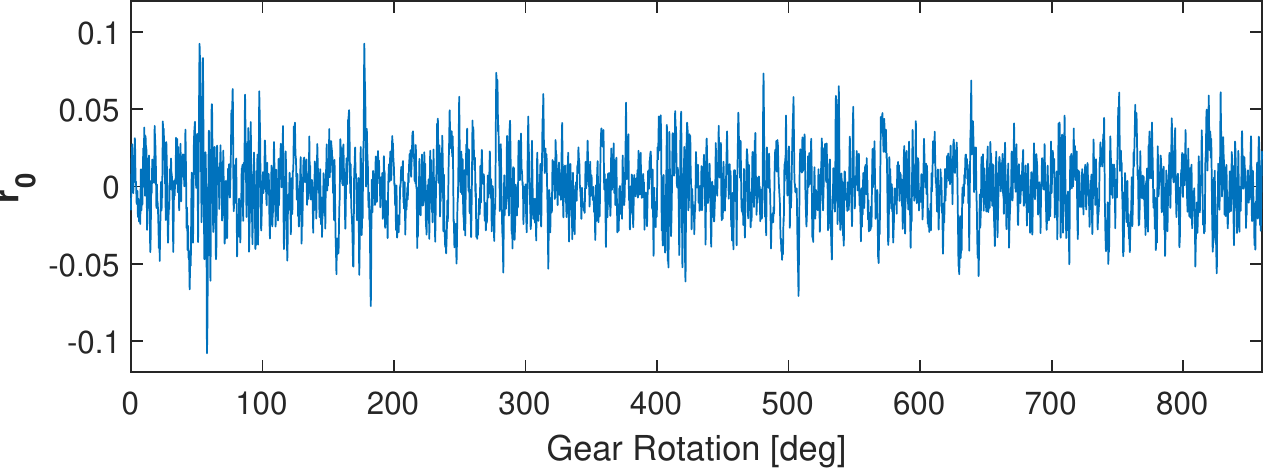}
  \caption{undamaged case}
\end{subfigure}
\begin{subfigure}{0.5\textwidth}
\includegraphics[width=\textwidth,height=3cm]{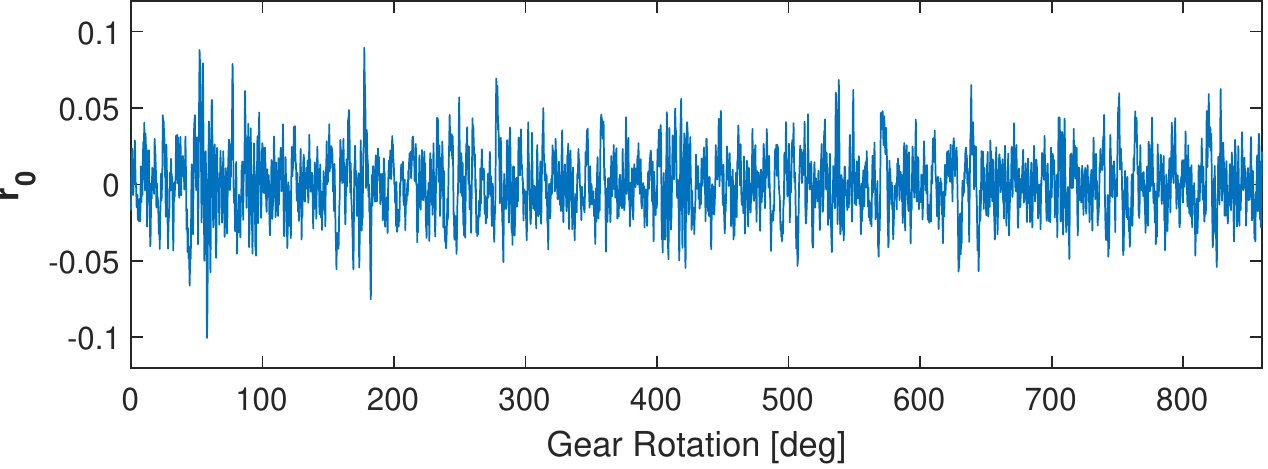}
\caption{damaged case}
\end{subfigure}
\caption{Residual of a simulation under varying load conditions with wind speed 13 m/s, computed calculating $L=7$ decomposition levels. Using these lower frequencies the damage cannot be detected.}
\label{residuals_l7}
\end{figure}

We remark that the choice of the number of decomposition levels $L$ to compute
strongly affects the effectiveness of our analysis. Specifically, if the number
of calculated decomposition levels is not high enough, the residual would
consist of structures associated with a frequency band that would also include
those low-frequencies not affected by the damage. As we will see shortly,  the
presence of the cracked tooth has a negligible effect on the structures
associated with lower frequencies. Therefore, the residual would also consist of
components that do not provide any meaningful insight  for the damage
identification process, which would dilute the valuable information.
Consequently, the residual would not be able to highlight the presence of
damage. To observe such a phenomenon, see Fig.~\ref{residuals_l7}, which represents the residuals computed
applying the proposed procedure on vibration signals in Fig.~\ref{wind_1080} associated with wind
speed 13 m/s, considering a delay $d=32000$ and calculating  $L=7$  decomposition levels. It can be clearly seen
that, in this case, $L=7$  decomposition levels are not enough to encode in the residual only the
relevant information associated with the highest frequencies and related to damage. There is no obvious
difference between the two residuals in Fig.~\ref{residuals_l7}. Thus, they can not be employed to perform
damage detection.

To correctly choose the number of decomposition levels to compute
and successfully apply the proposed procedure for damage detection, we need
information about the frequencies affected by the damage we want to identify.
Fortunately, in condition monitoring, the damage we want to identify and the
frequencies affected by it can often be assumed to be known~\citep{anto}.

\paragraph{Modal analysis with mrDMD on simulation data}
We now investigate more closely the modes' properties and how the proposed
procedure works. To this aim, we analyze the modes computed via mrDMD,
calculating $L=11$ decomposition levels, and used to obtain the residuals $\mathbf{r}_0$
in Fig.~\ref{residuals2}. We mainly focus on the scenario associated with the
wind speed of 13 m/s because signals generated employing the same gearbox model
we use and considering a similar wind condition were already simulated and
studied in~\cite{anto}. Thus, we can exploit the knowledge developed there about
the signals' physical characteristics to better interpret and understand the
properties of the modes generated by mrDMD.
 
Recall that we can compute the residual $\mathbf{r}_0$ associated
with a signal by subtracting from the first time-delay snapshot, $\mathbf{\tilde{y}}(t_{0})$, its mrDMD
approximation $ \mathbf{\hat{y}}(t_{0})$, i.e., $\mathbf{r}_0 = \mathbf{\tilde{y}}(t_{0}) - \mathbf{\hat{y}}(t_{0})$.
According to (\ref{formal_version}), if we set $t_0=0$, the mrDMD approximations of the first time-delay
snapshots can be explicitly written as 
\begin{equation}
  \mathbf{\hat{y}}(t_{0})= \sum_{l=1}^L\sum_{k=1}^{m_L}b_{k}^{(l,1)}\pmb{\phi}_{k}^{(l,1)}.
\end{equation}
Specifically, the linear combination of the modes $\phi_{k}^{(l,1)}$, computed
at the first time bin of each decomposition level, gives the mrDMD approximation
of the first time-delay snapshot representing the evolution of the analyzed
signal for a time-span as large as the delay. It is important to notice that to
obtain the residuals $\mathbf{r}_0$ only the modes computed at the first time bin of each
decomposition level are needed, and that each one of the modes
$\pmb{\phi}_{k}^{(l,1)}$ is associated with 
a frequency $\omega_{k}^{(l,1)}$.
 
In the scenario associated with wind speed 13 m/s, we know that the wind
turbulence mainly affects the signal at its highest frequencies, as it was
observed in~\cite{anto} for the similar wind speed of 12 m/s.
\begin{figure*}
\begin{center}
\includegraphics[width=\textwidth,height=9cm]{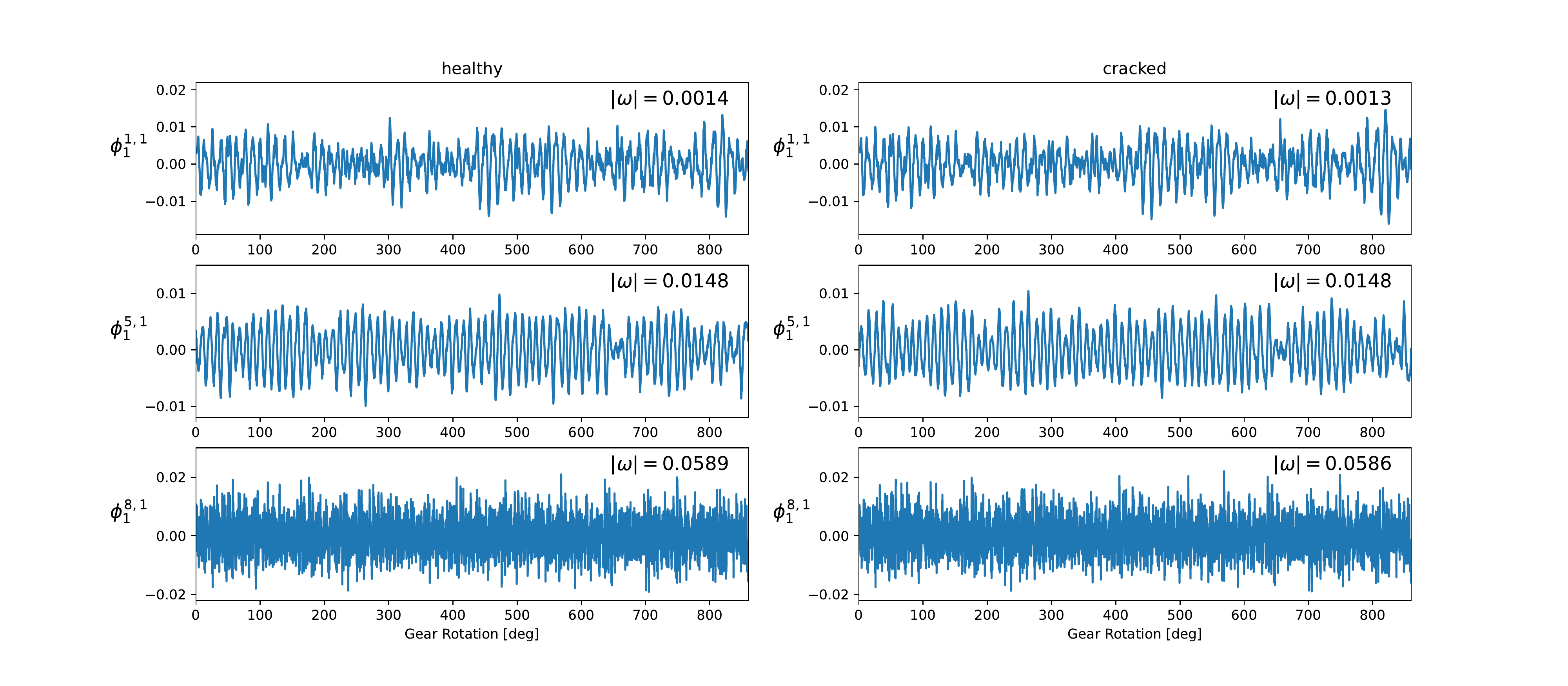}
\end{center}    
\caption {First modes computed at the first time bins of the decomposition
levels $l=1,5,8$ obtained applying the proposed analysis on simulation under varying load
conditions with wind speed 13 m/s and calculating $L=11$ decomposition levels} 
 \label{modes_mrdm}
\end{figure*}
Fig.~\ref{modes_mrdm} depicts 3 of those modes used to compute the
reconstructed time-delay snapshots $\mathbf{\hat{y}}_0$ required to obtain the residuals
$\mathbf{r}_0$  in Fig.~\ref{residuals2}. The modes in Fig.~\ref{modes_mrdm} have
been computed applying the proposed analysis on vibration signal associated with
wind speed 13 m/s in the healthy and damaged scenario (Fig.~\ref{wind_1080}).
Specifically, they are the first modes computed at the first time bins of
the decomposition levels $l=1,5,8$. Looking at Fig.~\ref{modes_mrdm}, we can
make a number of observations that give us a better understanding of the modes'
properties and that generally hold for modes computed at different time bins and
decomposition levels. Analyzing Fig.~\ref{modes_mrdm}, it is clear that the
higher the decomposition level, the higher is the energy content $|\omega|$
associated with the modes computed in it. Here $\omega$ is the frequency defined
in Algorithm~\ref{alg:DMD}. This fact is empirical proof of what we already
saw from a theoretical perspective in Section~\ref{multiresolution_dmd}, where
the mrDMD has been explained, and such a property of the modes was already
introduced. Another consideration is that modes computed at the highest
decomposition level are much noisier than those calculated at lower levels. This
can be explained by the fact that high-energy modes incorporate the effects of
the wind turbulence on the signal, which as we know from~\cite{anto} are more
evident at high-frequencies. Thus, it further confirms that high-energy modes
contain structures associated with high-frequencies, and that the residuals
consist of those structures associated with the highest frequencies. Finally, we
observe that in Fig.~\ref{modes_mrdm} the modes computed in the healthy
scenario and their associated energies ($|\omega|$) are qualitatively the same
as those calculated from signals that include the damage. This, can be motivated
by the fact that the modes depicted are associated with frequencies not affected
by the presence of the cracked tooth. More in general, we found that in this
scenario, the effects of damage can be effectively identified and discerned from
those related to wind turbulence only in those structures associated with the
highest frequencies.

\paragraph{Automation}
An advantage of the procedure we propose, compared to EMD, is that it may allow
the automation of the damage detection process. The main reason for that is that
it localizes the information related to damage. 
 
Considering how EMD works, here, after a certain number of IMFs have been
produced, a user made choice has to be performed to pick out all those IMFs that
may contain damage features, and therefore need to be further analyzed. In
general, we know that information related to the damage we are considering
should be in the first IMFs, which are those associated with higher frequencies
in the signal. However, this is only a rule of thumb, and because of the mode
mixing problem, there is no principled way to perform the IMFs' choice.
Therefore, using EMD, there is no obvious method to automatize the damage
identification process.
 
On the contrary, the procedure we propose localizes the information related to
damage in the residual. Thus, there is no need for a modes' selection, and the
cracked tooth detection task reduces to identifying sudden increases of
magnitude and anomalies within the residual. A broad spectrum of peak detection
methodologies is available and should be applicable to automatize the
damage identification process~\citep{Du2006,Harmer2008,Jordanova}. Automatizing
the damage identification process would be very advantageous in practice because
it would allow to monitor the health condition of several gearboxes
simultaneously and continuously, but it is out of the scope of this paper.


\subsection{Results with experimental data}
There is a lack of real or experimental data for the considered damage scenario
under the presence of wind fluctuations. Instead, we now consider experimental
data under controlled operating situations that offer other non-trivial
challenges than the previously studied simulation data. Specifically, in the
experimental data, the acceleration signals are mixed with other signals from
different interacting gearbox components and are significantly attenuated by the
dynamics of the case and other elements in the transmission path.
Note that with the results on this experimental data, we want to show that the
proposed mrDMD approach can overcome these additional challenging effects. We do
not claim that for these controlled operating situations, our method can detect damages
more effectively than other gear diagnostic techniques. 

\begin{figure} [h]
  \begin{center}
  \includegraphics[width=6cm,height=4cm]{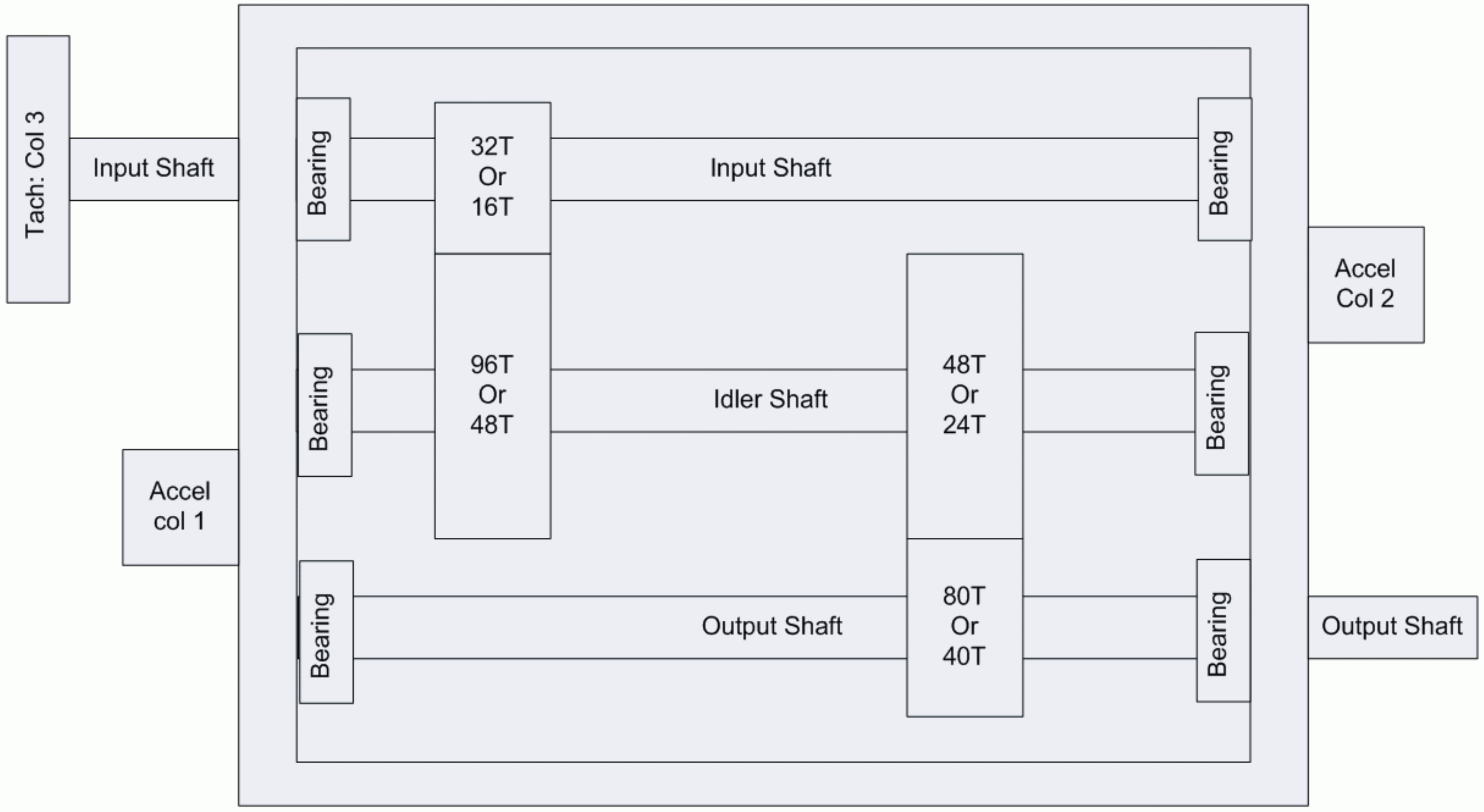}
  \caption{Schematic of the gearbox~\citep{PHM2009}}
  \label{schematic_gearbox}
\end{center}
\end{figure}

\begin{table}
  \caption{Geometry double stage reduction gearbox with spur gears set-up and its operating conditions}

      \begin{tabularx}{\textwidth}{cc@{\hspace{0.1cm}}cccccc}
        \hline
        \multicolumn{3}{c}{Geometry}& \multicolumn{5}{c}{ Operating conditions (angular speed)} \\
        \cline{1-2} \cline{4-8}
          components & Spur gear  & & 1st-HL  & 2nd-HL   & 3rd-LL  &  4th-HL &  5th-LL \\
        \hline
        Input shaft: 1-Input Pinion & 32 teeth & & 30 Hz & 35 Hz & 40 Hz & 45 Hz & 50 Hz \\
        Idler shaft: 1st idler gear & 96 teeth& &10 Hz & 11.7 Hz & 13.3 Hz & 15 Hz & 16.7 Hz \\
        Idler shaft: 2nd (output) idler gear & 48 teeth& & 10 Hz & 11.7 Hz & 13.33 Hz & 15 Hz & 16.7 Hz \\
        Output shaft: output pinion & 80 teeth& &6 Hz & 7 Hz & 8 Hz & 9 Hz & 10 Hz \\
        \hline
        \multicolumn{8}{c}{HL = High load condition; LL = Low load condition}
        
      \end{tabularx}
      \label{table: geometry}
  \end{table}

  \begin{table}
  \caption{Health conditions of the double stage reduction gearbox with spur gears set-up}
       \label{Health_conditions_gearbox}
       \begin{tabularx}{\textwidth}{ X@{\hspace{0.1cm}}c@{\hspace{0.1cm}}c@{\hspace{0.1cm}}c
        @{\hspace{0.1cm}}c@{\hspace{0.1cm}}c@{\hspace{0.1cm}}c@{\hspace{0.1cm}}c@{\hspace{0.1cm}}c@{\hspace{0.1cm}}c@{\hspace{0.1cm}}c@{\hspace{0.1cm}}c@{\hspace{0.1cm}}
        c@{\hspace{0.1cm}}c@{\hspace{0.1cm}}c@{\hspace{0.1cm}}c}
          \hline
          \multicolumn{1}{c}{  \makecell{health \\ condition} }&\multicolumn{5}{c}{Gears} & \multicolumn{8}{c}{ Bearings} & \multicolumn{2}{c}{Shaft} \\
          \cline{3-6} \cline{8-13} \cline{15-16}

          \centering{(file name)}& & 32T  &  96T & 48T   & 80T  & & IS:Is & ID:Is & OS:Is &  IS:Os & ID:Os & OS:Os & & Input & Output \\
          \hline
          Healthy (Spur 1)  & & Good   & Good & Good & Good & & Good &  Good & Good & Good & Good & Good & & Good & Good \\
          Broken tooth (Spur 4) & & Good & Good  & Eccentric & Broken & & Ball  & Good & Good & Good & Good & Good & & Good & Good \\
          \hline
          \multicolumn{16}{c}{ ID = Idler shaft; IS = Input shaft; Is = Input side; OS = Output shaft; Os = Output side; T = teeth}
        \end{tabularx}
        \label{table: health}
    \end{table}

\paragraph{Experimental dataset}
We consider accelerations signals from the
double stage reduction gearbox described in~\cite{PHM2009}\footnote{The labelled
experimental dataset we use in this work can be downloaded from
\url{https://data.mendeley.com/datasets/fkp3nn4tp7/1}.} and shown in Fig.
~\ref{schematic_gearbox}. Data were sampled synchronously from accelerometers
mounted on both the input and output shaft retaining plates, and were collected
at 30, 35, 40, 45 and 50 Hz input shaft speed, under high and low loading. The
sampling frequency is set to 66666.67 Hz and the acquisition time is 4 seconds.
Additionally, the runs were repeated twice for each load and speed. Both spur
and helical gears are used separately in the setup of the gearbox to obtain
different datasets. 

In this initial study, we analyze the data from the accelerometer
mounted on the output shaft retaining plate of the spur gearbox. To demonstrate our method's capability of detecting
signs of damage, we compare how it operates under five
different operating conditions given by different input shaft speeds and loading.
The geometry of the gearbox and the operating conditions we consider are
summarized in Table~\ref{table: geometry}. As can be
seen in Table~\ref{table: health}, we consider two different health conditions.
Either all the components of the gearbox are healthy, or there is a broken tooth together with an eccentric gear and a
damaged bearing's ball in the input shaft.

\begin{figure}[h!]
  \begin{subfigure}{0.5\textwidth}
    \includegraphics[width=\textwidth,height=3cm]{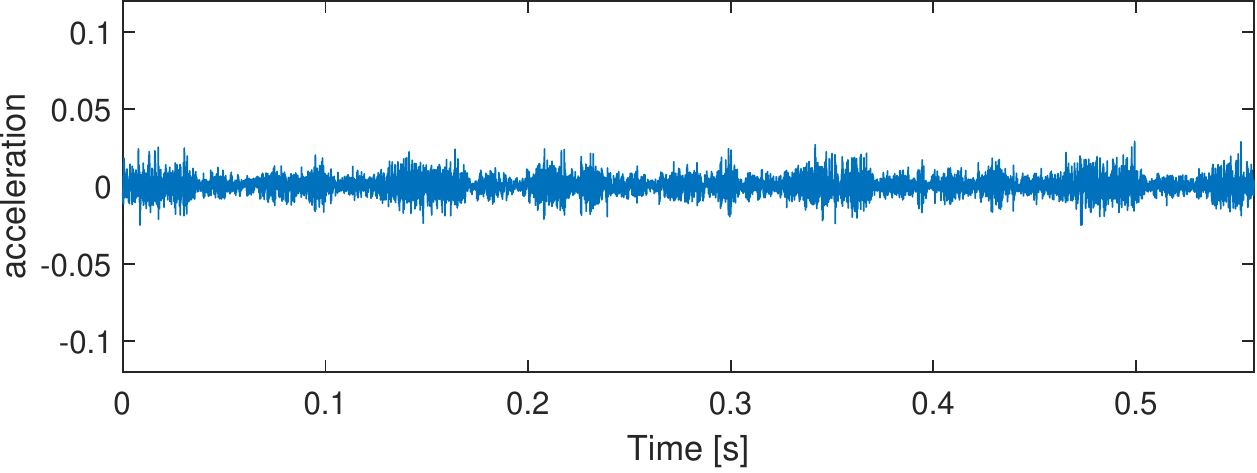}
    \caption{undamaged case}
    \label{experimental_acc_45a}
  \end{subfigure}
  \begin{subfigure}{0.5\textwidth}
    \includegraphics[width=\textwidth,height=3cm]{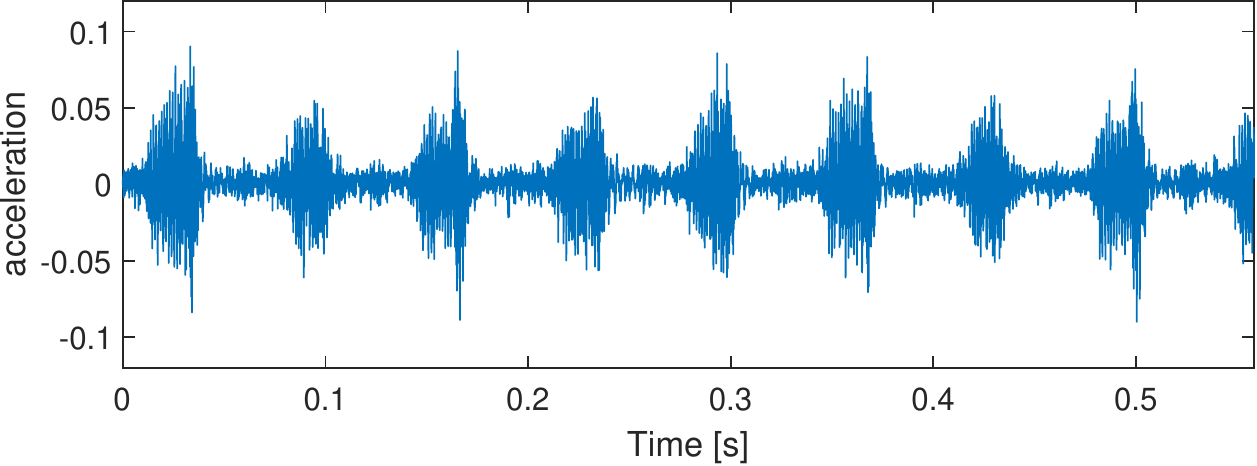}
    \caption{damaged case}
    \label{experimental_acc_45b}
  \end{subfigure}
  \begin{subfigure}{0.5\textwidth}
    \includegraphics[width=\textwidth,height=3cm]{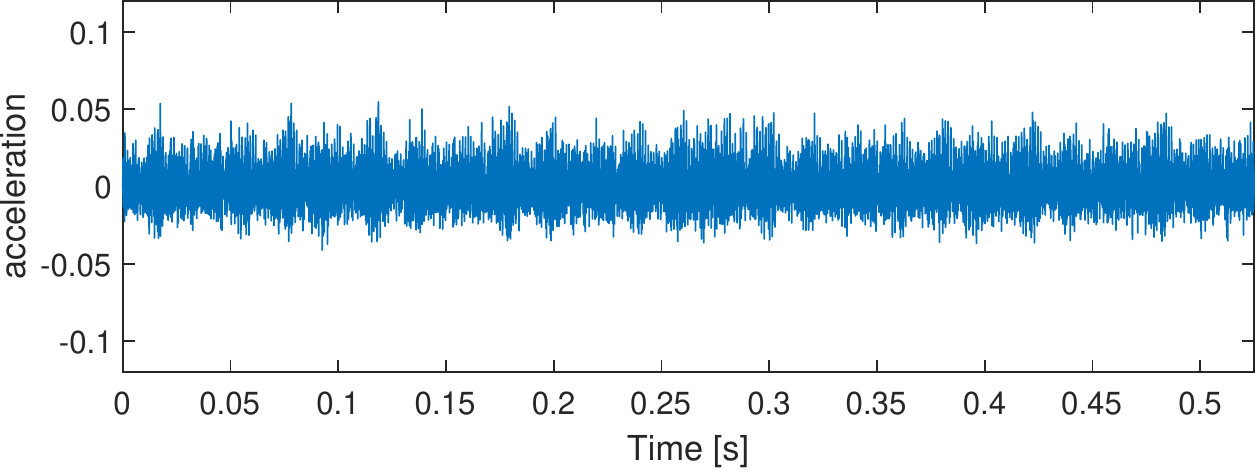}
    \caption{undamaged case}
    \label{experimental_acc_50c}
  \end{subfigure}
  \begin{subfigure}{0.5\textwidth}
    \includegraphics[width=\textwidth,height=3cm]{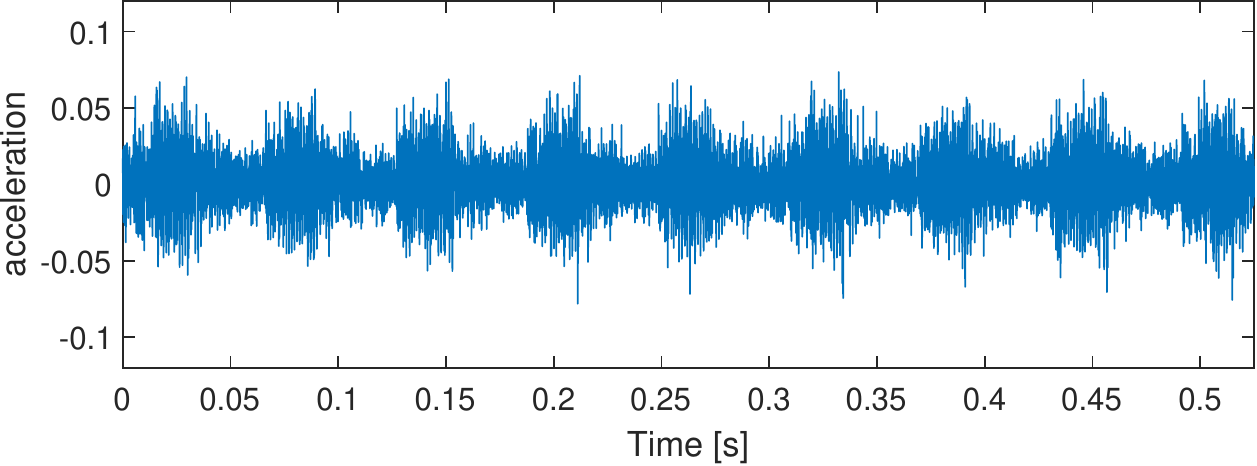}
    \caption{damaged case}
    \label{experimental_acc_50d}
  \end{subfigure}
\caption{Acceleration signals from the double stage reduction gearbox with spur gears set-up operating at different conditions. (a) and (b) 45Hz input shaft speed with high load. (b) and (c) 50Hz input shaft speed with low load}
\label{experimental_acc_45hz_50hz}
\end{figure}

Fig.~\ref{experimental_acc_45hz_50hz} shows representative acceleration signals,
from the experimental dataset, generated with different load and speed
conditions. Comparing the acceleration signals in the damaged and undamaged
cases, the damage effects are plainly visible as sudden periodical magnitude
increases in the signals. This is a characteristic shared by all the
acceleration signals we analyze, independently of the operational condition
considered. Notice that, the x-axis in Fig.~\ref{experimental_acc_45hz_50hz} are
labeled with time and not degrees of rotation as previously done with simulation
data. This is because in the damaged scenarios the eccentric gear on the Idler
shaft's output side may modulate the speed of the gear on the output
shaft~\citep{Mba2017}, which is the one with the broken tooth. Consequently, in
the damaged scenario, the actual angular speed of the gear on the output shaft
may differ from the theoretical angular speed reported in Table~\ref{table:
geometry}. Since we do not know the entity of the damage it is not possible to
determine a priori degrees of rotation of the output gear for a fixed time
interval.

For each operating condition we analyse, we consider a portion of the signal
associated with three rotations of the output shaft in healthy condition plus
the following 15000 snapshots. This will ensure that when
we apply our method with a time delay embedding associated with three rotations
of the healthy output shaft we have 15000 available time-delay snapshots in each
scenario we consider. 

  \subsubsection{mrDMD-based approach for damage detection on experimental data }
 We now apply the proposed procedure to acceleration signals from the double
 stage reduction gearbox with spur gears set-up operating at 30, 35, 40, 45 and
 50 Hz input shaft speed, under high and low loading. Our goal is to show that
 the anomalies caused by the broken tooth and other damages, that affect high
 frequency structures of the signal, can be detected by analyzing the residuals
 even in a scenario in which the acceleration measurements include the dynamics
 of the case and other elements in the transmission path. In each operating
 condition we compute $L=8$ decomposition levels, and consider a delay $d$ associated with three rotations of the
 healthy output shaft. Thus, looking at the residuals instantaneous amplitudes,
 we are analyzing a portion of the signal representing its temporal evolution
 for the time needed from the healthy output shaft for a rotation of
 $1080^{\circ}$. Since each operating condition has a different output shaft
 speed, the delay $d$ will be given by different amounts of snapshots. The
 parameter choices have been summarized  in Appendix B, Table~\ref{table:
 parameters mrDMD}.

\begin{figure}[h]
  \begin{subfigure}{0.5\textwidth}
  \includegraphics[width=\textwidth,height=3cm]{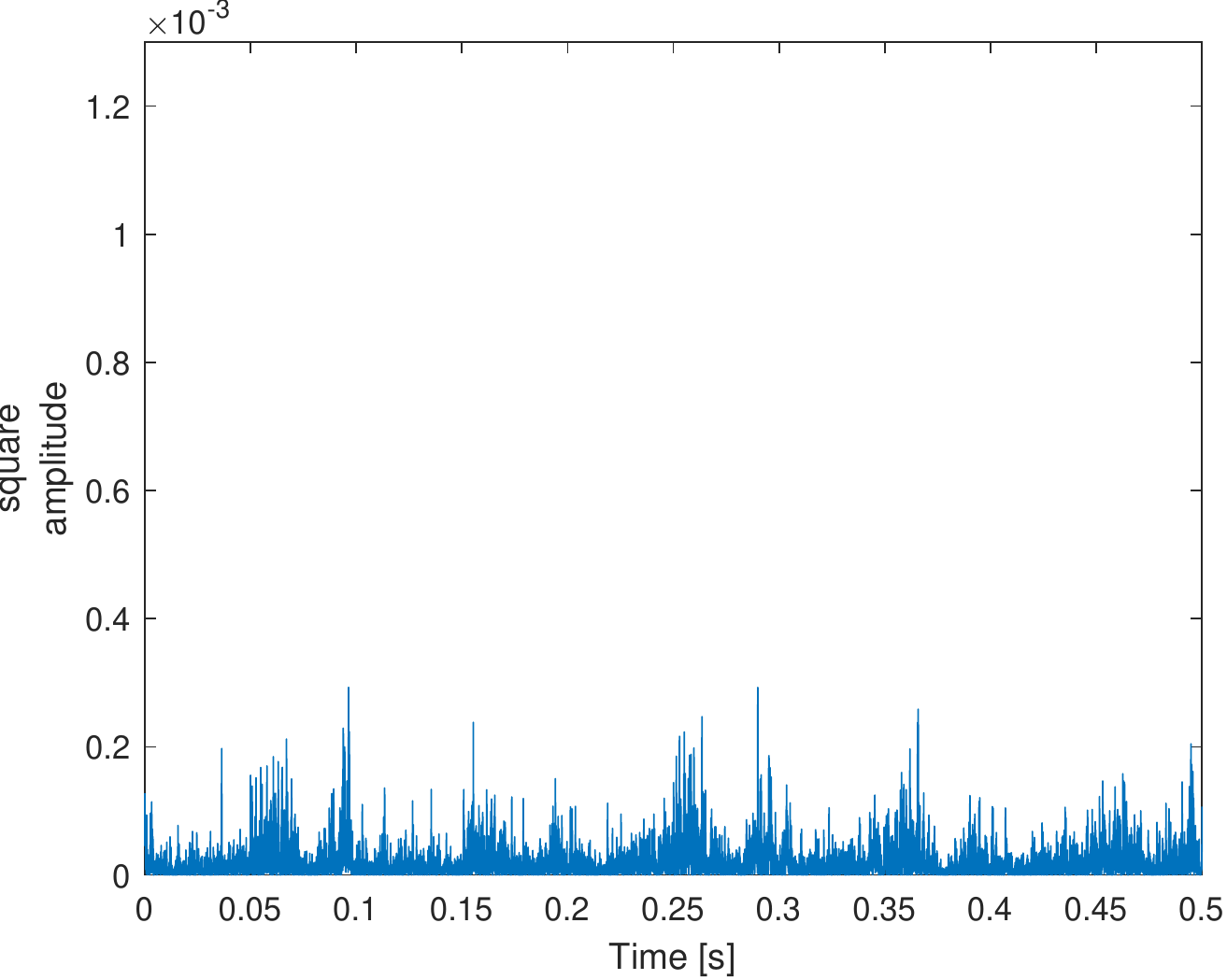}
  \caption{undamaged case}
  \label{experimental_30hz_a}
\end{subfigure}
\begin{subfigure}{0.5\textwidth}
\includegraphics[width=\textwidth,height=3cm]{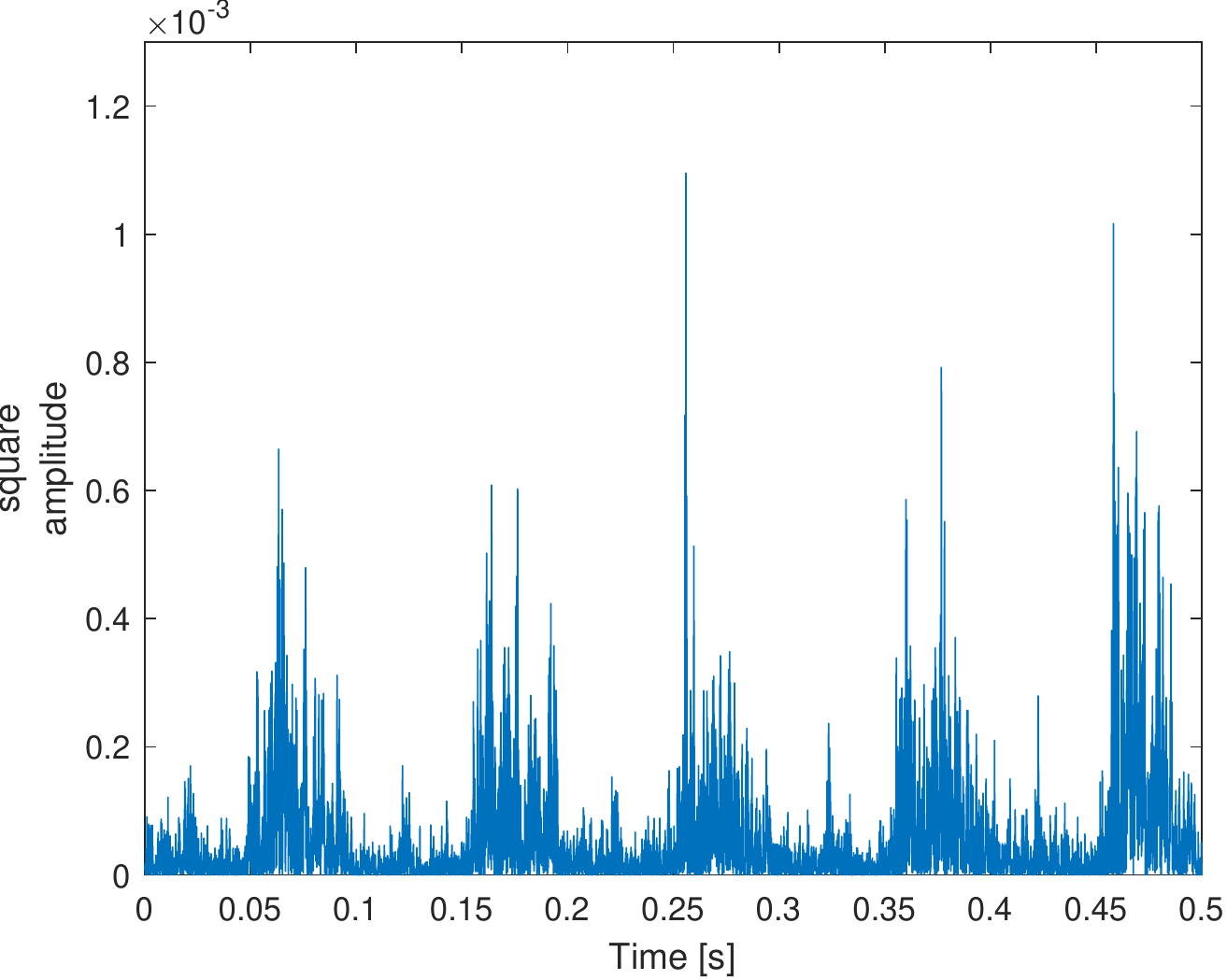}
\caption{damaged case}
\label{experimental_30hz_b}
\end{subfigure}
\caption{Square instantaneous amplitudes of the residuals
$\mathbf{r}_0$ (L=8) obtained from acceleration signals from the
double stage reduction gearbox with spur gears set-up operating at 30Hz input
shaft speed with high load}
\label{experimental_30hz}
\end{figure}

\begin{figure}[h]
  \begin{subfigure}{0.5\textwidth}
  \includegraphics[width=\textwidth,height=3cm]{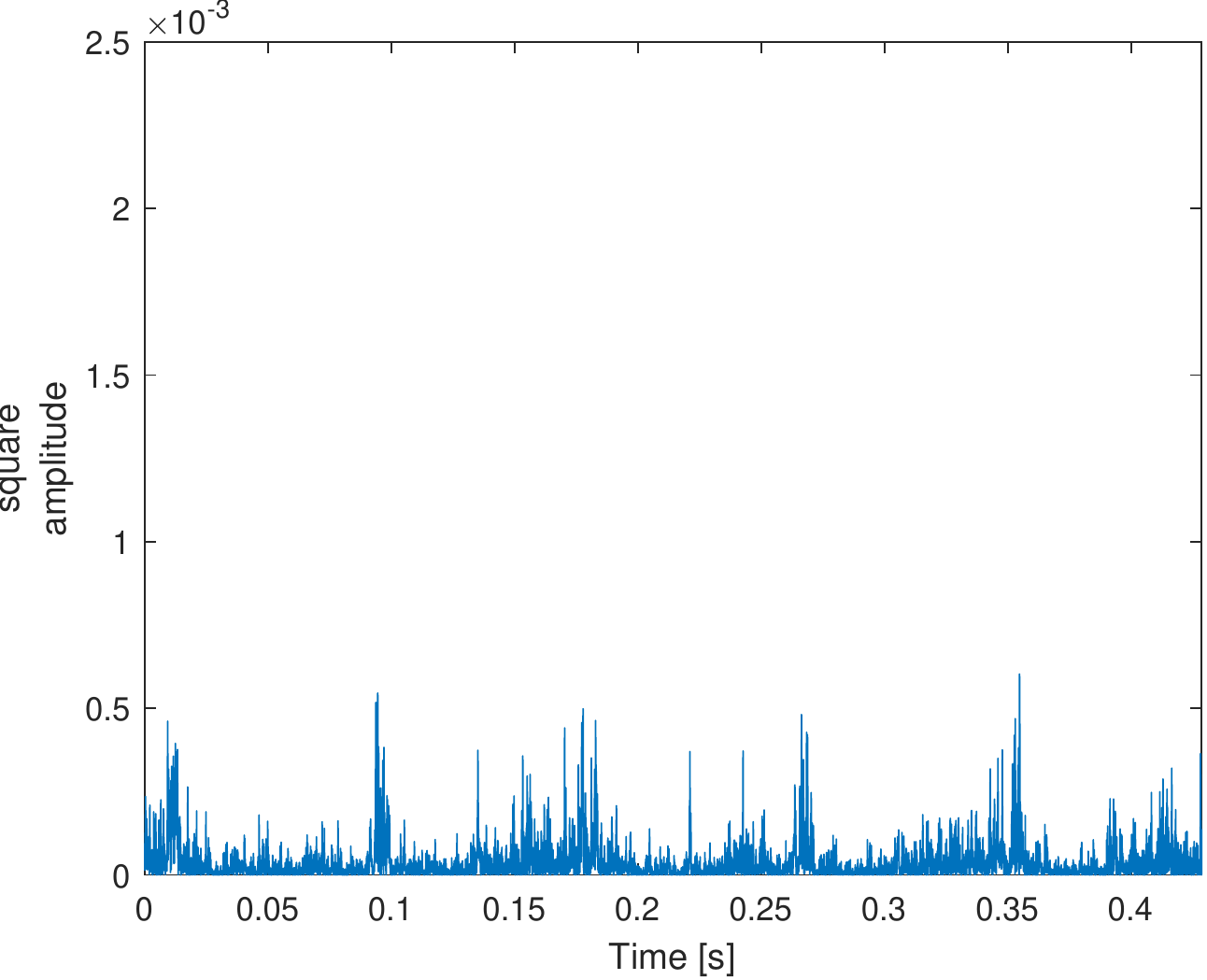}
  \caption{undamaged case}
\end{subfigure}
\begin{subfigure}{0.5\textwidth}
\includegraphics[width=\textwidth,height=3cm]{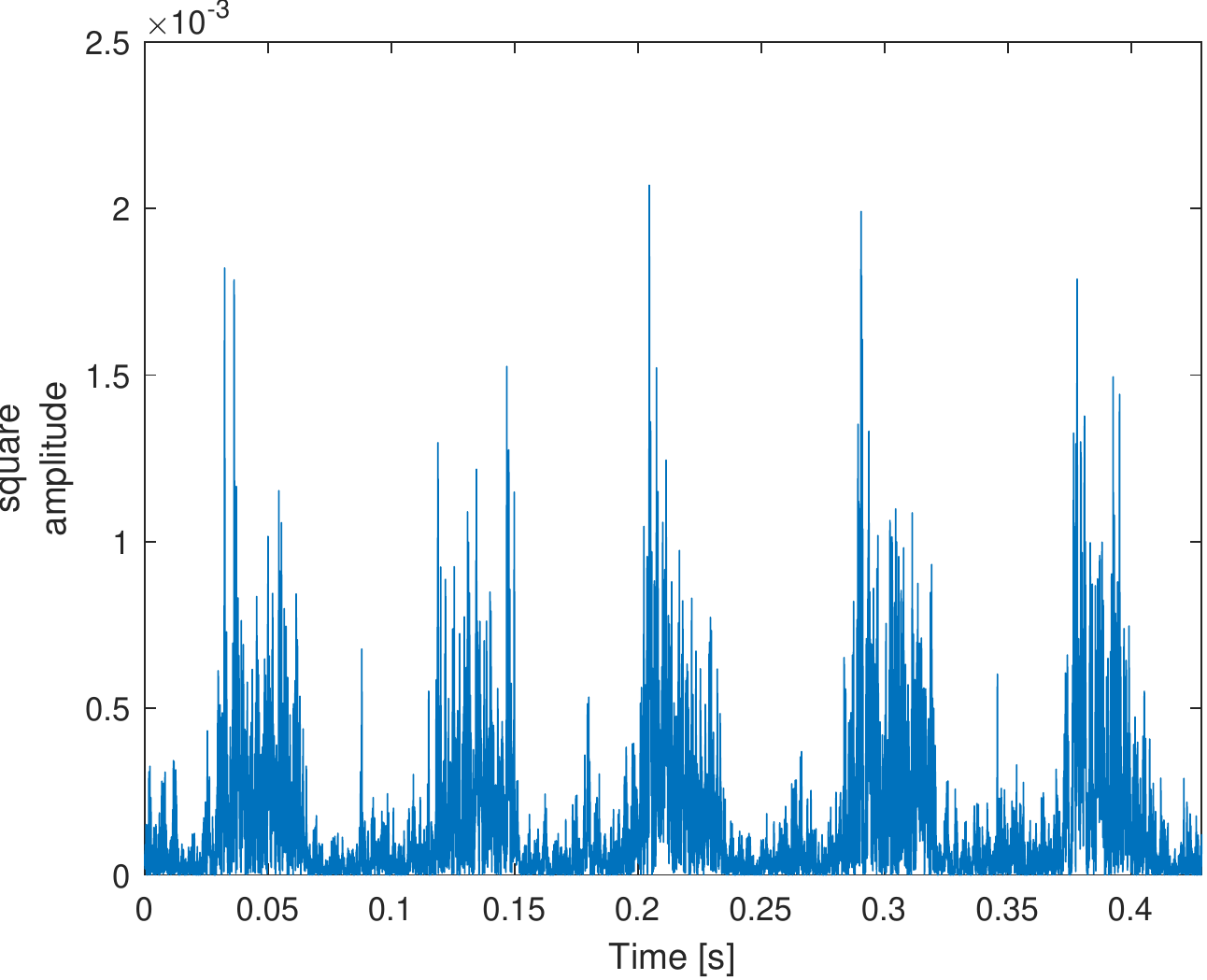}
\caption{damaged case}
\end{subfigure}
\caption{Square instantaneous amplitudes of the residuals
$\mathbf{r}_0$ (L=8) obtained from acceleration signals from the
double stage reduction gearbox with spur gears set-up operating at 35Hz input
shaft speed with high load}
\label{experimental_35hz}
\end{figure}

\begin{figure}[h]
  \begin{subfigure}{0.5\textwidth}
  \includegraphics[width=\textwidth,height=3cm]{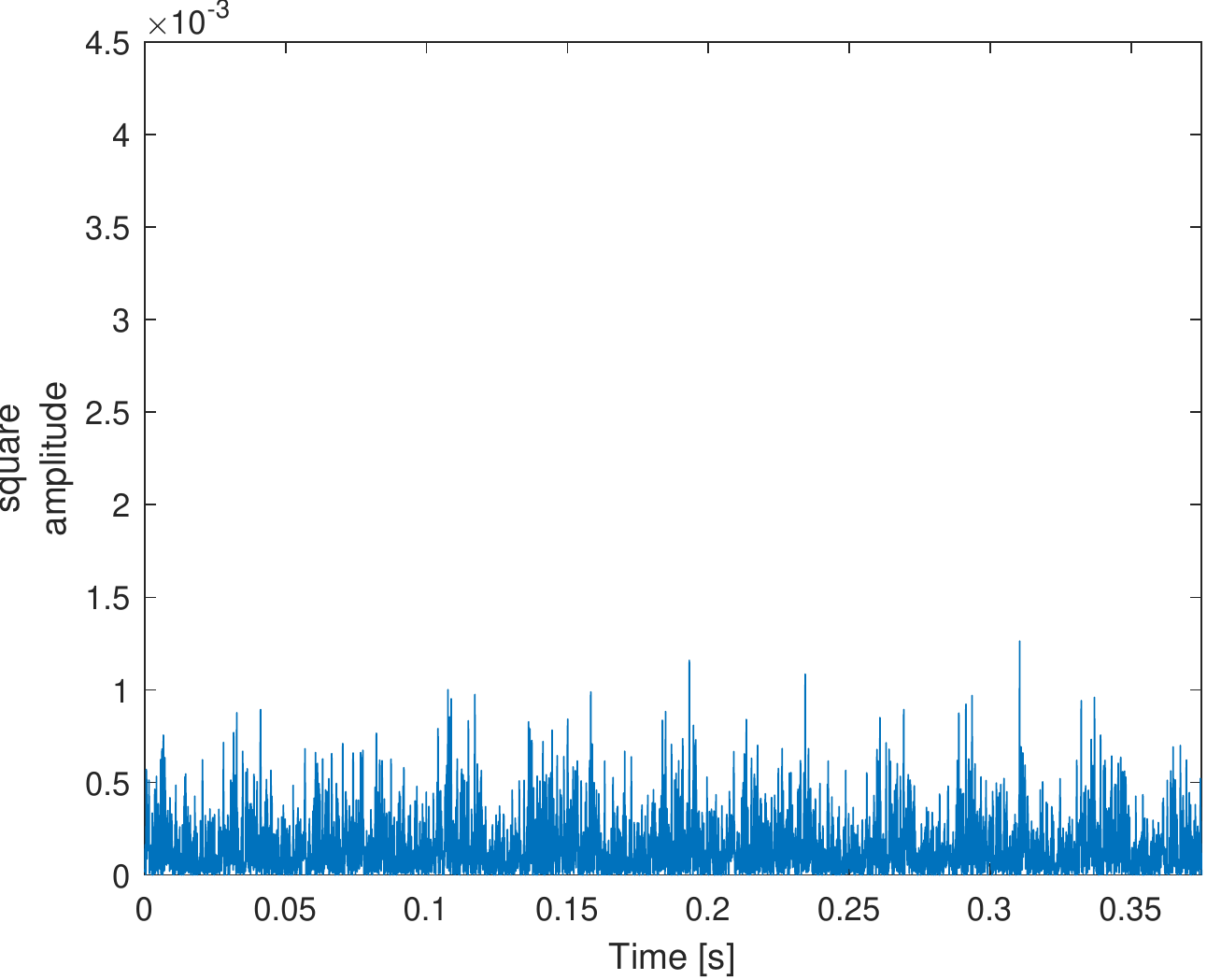}
  \caption{undamaged case}
\end{subfigure}
\begin{subfigure}{0.5\textwidth}
\includegraphics[width=\textwidth,height=3cm]{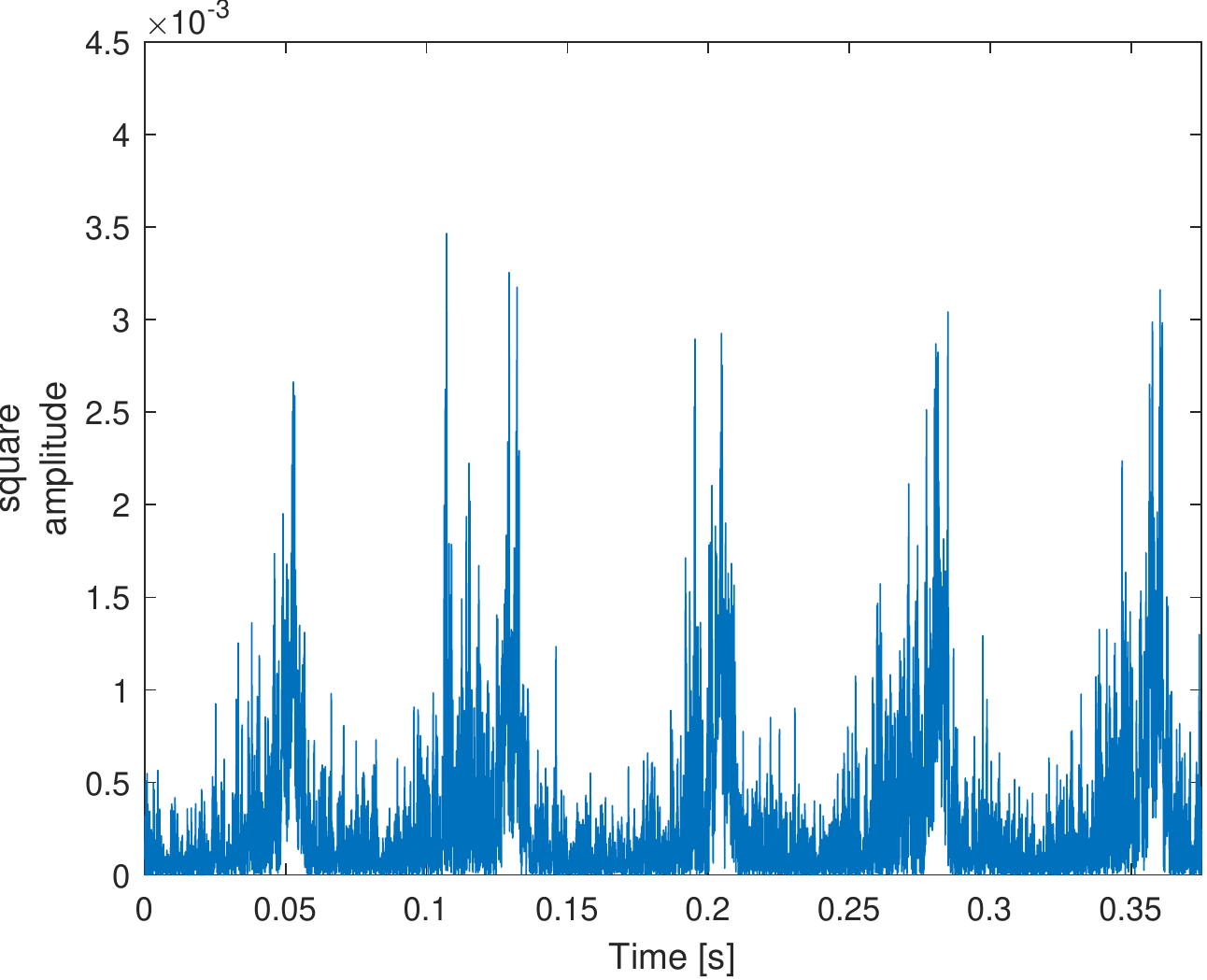}
\caption{damaged case}
\end{subfigure}
\caption{Square instantaneous amplitudes of the residuals
$\mathbf{r}_0$ (L=8) obtained from acceleration signals from the
double stage reduction gearbox with spur gears set-up operating at 40Hz input
shaft speed with low load}
\label{experimental_40hz}
\end{figure}

\begin{figure}[h]
  \begin{subfigure}{0.5\textwidth}
  \includegraphics[width=\textwidth,height=3cm]{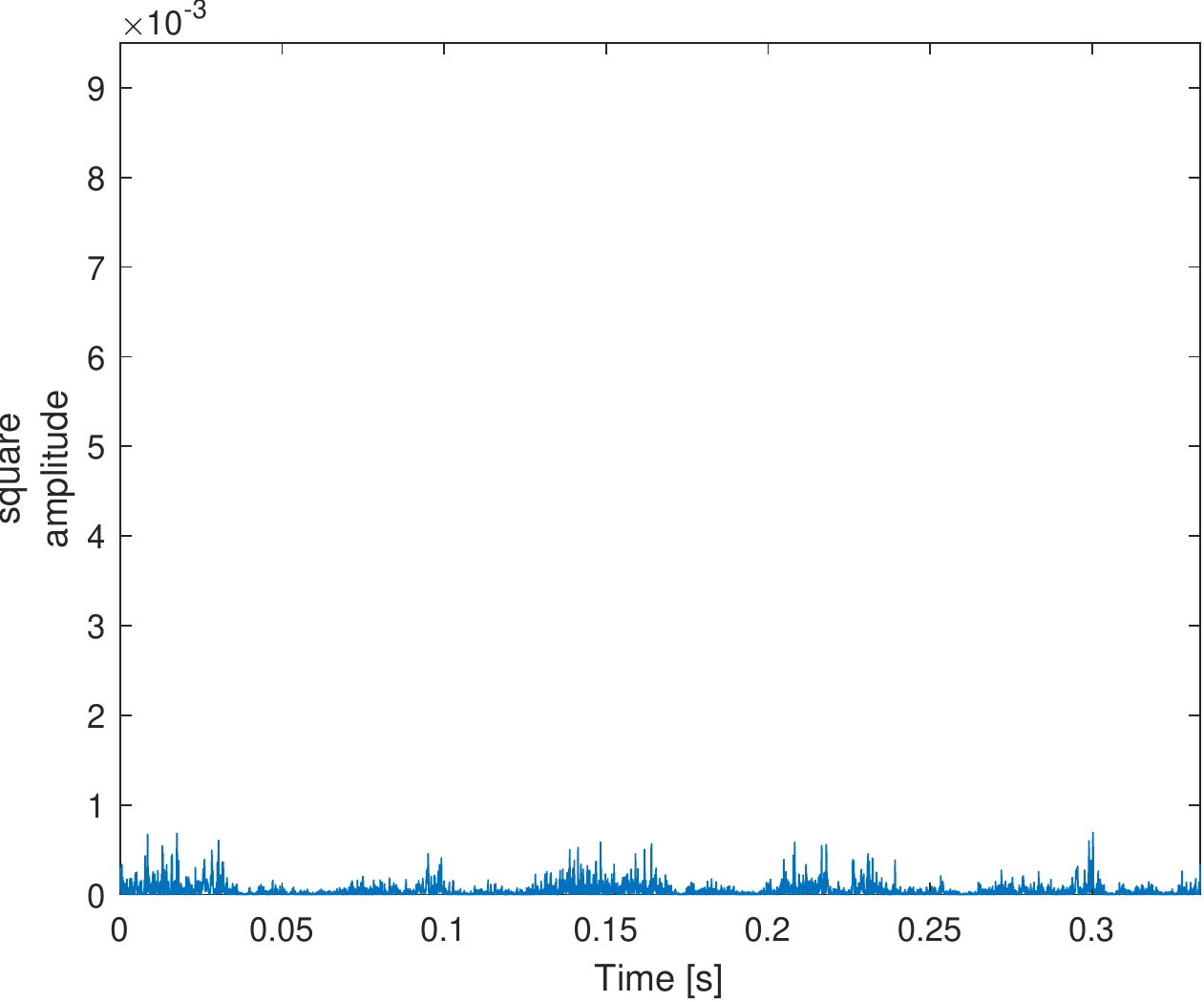}
  \caption{undamaged case}
\end{subfigure}
\begin{subfigure}{0.5\textwidth}
\includegraphics[width=\textwidth,height=3cm]{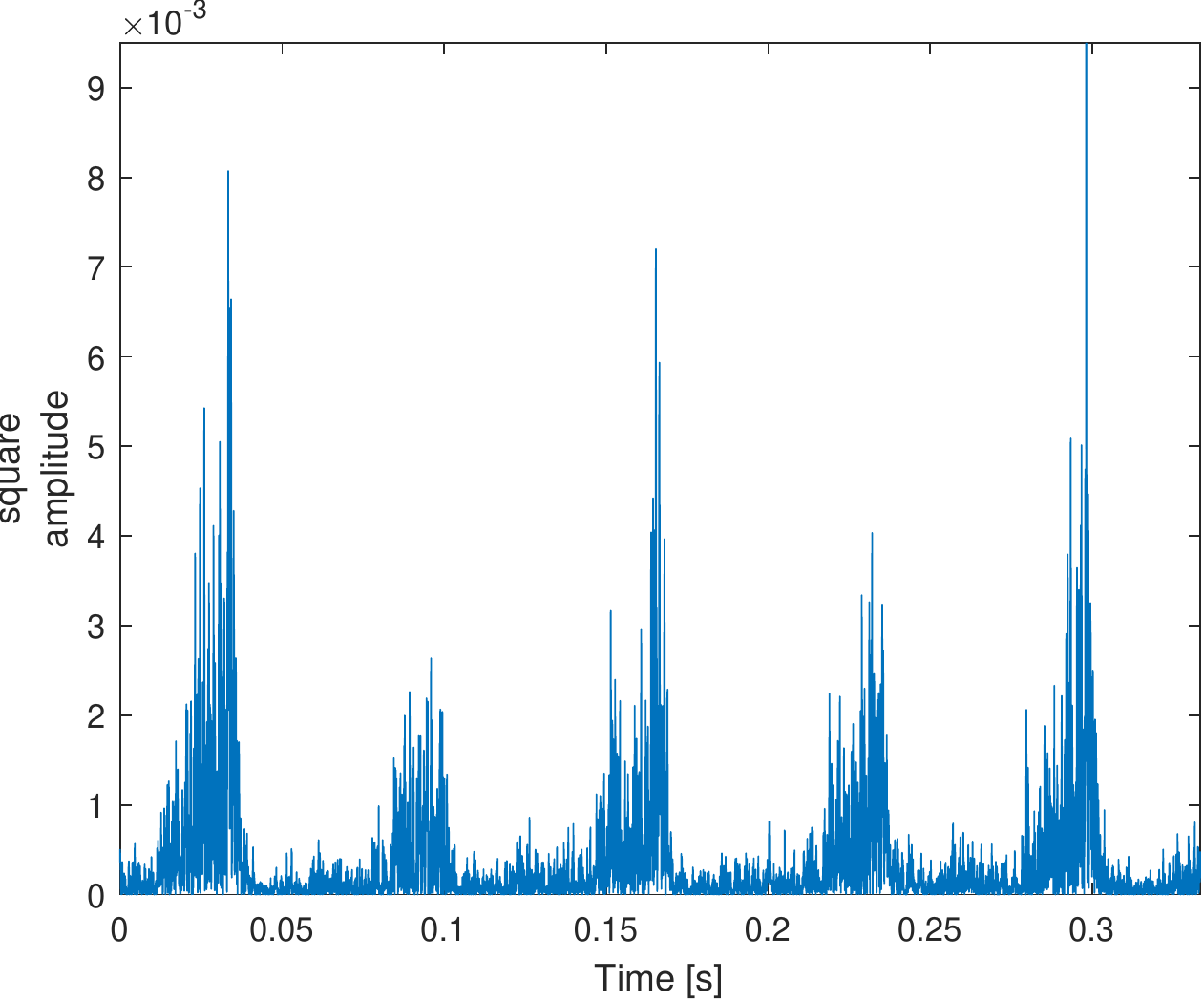}
\caption{damaged case}
\end{subfigure}
\caption{Square instantaneous amplitudes of the residuals
$\mathbf{r}_0$ (L=8) obtained from acceleration signals from the
double stage reduction gearbox with spur gears set-up operating at 45Hz input
shaft speed with high load}
\label{experimental_45hz}
\end{figure}

\begin{figure}[h!]
    \begin{subfigure}{0.5\textwidth}
    \includegraphics[width=\textwidth,height=3cm]{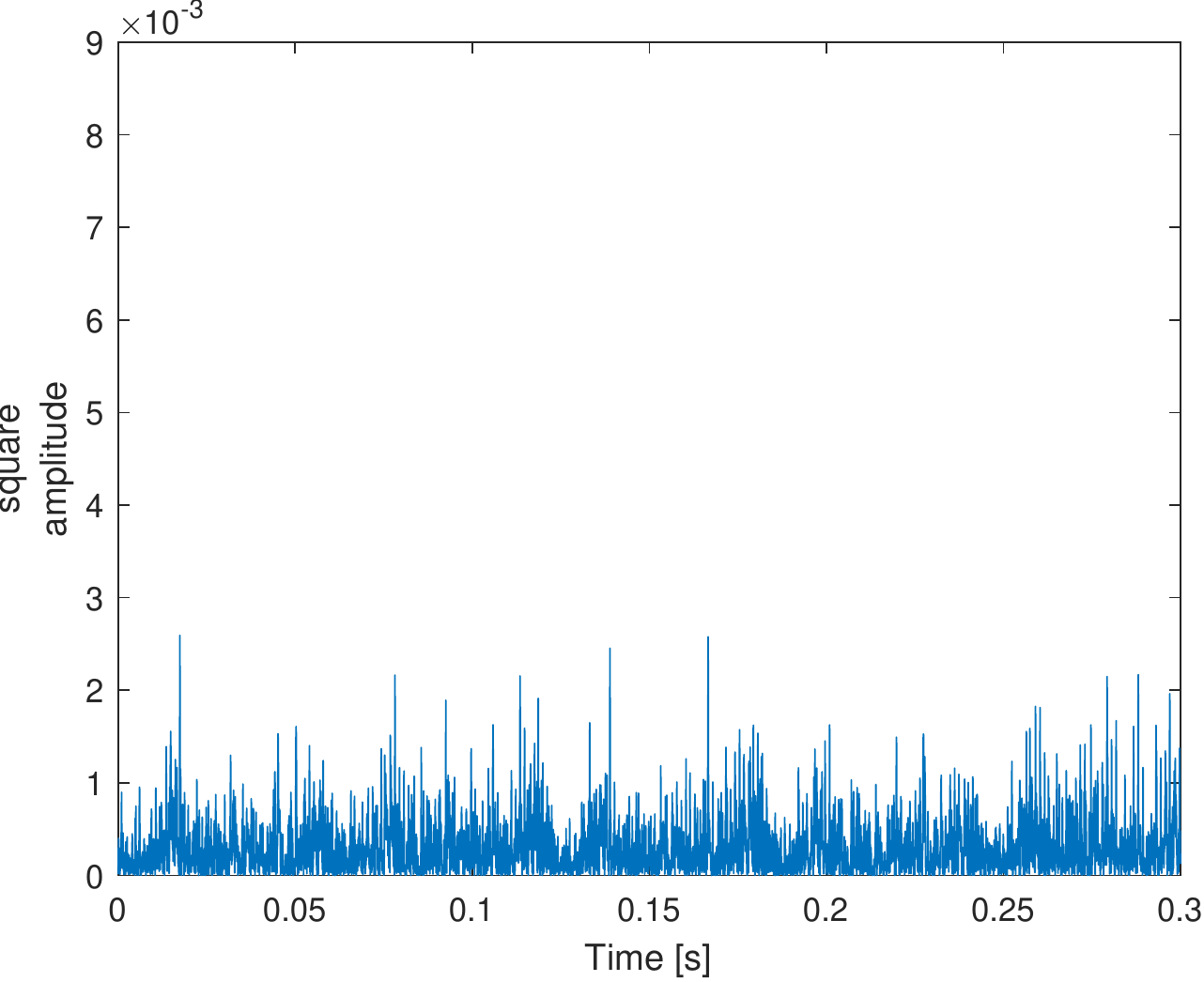}
    \caption{undamaged case}
    \label{experimental_50hz_a}  
  \end{subfigure}
  \begin{subfigure}{0.5\textwidth}
  \includegraphics[width=\textwidth,height=3cm]{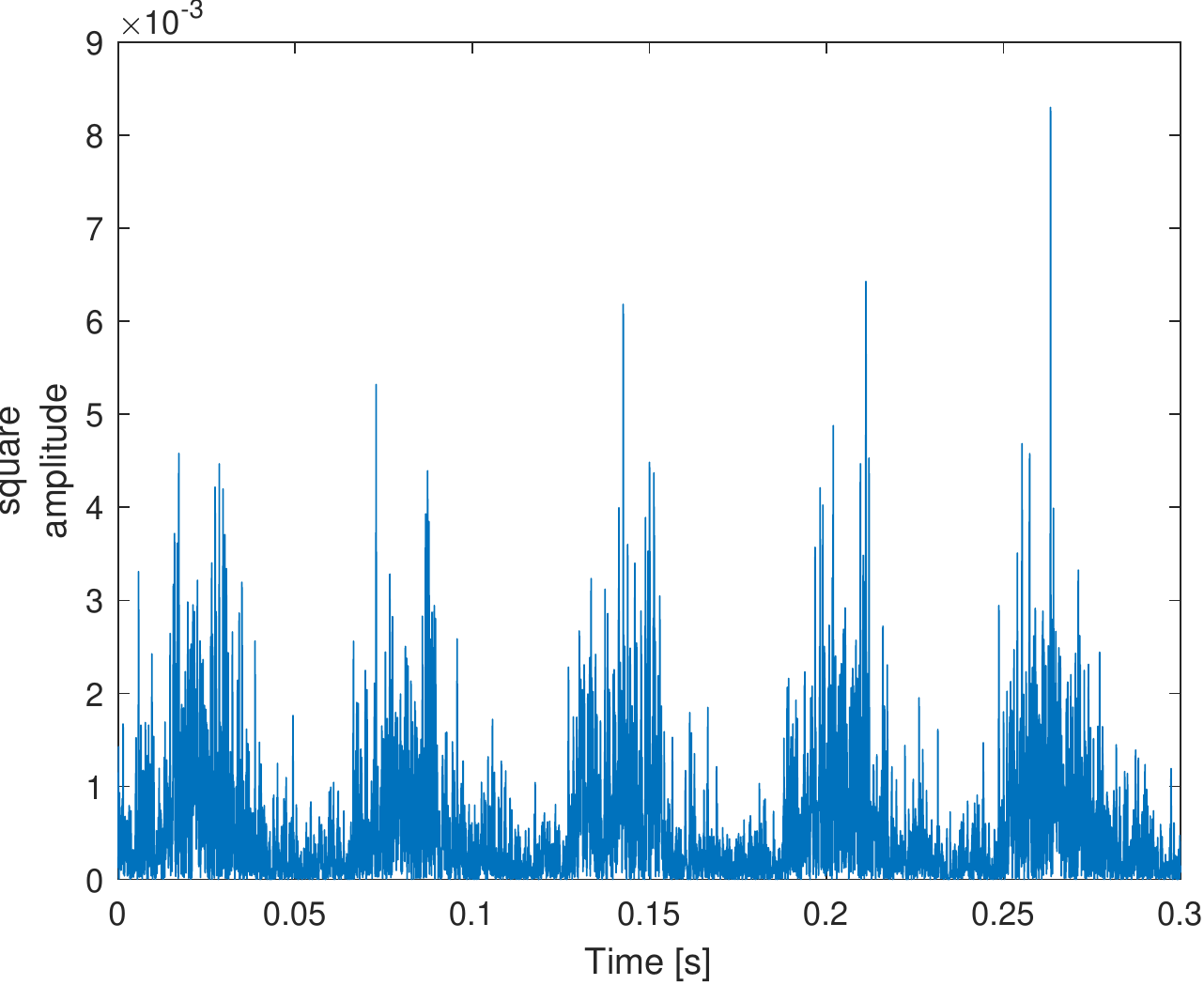}
  \caption{damaged case}
  \label{experimental_50hz_b}  
  \end{subfigure}
  \caption{Square instantaneous amplitudes of the residuals
  $\mathbf{r}_0$ (L=8) obtained from acceleration signals from the
  double stage reduction gearbox with spur gears set-up operating at 50Hz input
  shaft speed with low load}
  \label{experimental_50hz}  
  \end{figure}

\paragraph{MrDMD-residual analysis on Experimental data}
  Fig.~\ref{experimental_30hz}-\ref{experimental_50hz} represent the results of
  the mrDMD-based approach on the experimental data. Specifically, they show the
  square instantaneous amplitudes of the residuals obtained by applying the
  proposed procedure to acceleration signals operating at 30, 35, 40, 45 and
  50 Hz input shaft speed, under high and low loading. 

  As we have previously seen, independently of the operational condition, in the
  acceleration signal associated with the damaged scenario, the damage effects
  are visible as sudden periodical magnitude increases
  (Fig.~\ref{experimental_acc_45hz_50hz}). Interestingly, such behavior is
  reflected in the high-frequency components of the signals represented by the
  residuals, as it can be seen by looking at their instantaneous amplitudes
  (Fig.~\ref{experimental_30hz}-\ref{experimental_50hz}). More precisely,
  comparing the residuals instantaneous amplitudes obtained in the damaged and
  undamaged scenarios in the different operating conditions the effects of
  damage can be clearly detected as periodical sudden increases of magnitude.
  
  From the mrDMD analysis on simulation
  data, we would expect only three main increases of magnitude in the
  instantaneous amplitudes of the residual associated with the damaged cases.
  Specifically, we would expect the magnitude increases to arise when the
  broken tooth part of the output gear interacts with the gears mounted on the
  Idlers shaft. Notice that, the residuals instantaneous amplitudes associated with the
  damaged scenarios show more than three prominent magnitude increases.
  This can be motivated by the presence of the other damages. In particular, the
  eccentric gear in the Idler shaft's output may modulate the speed of the gear
  on the output shaft, which is the one with the broken tooth. Consequently, the
  broken tooth part of the output gear interacts with the eccentric gear more
  than three times during the time-span we are considering, generating more
  magnitude increases.

  Nevertheless, also in the more challenging scenario offered by the experimental
  data, the developed procedure can highlight the effects of damage on the
  signal's high-frequency components, providing a tool that enables us to
  clearly distinguish the healthy scenarios from the damaged ones in various
  operating conditions.

  \paragraph{Modal analysis with mrDMD on experimental data}
  \begin{figure*}
      \includegraphics[width=\textwidth,height=9cm]{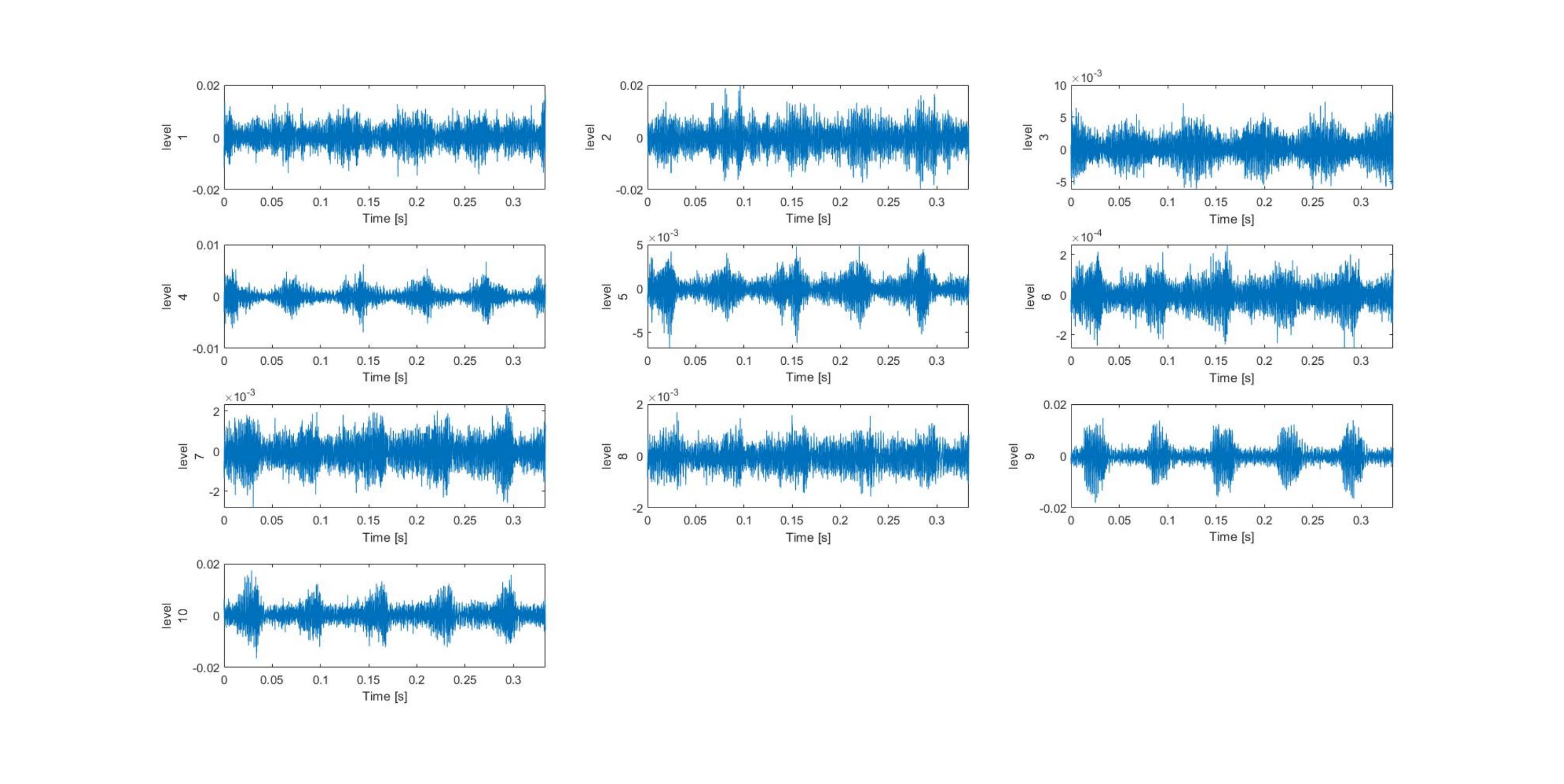}
  \caption {Signal components $\mathbf{p}_0^l$, $l=0,1,2,\dots,10$, computed from acceleration signal of double stage reduction gearbox with spur gears set-up operating at 45Hz input
  shaft speed with high load} 
  \label{mrDMDcrcomponents}
  \end{figure*}


 
  For the simulated data we performed a modal analysis where we in particular investigated the first mode for several decomposition levels. 
  Here, for additional insight, we instead consider at each level `` $l$ " the signal's
  reconstruction, $\mathbf{p}^l_0$ provided by the modes computed at that level, i.e.,
 \begin{equation}
  \mathbf{p}^l_0 = \sum_{k=1}^{m_l}b_{k}^{(l,1)}\pmb{\phi}_{k}^{(l,1)}.
 \end{equation} 
 The subscript `` 0 " is due to the fact that we are assuming we want to analyze
 the first time-delay snapshot, the one computed at time $t_0$, as we did in the
 mrDMD-residual analysis. Consequently, we are interested in the modes computed
 in the first time-bin of each level.

 Each signal component $\mathbf{p}^l_0$ summarizes information of the modes
 computed at level $l$. Note that, each mode is associated with a specific
 frequency. Therefore, each signal component $\mathbf{p}^l_0$ represents
 information related to a specific frequency band. Once the components
 $\mathbf{p}^l_0$ that highlight damage features have been identified, and the
 frequency band or the multiple frequency bands affected by damage have been
 assessed, one can focus only on the signal's structures of interest, either
 through the residual or by considering the components $\mathbf{p}^l_0$ associated with the
 frequecy bands of interest.

\begin{figure*}[h!] 
  \begin{center}
    \includegraphics[width=0.7\linewidth, height=4cm]{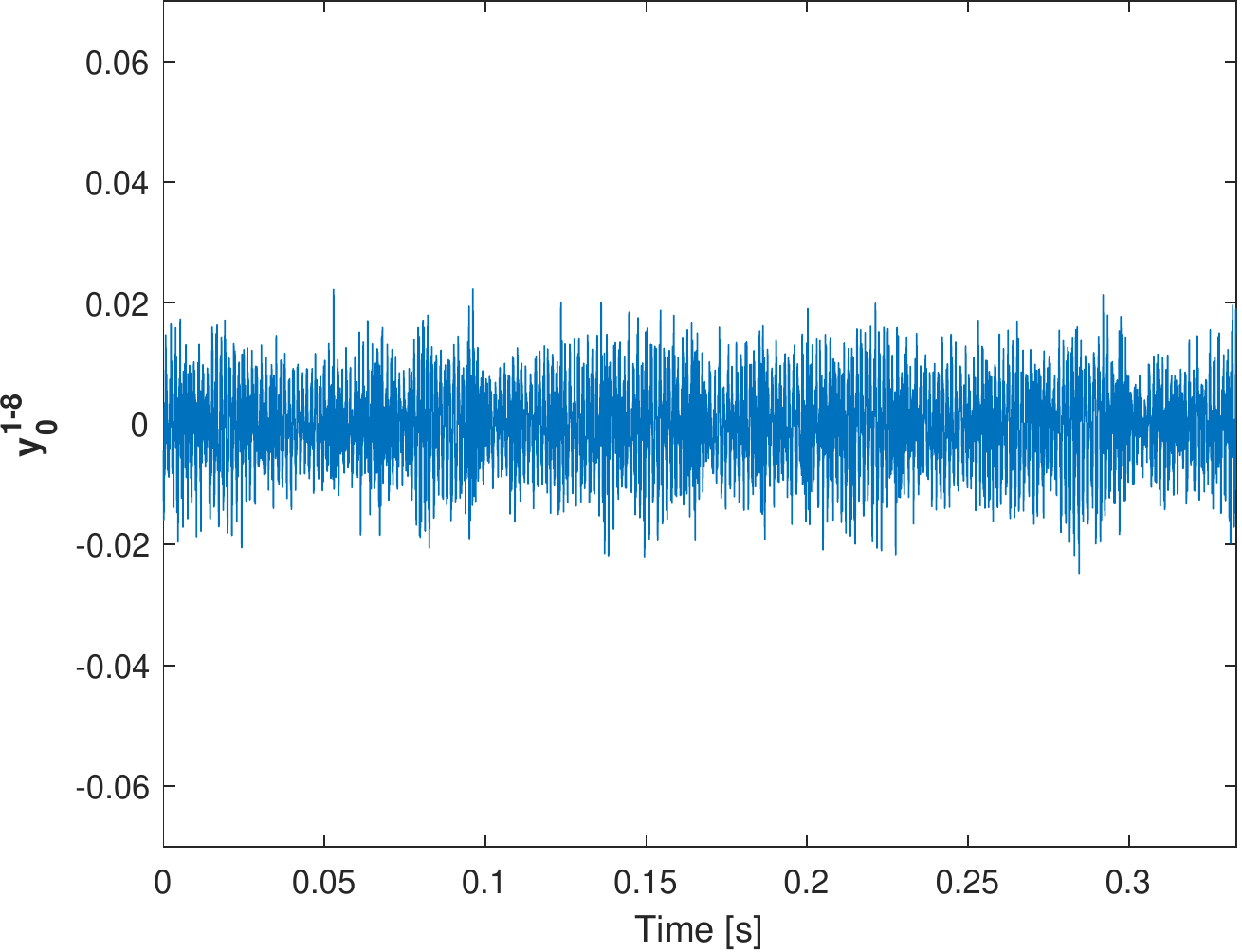}
  \end{center}
 \caption{Partial reconstruction, $\mathbf{y}_0^{1-8}$, of acceleration signal from double stage reduction gearbox with spur gears set-up operating at 45Hz input
 shaft speed with high load. The partial reconstruction is obtained with the first eight signal components, i.e.,
 $\mathbf{y}_0^{1-8} = \sum_{l=1}^{8}\mathbf{p}^l_0$}
\label{partial_rec}
 \end{figure*}

 Let us give an example. Fig.~\ref{mrDMDcrcomponents} illustrates ten components
 $\mathbf{p}^l_0$, $l=0,1,\dots,10,$ obtained applying the mrDMD approach we
 propose to the acceleration signal in Fig.~\ref{experimental_acc_45b} generated
 in high load conditions at 45hz input shaft speed. The delay embedding $d$ is
 the same we previously considered. 
 In the scenario we are considering, the first four signal
 components can be assumed to be mainly associated with the core part of the signal, and the
 effects that damages have on them are negligible. Thus, the relevant signal
 components for our analysis are those computed at higher decomposition levels.
 It can be seen that the characteristics of the last two signal components are
 very different from the previous ones. In particular, both the components computed at the eighth and the
 ninth  decomposition level show a much more prominent spiky behaviour than the
 previous ones, and the magnitude of the spikes is higher as well. Moreover, after the eighth level of
 decomposition, there is a sudden increase in the magnitude of the computed
 signal components. These two
 factors lead us to conclude that the signal components computed at the eighth
 and ninth decomposition levels are associated with features related to damage.
 To provide additional evidence, we analyze the signal reconstruction,
 $\mathbf{y}_0^{1-8}$ (Fig.~\ref{partial_rec}), obtained using the components
 computed at the first eight levels,
it can be seen that the sudden increases of magnitude that characterize the
presence of damage in the original acceleration signal (Fig.~\ref{experimental_acc_45b}) are not represented. This shows us that the features
related to damage are mainly highlighted in higher frequency structures, among
which there are the structures represented by the eighth and ninth signal components, i.e., 
$\mathbf{p}^9_0$ and $\mathbf{p}^{10}_0$. Recall that we computed eight decomposition levels in our mrDMD-residual
 analysis on experimental data. Thus, the signal components $\mathbf{p}^9_0$, and $\mathbf{p}^{10}_0$, were included
 in the residual. 

\section{Conclusion}
In this work, we proposed a numerical procedure for damage detection, which
exploits DMD's strengths of being equation-free and data-driven. From our
results with simulation data, we observed that performing damage detection with
Fourier analysis, TSA or EMD can be very challenging in a scenario that includes
varying load conditions. We also saw that both, our mrDMD-based method and the
EMD strategy, decompose the original signal into modes that can be used to
extract characteristics of the signal associated with different frequencies. In
particular, both methods can highlight those high-frequency structures affected
by wind turbulence and change in stiffness caused by the cracked tooth and that
can be effectively used to identify damages. On the one hand, we saw that the
IMFs, produced by EMD, represent structures in the signal that are not
associated with specific frequencies but rather with frequency bands. Moreover,
the information contained in an IMF can be related to a too broad frequency
band, or the same frequency information can be contained in different IMFs.
These facts affect the interpretability of the IMFs and the effectiveness of EMD
to perform fault detection because the information related to damage is not
localized. On the other hand, mrDMD associates each mode with a specific
frequency. Exploiting this fact, the proposed numerical procedure was able to
identify those structures in the signal where features related to damage were
highlighted. In particular, differences in the residuals' magnitude
(Fig.~\ref{residuals1} and Fig.~\ref{residuals2}) as well as in their
instantaneous amplitude (Fig.~\ref{amplitude_residuals5} and Fig.
~\ref{amplitude_residuals13}), allowed the identification of the cracked tooth
visually, independently of the wind condition considered. We also saw, from the
results with experimental data, that while the proposed mrDMD-based strategy is influenced
by accelerations signals affected by additional vibration sources, it still is
a feasible approach for damage detection.
 
The first strength of the proposed analysis is that, as EMD, it does not
consider the signals' spectra, avoiding all the issues related to the
non-deterministic time variation of the signals spectral properties caused by
the varying load condition. The other advantage is that the mrDMD algorithm is
not affected by the mode mixing problem, as it is for EMD-based strategies.
Moreover, with the strategy here proposed, information related to the effects of
the damage in the signal is localized in the residual.

Note that, the strategy we propose can be further improved to perform an
effective analysis in more challenging scenarios. Specifically, several DMD
improvements could be applied in the second step of Algorithm~\ref{alg:PNP}
~\citep{ optDMD, debiasingDMD, leastDMD}. A particularly beneficial DMD
improvement in this context was developed in~\cite{debiasingDMD}. It consists of
building unbiased and noise-aware DMD by explicitly accounting for noise in the
signal. Such modification could allow our method to distinguish between the
noise introduced by wind turbulence, transmission path vibrations and damage effects more easily. Furthermore,
additional improvements can be introduced working on the indicator function in
(\ref{indicator}). In general, it is possible to introduce various functional
forms for the indicator function that can be used advantageously. For instance,
one could consider the indicator function to take the form of different wavelet
bases i.e. Coiflet, Haar, Fejér-Korovkin, etc. 

In conclusion, this work shows the potentialities of a DMD-based strategy to
perform anomaly detection, analyzing simulated signals representing the vibration response
of a gearbox under varying load condition and experimental data generated with various speed
and load operating conditions. The results we reported here tell us that
this DMD-based strategy promises to provide high quality results by overcoming
spectral and mode mixing related problems in a scenario that includes the
effects of wind turbulence. 
Further research on mrDMD in the context of fault detection and condition monitoring
of wind turbines, and related machinery, is warranted.

\bibliographystyle{apalike}
    
\bibliography{second_paper_arxiv}

\begin{appendices} 
\section*{Appendices}
\section{}
\begin{figure}[h!]
    \begin{subfigure}{0.5\textwidth}
    \includegraphics[width=0.9\linewidth, height=6cm]{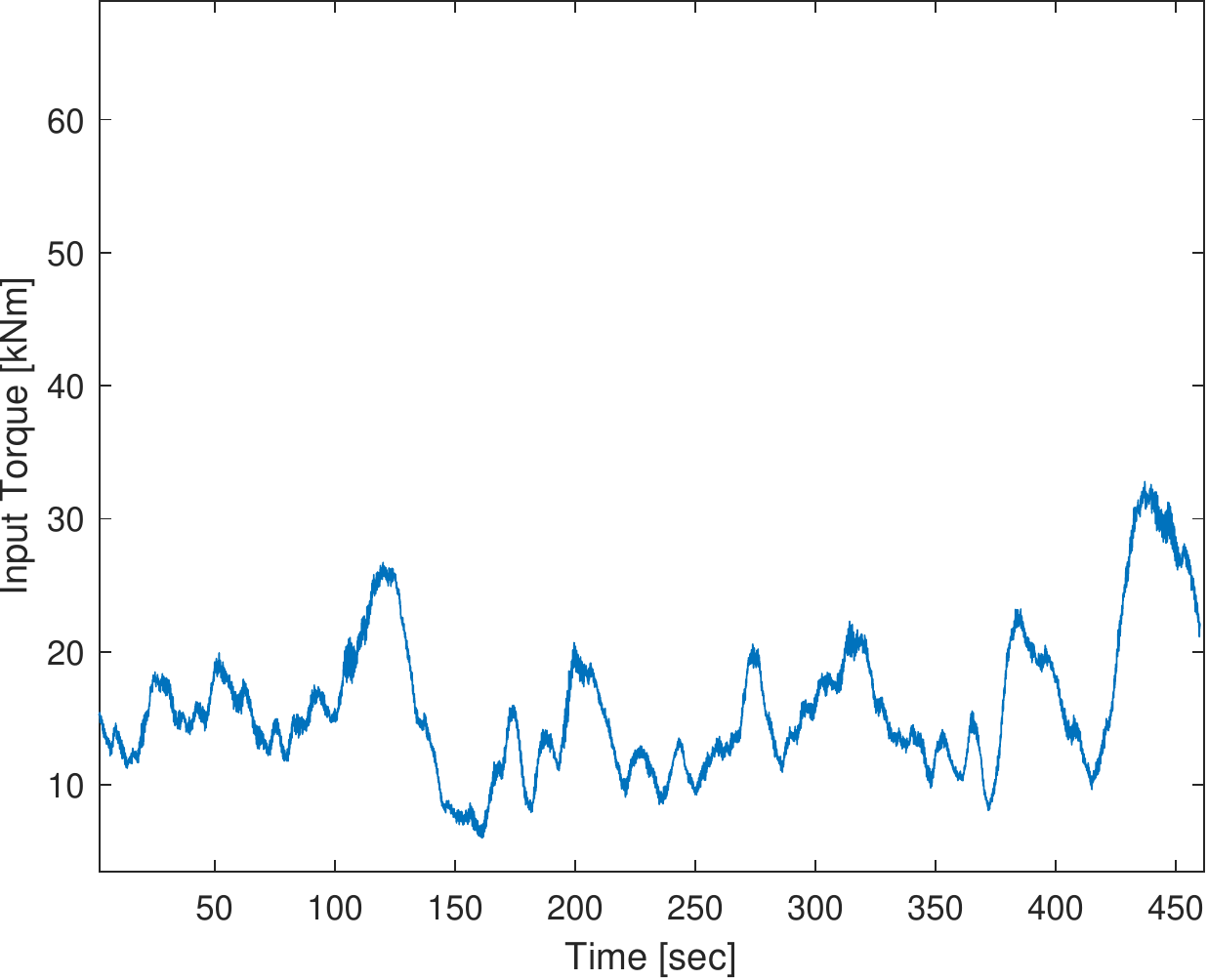} 
    \caption{wind speed 5 m/s}
    \label{input_torque5}
    \end{subfigure}
    \begin{subfigure}{0.5\textwidth}
    \includegraphics[width=0.9\linewidth, height=6cm]{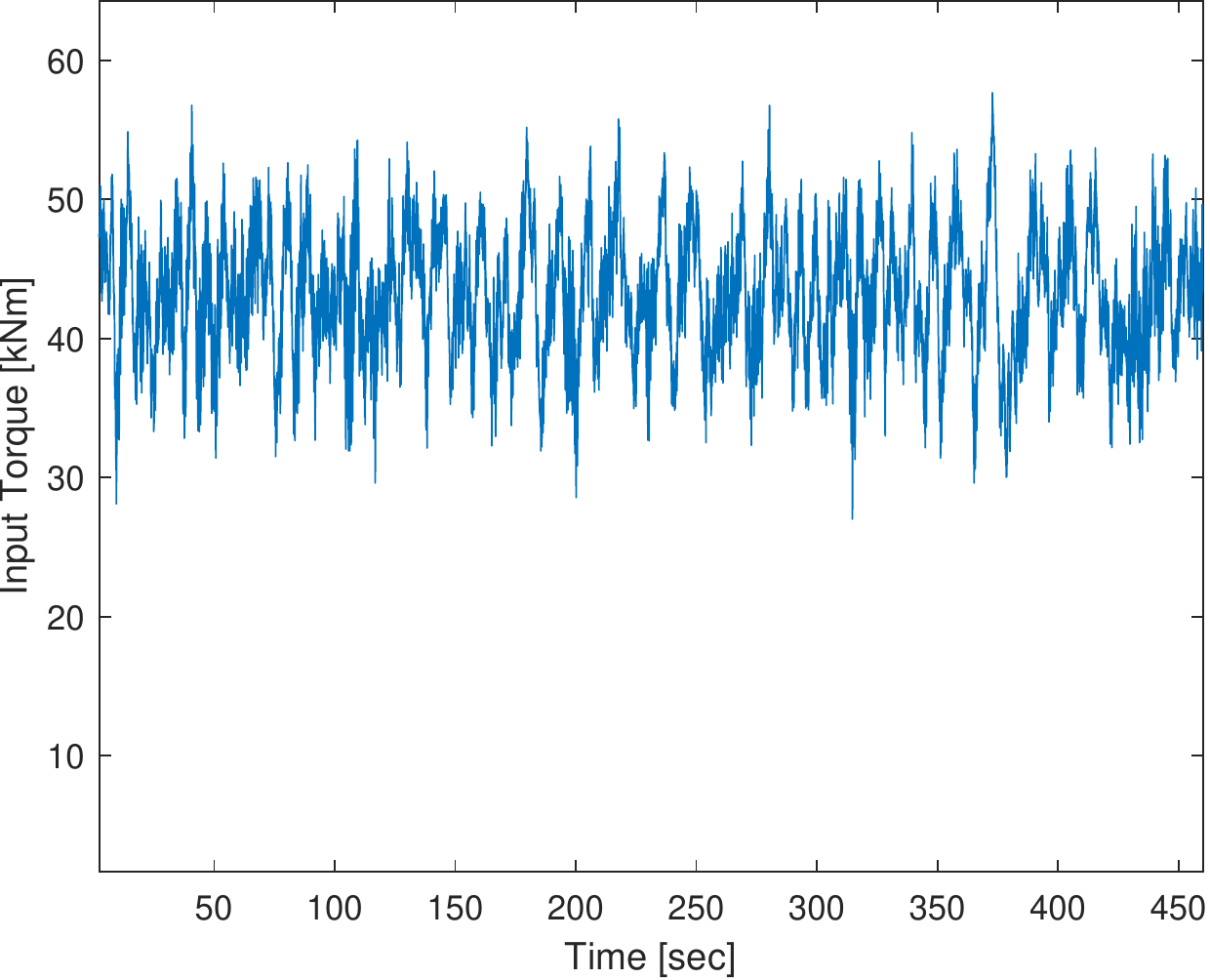}
    \caption{wind speed 13 m/s}
    \label{input_torque13}
    \end{subfigure}
    
    \caption{FAST simulation of the high speed shaft torque for different turbulent wind condition}
    \label{input_torques}
    \end{figure}

The aim is to build an equation of motion for the gearbox model considered in Section~\ref{gearboxmodelll} that takes into account
the rotational movements of the two gears. To do that, such an equation needs to include
the two main factors that affect the gears' movements, which are: the input torque that acts on
the driving gear and the effects of the interaction between the teeth of the two
gears. The area  where the interaction between the two gears takes place is
called \textit{gear mesh}.
\begin{figure}[h!]
  \begin{center}
    \includegraphics[scale=0.7]{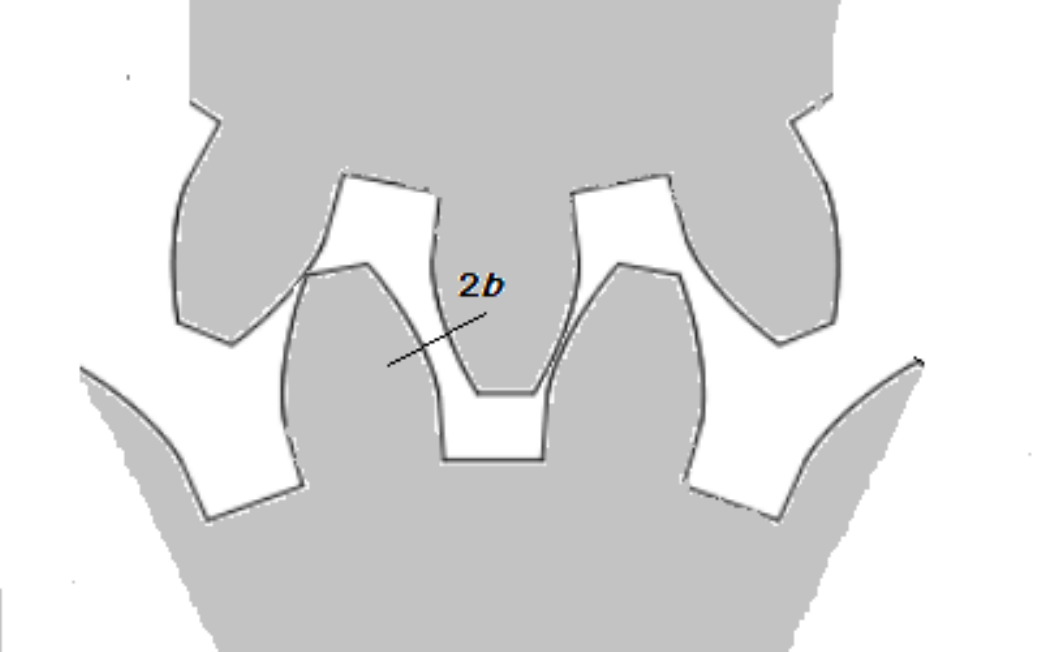}
    \caption{The backlash between gears}
     \label{backlash_pic}
    \end{center}
  \end{figure}
  \paragraph{The Gear mesh}
  To include the effects of the interaction between the teeth of the two gears we
  include into the equation of motion we are going to build the backlash,
  the time-varying mesh stiffness function $\bar{K}(\bar{t})$, the
  static transmission error $\bar{e}(\bar{t})$, and a linear viscous
  dumping parameter $\bar{C}$. The backlash is a clearance or lost motion of the
  gears caused by gaps between the teeth (Fig.~\ref{backlash_pic}). This phenomenon is incorporated in the model via the backlash function $B(\bar{x}(\bar{t}))$, where $\bar{x}(\bar{t})$
  is displacement function that will be introduced shortly. The time-varying mesh
  stiffness function, $\bar{K}(\bar{t})$, describes the varying stiffness of the
  teeth that are in the gear mesh, and is caused by the transition from single to
  double and double to single of pairs of teeth in contact. Finally, the static
  transmission error $\bar{e}(\bar{t})$, is caused by geometrical errors of the
  teeth profile and represents the difference between the actual position of the
  driven gear and the position it would occupy if the gears' edges were
  manufactured perfectly.
  The mesh stiffness and static transmission error functions are assumed to be
  periodic functions of time and, in the theoretical setting we are considering,
  can be expressed in the following Fourier form:
  \begin{equation}
  \bar{K}(\bar{t})= \bar{K}\left(\bar{t}+ \frac{2 \pi}{\bar{\Omega}_{mesh}}\right)= \bar{K}_m - \sum_{j=1}^{\infty}\bar{K}_j \cos\left(j \bar{\Omega}_{mesh}\bar{t} \right),
  \end{equation}
  
  \begin{equation}
  \bar{e}(\bar{t})= \bar{e}\left(\bar{t}+ \frac{2 \pi}{\bar{\Omega}_{mesh}}\right)=  \sum_{j=1}^{\infty}\bar{e}_j \cos\left(j \bar{\Omega}_{mesh}\bar{t} \right),
  \end{equation}
  where $\bar{K}_m$, $\bar{K}_j$ and $\bar{e}_j$ are constant Fourier coefficients
  of the respective signals~\citep{Kah}. $\bar{\Omega}_{mesh}$ is the meshing frequency defined as
  $\bar{\Omega}_{mesh}= n_1\bar{\Omega}_1= n_2 \bar{\Omega}_2,$ where $n_1$ and $n_2$ stand for the number of teeth of the first (driving)  and
  second (driven) gears  and $	\bar{\Omega}_i$ is the rotating frequency of the
  $i$-th gear.
  \paragraph{The torque.}
  In the scenario we consider, the input torque $\bar{T}_1(\bar{t})$ that
  ignites the movement and  keeps the gears moving, is not constant, but it
  rather fluctuates due to the fluctuations of the wind. Thus, the input torque
  is given by a constant part $\bar{T}_{1m}(\bar{t})$ and a fluctuating part
  $\bar{T}_{1var}$, i.e.,
  $\bar{T}_1(\bar{t})=\bar{T}_{1m}(\bar{t})+\bar{T}_{1var}(\bar{t})$. The output
  torque $\bar{T}_2(\bar{t})$ is considered to be constant, i.e.,
  $\bar{T}_2(\bar{t})= \bar{T}_{2m}(\bar{t})$, with $\bar{T}_{2m}$ being the
  mean output torque. In the simulation, the component of the input torque
  associated with the fluctuations of the wind, $\bar{T}_{1var}$, is derived from the torque
  computed via the FAST~\citep{FAST} design code, i.e., $\bar{T}_{1var}(\bar{t})=
  \bar{T}_{FAST}-\bar{T}_{FASTmean}$. The torques generated with FAST code, i.e.
  $\bar{T}_{FAST}(\bar{t})$, that have been used in this work to simulate the effect of the wind
  turbulence in different wind conditions, are shown in  Fig.~\ref{input_torques}. 
  \subsection{Equation of motion}
  In order to describe the model with a single equation, we consider the following
  coordinate 
  \begin{equation}
  \bar{x}(\bar{t})= r_1 \phi_1(\bar{t})- r_2 \phi_2(\bar{t})- \bar{e}(\bar{t}),
  \end{equation}
  where $\phi_i(\bar{t})$ stands for the torsional displacement of the $i$-th gear
  while $r_i$ is its radius. 

  The coordinate $\bar{x}(\bar{t})$ is given by the difference between the dynamic
  transmission error and the static transmission error. Moreover, through
  $\bar{x}(\bar{t})$ the model's equation of motion yields the following formulation~\citep{anto, Kah}
  \begin{equation}
  \label{motion_dimension}
  m_c\ddot{\bar{x}}(\bar{t}) + \bar{C}\dot{\bar{x}}(\bar{t})+ \bar{K}(\bar{t})B(\bar{x}(\bar{t}))= \bar{F}_m+ \bar{F}_{te}(\bar{t})+ \bar{F}_{var}(\bar{t}),
  \end{equation}
  where
  \begin{equation}
  m_c= \frac{I_1 I_2}{I_1 r_2^2+ I_2 r_1^2}, \quad \bar{F}_m= \frac
  {\bar{T}_{1m}}{r_1}=\frac{\bar{T}_{2m}}{r_2} ,
  \end{equation}
  \begin{equation}
  \bar{F}_{te}(\bar{t})=-m_c\ddot{\bar{e}}(\bar{t})= -\sum_{j=1}^{\infty} F_{tej} (j \bar{\Omega}_{mesh})^2 \cos(j  \bar{\Omega}_{mesh}\bar{t}),
  \end{equation}
  with $I_i$ mass moment of inertia of the $i$-th gear and $F_{tej}= \bar{e}_j$.
  Moreover, concerning the quantities that involve the torque obtained via the
  FAST code:
  \begin{equation}
  \label{fvar}
  \bar{F}_{var}(\bar{t})= \frac{\bar{T}_{1var}(\bar{t})}{Nr_1},
  \end{equation}
   where $N$ is a normalization constant. The normalization constant is included
  because the FAST simulations correspond to a different physical system. Thus,
  it is needed to adapt the magnitude of the fluctuations to our setting.
  $\bar{F}_m$ represents the mean force excitations while $\bar{F}_{te}$ and
  $\bar{F}_{var}$ pertain to internal excitations related to the static
  transmission error and external excitations related to wind turbulence,
  respectively. Lastly, the backlash function $B(\bar{x}(\bar{t}))$ is defined
  as
  \begin{equation}
     B(\bar{x}(\bar{t})) = \left\{\begin{array}{lr}
          \bar{x}(\bar{t})-b_g, & \text{for }  \bar{x}(\bar{t})\geq b_g\\
          0, & \text{for } -b_g< \bar{x}(\bar{t})< b_g\\
          \bar{x}(\bar{t})+b_g, & \text{for } \bar{x}(\bar{t})\leq -b_g
          \end{array}\right\}, 
  \end{equation}
  where $2b_g$ represents the total gear backlash. The backlash function controls
  the contact between teeth and incorporates in the model the fact that
  occasionally contact is lost (Fig.~\ref{backlash_pic}). 
  \subsubsection{Dimensionless equation of motion.}
  The equation of motion (\ref{motion_dimension}) can be written in a
  dimensionless form~\citep{anto, Kah} by setting :
  $x(\bar{t})=\frac{\bar{x}(\bar{t})}{b_g}$, $w_n=\sqrt{\frac{\bar{K}_m}{m_c}}$,
  $z= \frac{\bar{C}}{2\sqrt{m_c \bar{K}_m}}$, $K(\bar{t})=
  \frac{\bar{K}(\bar{t})}{\bar{K}_m}$, $F_{te}(\bar{t})=
  \frac{\bar{F}_{te}(\bar{t})}{m_cb_gw_n^2}=-m_c\frac{\ddot{\bar{e}}(\bar{t})}{m_cb_gw_n^2}$,
  $F_{var}(\bar{t})= \frac{\bar{F}_{var}(\bar{t})}{m_cb_gw_n^2}$ and $t=w_n
  \bar{t}$. The meshing frequency can be written in a nondimensional form as well, i.e.,
  $ \Omega_{mesh}=\frac{\bar{\Omega}_{mesh}}{w_n}$. The nondimensional form of the
  equation of motion (\ref{motion_dimension}) is
  \begin{equation}
  \ddot{x}(t)+2z\dot{x}(t)+K(t)B(x(t))=F_m+F_{te}(t)+F_{var}(t)
  \end{equation}
  where the dimensionless backlash function is defined in (\ref{backlash_function}).
  \begin{table}
    \caption{Simulation parameters}
       \label{simulation_parameters}
        \begin{tabularx}{\columnwidth}{X|X|X|X}
            \hline
            Parameter & Value  &  Parameter  & Value   \\
            \hline
     $I_{1,2}$ & $0.00115(\text{Kg}\;\text{m}^2)$   &$F_m$        & 0.1 \\
           $m_c$     & $0.23\text{(Kg)}$  &  $F_{te1}$        & 0.01 \\
            $r_{1,2}$        & 0.0 5(m)  &  $F_{te2}$        & 0.004 \\
            number of teeth         & 16 &   $F_{te3}$        & 0.002 \\
            $\bar{K}_{m}$    & $3.8\cdot 10^8
            \left(\frac{\text{N}}{\text{m}}\right)$      & $K_1$        & 0.2 \\
             $b_{g}$        & $0.1 \cdot 10^{-3}(m)$  & $K_2$        & 0.1 \\
             $z$        & 0.05  & $K_3$        & 0.05 \\
             $\Omega_{mesh}$        & 0.5  & $N$          & 10 \\
             \hline
        \end{tabularx}
        \label{table: simulation parameters}
    \end{table}

    \subsection{ Parameter settings}
    The parameters for the experiments reported in this paper are given in
    Table~\ref{table: simulation parameters}. We also performed additional
    experiments on signal simulated modifying the parameters related to the
    internal excitations and changing the severity of the damage.
    Table~\ref{table: simulation parameters2} shows the three different
    parameter settings we used in our simulation model to generate acceleration
    signals affected by different gearbox internal excitations. The parameters
    have been modified in a combined fashion, and their variation affects the
    simulated signals locally but not their global structure. For what concerns
    the severity of the damage, in the simulations analyzed and illustrated in
    this work, the crack of a gear's tooth is modelled by decreasing the
    dimensionless mesh stiffness function by 13 \% of its nominal stiffness.
    Over our experiments, we considered two additional severity conditions.
    Specifically,  we reduced the mesh stiffness function by 10 \% and 16 \%.
    Thus we considered two additional scenarios where the severity of the damage
    either increases or decreases. Overall we considered three conditions for
    the internal excitation parameters, three damage severities and two wind
    conditions. The combination of these allowed us to simulate eighteen
    different scenarios on which to test our
    method
    \begin{table}[h!]
      \caption{Parameter settings for gearbox internal excitations}
          \begin{tabularx}{\columnwidth}{X|X|X|X}
              \hline
              Parameter & first setting  &  second setting  & third setting   \\
              \hline
               $F_{te1}$        & 0.01 &   0.02  & 0.003\\
               $F_{te2}$        & 0.004 &  0.001 & 0.002 \\
               $F_{te3}$        & 0.002 &  0.009 & 0.004 \\
              \hline
          \end{tabularx}
          \label{table: simulation parameters2}
      \end{table}
      \section{}
    \begin{table}[H]
      \caption{Parameter choices for the different operating conditions of the experimental dataset}
          \begin{tabularx}{\textwidth}{ccccc}
            \hline
            \multicolumn{1}{c}{\makecell{Operating conditions \\ (Angular speed $-$ Load)} }&\multicolumn{1}{c}{\makecell{Delay `` $d$ " \\ (Snaphots $-$ Time [s])}}& \multicolumn{1}{c}{\makecell{Length acceleration signal \\ (Snaphots $-$ Time [s])}} & \multicolumn{1}{c}{Decomposition levels} \\
            \hline
            30Hz $-$ High &   33333 $-$ 0.5000 s &   48333 $-$ 0.7250 s  & 8 \\
            35Hz $-$ High &   28572 $-$ 0.4286 s &  43572 $-$ 0.6536 s  & 8 \\
            40Hz $-$ Low  &   24999 $-$ 0.3750 s &   39999 $-$  0.6000 s  & 8\\
            45Hz $-$ High &   22221 $-$  0.3333 s  &   37221 $-$ 0.5583 s  & 8 \\
            500Hz $-$ Low &   19998 $-$ 0.3000 s  &   34998 $-$ 0.5250 s  & 8\\
            \hline
          \end{tabularx}
          \label{table: parameters mrDMD}
      \end{table}

\end{appendices}

\end{document}